\documentclass {article}
\usepackage[left = 28mm, right = 28mm]{geometry}
\usepackage{graphicx} % Required for inserting images
\usepackage{lmodern}

\DeclareMathAlphabet{\mathdutchcal}{U}{dutchcal}{m}{n}

\usepackage{microtype}
\usepackage{amsmath,amsfonts,amssymb,amsthm}
\usepackage{mathrsfs}
\usepackage{graphicx}
\usepackage{tensor}
\usepackage{braket}
\usepackage{mleftright} \mleftright
\usepackage{array}
\usepackage{booktabs}
\usepackage{mathtools} % for mathclap
\usepackage{mdframed}
\usepackage{cite}

\usepackage{float}

\usepackage{eucal}
\usepackage{slashed}
\usepackage[normalem]{ulem}
\usepackage{xcolor}
\usepackage{subcaption}
\usepackage[
singlelinecheck=false 
]{caption}
\usepackage{lmodern}
\usepackage{subcaption}
\usepackage{mathrsfs}
\usepackage{tikz}
\usepackage{upgreek}
\usepackage{rotating}

\usetikzlibrary{patterns}
\usetikzlibrary{decorations.pathmorphing}
\usepackage{subcaption}
\newcommand*{\rmII}{\bs{I}\hspace{-0.25em}\bs{I}}
\newcommand*{\rmIII}{\bs{I}\hspace{-0.25em}\bs{I}\hspace{-0.25em}\bs{I}}
\usepackage{hyperref}
\usepackage{color}\usepackage{graphicx}
\newcommand*{\scri}{\ensuremath{\mathscr{I}}}
\newcommand{\ft}[2]{{\textstyle\frac{#1}{#2}}}
\newcommand{\RomanNumb}[1]{\MakeUppercase{\romannumeral #1}}
\usepackage{subcaption}
\usepackage{braket}

\usepackage{color}
\usepackage{colortbl}
\usepackage{multirow}
\usepackage{hhline}
\usepackage{adjustbox}
\usepackage{calc}
\newcommand{\bs}[1]{\boldsymbol #1 }
\newcommand{\dsig}{{ \uuline{\gamma} }}
\usepackage{hyperref}
\newcommand{\LR}{{L_{\text{\tiny {(R)}}}}}
\newcommand{\LL}{{L_{\text{\tiny {(L)}}}}}
\newcommand{\epsR}{{\epsilon_{\text{\tiny {(R)}}}}}
\newcommand{\epsRsq}{{\epsilon^2_{\text{\tiny {(R)}}}}}
\newcommand{\epsL}{{\epsilon_{\text{\tiny {(L)}}}}}
\newcommand{\xiL}{{\xi_{\text{\tiny {(L)}}}}}
\newcommand{\xiR}{{\xi_{\text{\tiny {(R)}}}}}
\newcommand{\vxiR}{{{\vec\xi}_{\text{\tiny {(R)}}}}}
\newcommand{\vxiL}{{{\vec\xi}_{\text{\tiny {(L)}}}}}
\newcommand{\vnH }{{\vec n}_{\text{\tiny ($\mathcal H$)}}}

\newcommand{\kCTCp}{{\kappa^+_{\text{\tiny CTC}}}}
\newcommand{\kCTCm}{{\kappa^-_{\text{\tiny CTC}}}}
\newcommand{\kCTCpm}{{\kappa^\pm_{\text{\tiny CTC}}}}
\usepackage{ulem}
\newcommand{\uform}{\bgroup\markoverwith
{\textcolor{black} {\rule[-0.6ex]{1pt}{0.8pt}}}\ULon}
\newcommand{\oneform}[1]{\uform #1\vphantom{#1}}

\newcommand{\Tr}{{\mbox{\rm Tr}}{}} 
 
\newcommand{\Ad}{{\mbox{Ad}}{}} 
\numberwithin{equation}{section}

\newcommand{\fin}{\end{document}}
\newenvironment{equationInItem}{\abovedisplayskip=-\baselineskip \abovedisplayshortskip=-\baselineskip}{}
\usepackage{mciteplus}
\usepackage{soul}
\usepackage{calligra}

\newcommand{\proj}{\mathdutchcal{p}}
\begin{document}
%\linenumbers
\begin{titlepage}
\hfill {ULB-TH/24-10}
			\begin{center}
				
				\hfill \\
				\hfill \\
				\vskip 0.75in
				
				{\Large \bf Aspects of Warped AdS$_3$ Geometries}\\
				
				\vskip 0.4in
				
				{Pierre Bieliavsky$^a$, Philippe Spindel$^{b,\,c}$, Raphaela Wutte$^d$$^*$}

				\vskip 0.3in
				
			${}^{a}${\it Institut de Recherches en Math\'ematiques et Physique, UCLouvain, Chemin du Cyclotron, 2,  1348 Louvain-La-Neuve, Belgium\vskip .5mm
				
${}^{b}${\it \ Physique  th\'eorique, Universit\'e Libre de Bruxelles, Campus Plaine - CP 225, 1050 Bruxelles, Belgium, } \vskip .5mm
${}^{c}${\it Service de Physique de l'Univers, Champs et Gravitation, Universit\'e de Mons, Facult\'e des Sciences,20, Place du Parc,  7000 Mons, Belgium, } \vskip .5mm
${}^{d}${\it Department of Physics and Beyond: Center for Fundamental Concepts in Science, Arizona State University, Tempe, Arizona 85287, USA} \vskip .5mm
${}^*$Corresponding Author}

				\texttt{bieliavsky@math.ucl.ac.be, philippe.spindel@ulb.be,   rwutte@hep.itp.tuwien.ac.at }
				
			\end{center}
			
			\vskip 0.35in
			
			\begin{center} {\bf ABSTRACT }
 
We discuss the geometry of three-dimensional warped Anti-de Sitter spaces  
and quotients thereof, paying special attention to their underlying group manifold nature. 
We perform a systematic analysis of warped Anti-de Sitter geometries, focusing on their global properties
and illustrating their occurrence as special solutions of various three-dimensional gravity theories.

   \end{center}

			\vfill

			\noindent \today\\
   keywords: warped geometries, warped AdS$_3$ black holes; projection diagrams

		\end{titlepage}

\tableofcontents
\section{Introduction and summary}
The r\^ole warped AdS$_3$ geometries play in the framework of massive gravity theories, or as an element of the near-horizon extremal Kerr geometry \cite{Guica:2008mu} is discussed at length in Ref.\ \cite{Anninos:2008fx}. We will not repeat this discussion here. Let us only mention that, for instance, they already occurred in the framework of string theory thirty years ago (see Refs \cite{horowitz1995new,kiritsis1995infrared}).

This work mainly focuses on the geometry of these spaces. Our starting point is that AdS$_3$ can be seen as a group manifold:  the $SL(2,\,\mathbb R)$ group. Warped geometries result from a symmetry breaking 
\cite{Rooman:1998xf, Israel:2003cx, Israel:2004vv, Detournay:2015ysa}
of the usual (completely) symmetric geometry induced by the Cartan-Killing metric on the group manifold (see Sec. \ref{warpedAdS3intro}, where the process is detailed). 
\\
An advantage of this point of view, compared to other approaches (see for instance Ref.\ \cite{Bengtsson:2005zj}), is that  properties of these geometries are more easily obtained (at least according to the authors' opinion) by exploiting the group structure of the underlying manifold. For instance, the relationships between various coordinate systems used on these spaces can be immediately deduced by comparing the expressions of the Killing vector fields of the underlying AdS$_3$ geometry. The integration of geodesic equations becomes elementary, once written in the flat embedding space coordinates. The group quotients leading to black holes, as discussed by Ba\~{n}ados et al. \cite{Banados:1992gq},
 naturally extend to warped spaces. 
 
One of the main results of this paper is the construction of  warped AdS$_3$ quotients with $\mathbb R\times SO(2)$ residual isometry group from first principles. These quotients include the warped AdS$_3$ black holes of \cite{Anninos:2008fx, Nutku:1993eb, Gurses:1994bjn, Bouchareb:2007yx}, but also other geometries whose causal structures we discuss in detail in section \ref{sec: proj}. 

The outline of the paper is as follows.   {In section \ref{BasicAdS3},} we consider AdS$_3$ space, the $SL(2,\mathbb{R})$ group manifold, that we describe in detail. First, we introduce a (redundant) system of coordinates, called {\it null coordinates}, directly related to the underlying $SL(2,\mathbb{R})$ group manifold structure. It drastically simplifies numerous calculations, in particular the description of various one-parameter subgroups and the invariant vector and one-form fields linked to them. These objects are essential ingredients to understand the structure of the warped geometries. They also are crucial to obtain the various quotients of the warped AdS$_3$ geometries and to distinguish between them and may be used to easily obtain solutions of various three-dimensional gravity theories, see section \ref{subsec:gravtheories}.  On the other hand, as shown in section \ref{ExpGeodWadSSol}, they  provide a very simple way to integrate the geodesic equations on these spaces. Returning to warped geometries, we construct in section \ref{sec:warpedAdS3BH1}, from a preferred coordinate system built in section \ref{AdcanlocCoord}, quotients arising from spacelike, timelike or null warping. 
 We also briefly discuss the possibility of a local warping (see section \ref{locwarp}) and connect with the solutions found in \cite{Anninos:2010pm}.
 We
provide local coordinate systems for these spaces and 
 classify them (Table \ref{TableItoIII}). Depending on the identification vector \eqref{vecpart} used to quotient the geometry, we obtain metric expressions (Eqs \eqref{wmetcanPhi} or \eqref{wmetcanPhisd})   written in a local coordinate system adapted to the direction of identification.
In subsection \ref{subsec:decrypt}, we provide an algorithm to obtain the coordinate transformations between different local coordinate systems that we illustrate, for the warped black hole geometry in Appendix \ref{loctonullcoord}. Additionally, we discuss the existence of Killing spinors and occurrence of topological as well as chronological singularities for the quotient geometries.\\ 
In section \ref{sec: proj}, we use the projection diagram technique developed in \cite{Chrusciel:2012gz} and provide the causal structure of the quotients discussed in section \ref{sec:warpedAdS3BH1}. Our results coincide with the analysis of \cite{Jugeau:2010nq}, which provided causal diagrams for the warped black hole solutions of \cite{Anninos:2008fx}, in the regime of overlap.
We end the discussion of the causal structure with subsection \ref{ProjMICS} where we discuss an
alternative description of the embedding of warped AdS black holes in  $\widetilde{SL(2,\,\mathbb R})$ (the Einstein cylinder) and  a complementary way to visualise their causal structure.
Appendix \ref{Notations} contains some remarks about the notations used in this work. Appendix \ref{ProofsEqs} is devoted to intrinsic geometrical   proofs  of the expressions of several one-forms needed for the calculations leading to 
preferred 
coordinate systems on warped AdS$_3$ spaces and their quotients.
Appendix \ref{Hparam}
gives further comments on the coordinate system of warped AdS$_3$. In \ref{wAdS3bhmet}, we relate the coordinates used here to those employed in \cite{Anninos:2008fx, DKLSW2024} for describing warped black holes, to those used in \cite{Guica:2008mu} for the near-horizon extremal Kerr (NHEK) geometry at fixed azimuthal angle, and to those used in \cite{Banados:1992gq} for BTZ black holes in the limit of vanishing warping parameter. 
Throughout this work, the AdS$_3$ radius is set to $\ell = 1$. To restore it, it suffices to replace everywhere the line element $ds^2$ by $ds^2/\ell^2$.

\section{   {\texorpdfstring{AdS$_3$}{AdS3} and Warped \texorpdfstring{AdS$_3$}{AdS3} Spaces}}
\label{BasicAdS3}
      \subsection{Null Coordinate Systems on \texorpdfstring{AdS$_3$}{AdS3}}\label{Nullcoord}
   In this section, we summarise, for the reader's convenience, some basic properties of AdS$_3$.
Anti-de Sitter spaces (AdS$_d$)
can be embedded as $d$-dimensional hyperboloids in a $(d+1)$ -dimensional flat spaces with ultra-hyperbolic metric of signature  $(-,\,-\,\underbrace{+,\,\dots\,+}_{d-1})$. 
As symmetric spaces they appear as the coset $SO(2,\,d-1)/SO(1,\,d-1)$ (see Ref. \cite{gilmore2008lie}, section {\bf 12}). 
Since there are closed causal lines in these spaces of topology $S^1\times \mathbb R^{d-1}$, we unroll them and consider their universal covering manifolds.   In this paper,
we call Anti-de Sitter the space that contains closed causal curves. 
We denote the universal cover of Anti-de Sitter as $\widetilde{\text{AdS}}_d$.
\footnote{See Ref.\cite{gilmore2008lie}, section {\bf 6} or Ref.\cite{gilmore1974lie}, section {\RomanNumb {5}}, page (108); also see Ref. \cite{rawnsley2012universal} 
where a clear and intuitively transparent exposition of the covering group $\widetilde{SL(2, R)}$ is given.} Geometrically, these spaces can be seen as piles of AdS$_d$ spaces, each one cut along a spacelike section and glued together. Concerning local geometrical properties, once established on AdS$_d$, they remain true on $\widetilde{\text{AdS}}_d$.\\
In three dimensions, the situation drastically simplifies thanks to the 2--1 group homomorphisms: $ SL(2,\,\mathbb R)\times SL(2,\,\mathbb R)\mapsto SO(2,\,2)$ and $ SL(2,\,\mathbb R) \mapsto SO(2,\,1)$ (see Ref. \cite{garrett}).  The three-dimensional Anti-de Sitter space has the particular property of being isomorphic to the $SL(2,\,\mathbb R)$ group equipped with its Killing metric (see, for instance, Ref. \cite{Bieliavsky:2002ki} for an extended discussion). Hereafter we limit ourselves to the essential points leading to this special picture and fix the notation.  \\
Thus the  AdS$_3$ space (of unit radius) can be seen as the hyperboloid $\mathcal H$ of equation:
\begin{equation}
  \label{AdS30hyp0}
{\mathcal H}\equiv  U_+\,U_- + V_+\,V_- = 1\qquad,
\end{equation}
embedded into the four-dimensional ultra-hyperbolic space 
    $M^{2, 2}$ endowed with the metric induced by the quadric $\mathcal H$. We parametrize this space using null pseudo-Lorentzian linear coordinates $\{U_+,\,U_-,\,V_+,\,V_-\}$ that, from now on, we call {\it null coordinates}. Then   the     $M^{2, 2}$ metric reads:  
  \begin{equation}
  \label{M22met}
\upeta =  - dU_+\,dU_--dV_+\,dV_-\qquad  .
\end{equation}
AdS$_3$ 
is in one-to-one correspondence with the simple real Lie  group of two by two matrices 
\begin{align}
   {\boldsymbol z} :=\left(\begin{array}{rr}
U_+&V_+\\
-V_-&U_-
\end{array}\right)\qquad \label{SL2Relem}
\end{align}
of unit determinant: 
\begin{align}
  SL(2,\mathbb R)=\left\{ {\boldsymbol z}\, \vert\, \text{det}({\boldsymbol z}):=U_+\,U_- + V_+\,V_- = 1
\right\}\qquad .\label{SL2Rmat}
\end{align}
Using the parametrization\footnote{This parametrization covers (almost) the whole hyperboloid, except for their trivial failures, related to the use of polar coordinates. }
\begin{subequations}
\begin{align}
    &U_\pm= \cosh(\upchi)\,\cos(\uptau)\pm\sinh(\upchi)\,\cos(\upvartheta )\qquad,\label{cylUpm}\\
    &V_\pm= \cosh(\upchi)\,\sin(\uptau)\pm\sinh(\upchi)\,\sin(\upvartheta )\qquad,\label{cylVpm}\\
    \intertext{with}
    &\upchi \in \mathbb R^+\quad,\quad \upvartheta \in [0,\,2\,\pi]\quad,\quad \uptau \in [0,\,2\,\pi]\quad,
\end{align}
\end{subequations}
 we easily infer from matrix multiplication the group composition law given in Appendix \ref{CovSLR} that defines the universal covering group $\widetilde{SL(2,\,\mathbb R)}$. 
It is this group, whose topology is $\mathbb R^3$, that constitutes the universal cover of three-dimensional Anti-de Sitter space, which does not possess closed timelike curves. In this work, we will use $SL(2,\,\mathbb{R})$ when discussing local properties and $\widetilde{SL(2,\,\mathbb{R})}$ when addressing the global structure.
 \\
Hereafter, we first consider tangent spaces on $\mathcal H$. 
To lighten some equations we make use of the shorthand notation: 
\begin{equation}
  \label{AdS30hyp}
H:= U_+\,U_- + V_+\,V_- \qquad.
\end{equation}
Tangent vectors consist of vectors $\ \vec v=u_+\,\partial_{U_+}+u_-\,\partial_{U_-}+v_+\,\partial_{V_+}+v_-\,\partial_{V_-}$ defined on     $M^{2, 2}$, such that on $\mathcal H$ they satisfy the condition: $\vec v(H)=0$. In terms of components, with respect to the natural basis linked to the null coordinate system, this  condition implies that:
\begin{align}
\text{If}\quad  U_+\,U_- + V_+\,V_- =  1\quad\text{then}\quad  u_+\,U_-+u_-\,U_++v_+\,V_- +v_-\,V_+=0\quad .\label{vHzero}
\end{align}
We also consider tangent one-forms whose
 contraction with the vector normal to the surface $\mathcal H$: 
 \begin{align}
\vnH =-2\,\left(U_+\,\partial_{U_+}+U_-\,\partial_{U_-}+V_+\,\partial_{V_+}+V_-\,\partial_{V_-}\right)\qquad \label{nH}
\end{align}
 vanishes on $\mathcal H$.\\
Let us now discuss the isometry group of AdS$_3$. It is isomorphic  to the  linear homogeneous transformation group $O(2,\,2)$ of the null-coordinates preserving the metric (\ref{M22met}). The one connected to the identity: ${Iso}_0(AdS_3)$ is, locally, isomorphic to $SL(2,\,\mathbb R)\times SL(2,\,\mathbb R)$ (see Ref.\cite{gilmore1974lie}, section {\RomanNumb {4}}.3) and easily described 
by using the matrix representation (\ref{SL2Rmat}) of $SL(2,\mathbb R)$.
Its action  is given by left and right multiplications of ${\boldsymbol z}$ by $SL(2,\,\mathbb R)$ matrices $g_L$ and $g_R$:
\begin{align}
    {\boldsymbol z}\mapsto g_L\, {\boldsymbol z}\,g_R\qquad,\qquad  g_{L} ,\, g_R\,\in SL(2,\,\mathbb R) \label{Iso0AdS3}\qquad .
\end{align}
We consider now the $SL(2,\,\mathbb R)$ fundamental representation of the Lie group generators: 
\begin{align}
    X_1:=\left(\begin{array}{cc}
1&0\\
0&-1
\end{array}\right)\qquad,\qquad  X_2:=\left(\begin{array}{cc}
0&1\\
1&0
\end{array}\right)\qquad,\qquad X_3:=\left(\begin{array}{cc}
0&1\\
-1&0
\end{array}\right)\qquad.\label{sl2Rgen}
\end{align} 
Their left and right actions on a group element ${\boldsymbol z}$ provide explicit expressions of the corresponding right- and left-invariant Killing vector fields\footnote{We refer the reader to Ref.\cite{Spindel:1986ic}, section {\RomanNumb {3}}, for elements of differential geometry on Lie groups. For a more in-depth treatment, for instance, see Ref.\cite{Helgason}.}:
\begin{align}
\vec r_{a}(\bs{z})=\left. \ft d{d\alpha}Exp[\alpha\,X_a]\,\bs{z}\right\vert_{\alpha=0}\qquad,\qquad \vec l_{a}(\bs{z})=\left. \ft d{d\alpha}\bs{z}\,Exp[\alpha\,X_a] \right\vert_{\alpha=0}\qquad .
\label{XRLxiRL}
\end{align}
Associativity of the group composition law makes left and right transformations commuting. At the level of infinitesimal transformations this translates into the vanishing of the Lie brackets:
\begin{align}
 &[\vec l_{a}, \vec r_{b}] = 0  \quad .
\end{align}
The  invariant vector fields obtained from the generators Eqs (\ref{sl2Rgen}) read:
\begin{subequations}
 \label{KVLR}
\begin{align}
\vec r_{1}
&= U_+\,\partial_{U_+}-U_-\,\partial_{U_-}+V_+\,\partial_{V_+}-V_-\,\partial_{V_-} &,\quad\vec l_{1}
&=U_+\,\partial_{U_+}-U_-\,\partial_{U_-}-V_+\,\partial_{V_+}+V_-\,\partial_{V_-}\quad, \\
\vec r_{2}
&= -V_-\,\partial_{U_+}+V_+\,\partial_{U_-}+U_-\,\partial_{V_+}-U_+\,\partial_{V_-} &,\quad\vec l_{2}
&=V_+\,\partial_{U_+}-V_-\,\partial_{U_-}+U_+\,\partial_{V_+}-U_-\,\partial_{V_-}\quad, \\
\vec r_{3}
&= -V_-\,\partial_{U_+}-V_+\,\partial_{U_-}+U_-\,\partial_{V_+}+U_+\,\partial_{V_-} &,\quad \vec l_{3}
&=-V_+\,\partial_{U_+}-V_-\,\partial_{U_-}+U_+\,\partial_{V_+}+U_-\,\partial_{V_-}\quad ,
\end{align}
\end{subequations}
and can obviously be extended to the whole     $M^{2, 2}$ space. They satisfy the condition (\ref{vHzero}) and the commutation relations:  
\begin{align}
 &[\vec l_{a}, \vec r_{b}] = 0 \quad ,\quad[\vec r_{a}, \vec r_{b}] =  + 2\,C^{c}_{ab}\, \vec r_{ c}\quad ,\quad[\vec l_{a}, \vec l_{b}] = - 2\,C^{c}_{ab}\, \vec l_{ c}\quad,\label{comRL}\\
\intertext{with}
&\eta^{(3)}=\text{diag}\ (1,1,-1)\quad,\quad   C^{c}_{ab}:=(\eta^{(3)})^{cp}\,\epsilon_{pab}\quad ,\quad(\epsilon_{123}=+1\ \text{and}\ a,\, b,\,c \in \{1, 2, 3\}) \quad .
\end{align}
The norms of these Killing vectors, with respect to the metric Eq.\ \eqref{M22met} (that on $\mathcal H$ reduces, up to a factor, to the Lie algebra Cartan-Killing metric: $g_{ab}=C^p_{aq}C^q_{bp}$ where $C^a_{bc}$ are the structure constants),  are given by: 
\begin{align}
\vec r_{m} . \vec r_{n} =\vec l_{m} . \vec l_{n}  = H\,\eta^{(3)}_{m n} \label{norm}\qquad.
\end{align}
They are constant on $\mathcal H$. The choice of normalisation Eqs (\ref{norm})  allows us to speak, by reference to the local light cone (see Ref. \cite{gilmore2008lie}, sections {\bf 4.8} and {\bf 7.2}), about spacelike,  timelike  or lightlike (null) vectors and one-forms as well as Lie algebra elements.\footnote{We use the terminology inspired by the Minkowskian nature of $sl(2,\,\mathbb R)$. Mathematicians call {\it timelike} generators {\it elliptic}, {\it spacelike} generators {\it hyperbolic} and {\it lightlike} generators {\it nilpotent} or {\it parabolic}. In order to obtain a metric signature $(-,\,+,\,+)$ we adopt a metric proportional to the opposite of the one obtained on the $SL(2,\,\mathbb{R})$ group from the usual Cartan-Killing metric. }\\
Let us emphasise that the vectors obtained in Eqs (\ref{KVLR}) rest on our (arbitrary) choice of the Lie algebra generators Eqs (\ref{sl2Rgen}). For instance the null coordinate transformation $\{U_+,\,U_-,\,V_+,\,V_-\}\mapsto \{U_+,\,U_-,\,-V_-,\,-V_+\}$, an $O(2,2)$ isometry  not connected to the identity, interchanges $\{\vec r_1,\,\vec r_2,\,\vec r_3\}$ with $\{\vec l_1,\,\vec l_2,\,-\vec l_3\}$.\\
Later on, we shall consider   one-forms that on $\mathcal H$ constitute a dual basis of the tangent space basis $\{\vec l_{a} \}$, or $\{\vec r_{a} \}$. We denote them by $\{\oneform{\lambda}^{a}\}$ and $\{\oneform{\rho}^{a}\}$:  
\begin{subequations}
\begin{align}
\oneform{\rho}^{1}
&=\frac 1{2\,H}( U_-\,d{U_+}-U_+\,d{U_-}+V_-\,d{V_+}-V_+\,d{V_-}) &,\quad \oneform{\lambda}^{1} 
&=\frac 1{2\,H}(U_-\,d{U_+}-U_+\,d{U_-}-V_-\,d{V_+}+V_+\,d{V_-})\quad, \label{IRLform1}\\
\oneform{\rho}^{2}
&= \frac 1{2\,H}(-V_+\,d{U_+}+V_-\,d{U_-}+U_+\,d{V_+}-U_-\,d{V_-}) &,\quad \oneform{\lambda}^{2} 
&=\frac 1{2\,H}(V_-\,d{U_+}-V_+\,d{U_-}+U_-\,d{V_+}-U_+\,d{V_-})\quad, \\
\oneform{\rho}^{3}
&= \frac 1{2\,H}(-V_+\,d{U_+}-V_-\,d{U_-}+U_+\,d{V_+}+U_-\,d{V_-}) &,\quad  \oneform{\lambda}^{3} 
&=\frac 1{2\,H}(-V_-\,d{U_+}-V_+\,d{U_-}+U_-\,d{V_+}+U_+\,d{V_-})\quad ,\label{IRLform3}
\end{align}  
\end{subequations}
They satisfy to the  
relations:
\begin{align}
\oneform{\lambda}^{a}( \vec l_{b})=\delta^a_b=\oneform{\rho}^{a}(  \vec r_{b})\label{duality} \quad , \quad \lambda^a_A=\frac 1H\,(\eta^{(3)})^{ab}\,\eta_{AB}l_b^B\quad,\quad \rho^a_A=\frac 1H\,(\eta^{(3)})^{ab}\,\eta_{AB}r_b^B\quad.
\end{align}

As shown by Eq.\ (\ref{Iso0AdS3}) the $O(2,\,2)$ component connected to the identity is isomorphic to $(SL(2,\mathbb R)_L \otimes SL(2,\mathbb R)_R)/\{Id,\,-Id\}$. The three other components are obtained by acting with the discrete transformations $\mathcal{P}$, $\mathcal{T}$ and their product \cite{Bieliavsky:2002ki}, where:
\begin{align}
   & \mathcal P: \left\{\begin{array}{l} \{U_+,\,U_-,\,V_+,\,V_- \}\leftrightarrow \{U_-,\,U_+,\,V_+,\,V_- \}\quad;\\
    \{\vec r_1,\,\vec r_2,\,\vec r_3\}\leftrightarrow \{-\vec l_1,\,\vec l_2,\,\vec l_3\}\quad .
   \end{array}
   \right .\label{P}\\
   &\nonumber\\
 & \mathcal T: \left\{\begin{array}{l} \{U_+,\,U_-,\,V_+,\,V_- \}\leftrightarrow \{U_+,\,U_-,\,-V_-,\,-V_+ \}\quad;\\
    \{\vec r_1,\,\vec r_2,\,\vec r_3\}\leftrightarrow \{\vec l_1,\,\vec l_2,\,-\vec l _3\}\quad .
   \end{array}
   \right .\label{T}
\end{align}

The $M^{2,2}$ submanifold $\mathcal{H}$ equipped with the metric induced by the flat metric $\upeta$ (Eq.\ (\ref{M22met})) defines the AdS$_3$ space. The intrinsic metric so obtained is invariant with respect to the isometry group $O(2,\,2)$, the subgroup of the $M^{2,2}$ isometry group that preserves $\mathcal{H}$. This metric, seen on $SL(2,\,\mathbb{R})$, is invariant with respect to both left and right actions of the group on itself (Eq.\ \eqref{Iso0AdS3}). To be, let's say, left invariant, a metric must be such that its components, with respect to a basis of left invariant one-forms are constant. But then, expressed in terms of right invariant one-form the metric components, in general, become point dependent. So a general left invariant metric is not invariant with respect to the right action of the group on itself. To be both invariant with respect to the left and the right actions the metric components have to be invariant with respect to the adjoint group, defining a Casimir operator in the Lie algebra (see for instance Ref.\cite{Spindel:1986ic}, section \RomanNumb{4}.5). On a simple Lie group there exists only one (up to a factor) bi-invariant metric whose components are those of the Cartan-Killing metric of the Lie algebra.\\
The choice of the generators Eqs (\ref{sl2Rgen}), that diagonalise the Cartan-Killing metric in the Lie algebra (see Ref. \cite{gilmore2008lie}, section {\bf 4.8}),  allows us to write the bi-invariant  metric as:
\begin{align}
  { \uuline{g}}\strut_{(0)}=\sum_{m,\,n}\eta^{(3)}_{m\,n}\,\oneform{\rho}^{m}\otimes \oneform{\rho}^{n}=  \sum_{m\,n}\eta^{(3)}_{m\,n}\,\oneform{\lambda}^{m}\otimes \oneform{\lambda}^{n} \qquad .\label{g0KV}
\end{align}
Here we have added an index $(0)$ to the metric symbol, having in mind the further consideration of a one-parameter family of metrics obtained from deforming the bi-invariant one.
\subsection{Warped \texorpdfstring{AdS$_3$}{AdS3} metrics} 
\subsubsection{Partially invariant metrics}
\label{warpedAdS3intro}
In this section we introduce warped AdS$_3$ geometries. They are obtained by adding to the maximally symmetric metric (\ref{g0KV}) an extra piece that partially breaks the six-parameter isometry group (\ref{Iso0AdS3}). The term added consists (up to a factor) of the tensorial product of an invariant one-form with itself. Such a form is completely defined by an element of the dual of the Lie algebra of the group and, as a consequence, by the choice of a Lie algebra generator $X$, thanks to the canonical duality defined by the Cartan-Killing metric between the Lie algebra itself and its dual. Of course, elements related by adjoint transformations (in terms of matrices such that $Y=g^{-1}\,X\,g$ with $g$ an element of the group)  yield isomorphic warped geometries.\\
In what follows let us assume that the extra term is right invariant. A single one-side (of right transformations)  $SL(2,\,\mathbb R)$ factor is maintained but the other one of left transformations is reduced to a one-parameter subgroup. The generator of this subgroup is $X$ itself. It defines the Lie algebra of the so-called stabilizer subgroup of the Lie algebra element $X$: the subgroup of adjoint transformations that fix the selected generator. Moreover, according to the norm of this generator with respect to the Cartan-Killing metric we will speak about spacelike, timelike or lightlike directions of warping.\\
\strut
The result of this construction of warped AdS$_3$ consists to write the  metric as:
\begin{equation}
\label{warpedAdS1}
   \uuline{g}\vphantom{g}_{(\lambda)}=\uuline{g} \vphantom{g}_{(0)}+ \lambda \,\oneform{\xi}_{\text{\tiny {(R)}}}\otimes\oneform{\xi}_{\text{\tiny {(R)}}}\qquad,
\end{equation}
where $\oneform{\xi}_{\text{\tiny {(R)}}}=\xi_{\text{\tiny R}a}\,\underline \rho^a $ is a right invariant one-form, musical dual of a normalised (to $\pm1$ or $0$) corresponding right invariant Killing vector field: $\vxiR=\xi_{\text{\tiny R}}^a\,\vec r_a$ obtained from the generator $X_L$ as in Eqs (\ref{XRLxiRL}). Using Eq.\ (\ref{g0KV}), we can express the deformed metric solely in terms of right-invariant one-forms:
\begin{equation}
\label{glamKV}
   \uuline{g}\vphantom{g}_{(\lambda)}=\sum_{m,\,n}\eta^{(3)}_{m\,n}\,\oneform{\rho}^{m}\otimes \oneform{\rho}^{n}+ \lambda \,\oneform{\xi}_{\text{\tiny {(R)}}}\otimes\oneform{\xi}_{\text{\tiny {(R)}}}\qquad.
\end{equation}
More precisely, its isometry group now consists of two components. 
The part connected to the identity that reduces to the one-parameter group of left transformations generated by $X$ times the remaining full subgroup of right transformations: $(\left\{Exp[\alpha\,X]\vert \alpha\in \mathbb R\right \}_L  \times   SL(2,\,\mathbb R)_R)/\mathbb Z_2$. The part disconnected from the identity is obtained from the composition of the first one with, for $\oneform{\xi}_R$ arbitrary, the involution $g^{-1}_L\,\mathcal {P\,T}\,g_L$ where $g_L$ is a left transformation that maps $\underline \xiR$ on $\underline \rho^1$ or $\underline \rho^3$ (Indeed $\mathcal {P\,T}$ maps $\{\underline \rho^1,\,\underline \rho^2,\,\underline \rho^3\}$ on $\{-\underline \rho^1,\,\underline \rho^2,\,-\underline \rho^3\}$).
\strut\\
The same construction may be driven by exchanging left invariant objects with right ones. 
Accordingly, using $O(2,\,2)$ transformations, we may restrict ourselves to right invariant objects and use them in the expressions of the metric  $\uuline{g}\vphantom{g}_{(\lambda)}$ with 
a warping one-form ($\oneform{\xi}_{\text{\tiny {(R)}}}$) (that is defined up to a sign):
\begin{subequations}
\begin{align}
&\oneform{\xi}_{\text{\tiny {(R)}}}=\underline \rho^1  &\text{(spacelike warping)}\quad &,\label{spwf}\\
&\oneform{\xi}_{\text{\tiny {(R)}}}=  \underline \rho^3  &\text{(timelike warping)}\quad &,\label{tiwf}\\  
& \oneform{\xi}_{\text{\tiny {(R)}}}=(\underline \rho^2+ \underline \rho^3)/\sqrt{2}  &\text{ (lightlike warping)}\label{liwf}\quad &.
\end{align}
\end{subequations}
\strut\\
\strut
For the metric (\ref{glamKV}) to be non-degenerate,  we impose that the parameter $\lambda$ satisfies the condition:
\begin{align}
 1+\lambda\,  (\xi_{\text{\tiny(R)}_\mu}\,g\strut_{(0)}^{\mu\nu}\,\xi_{\text{\tiny(R)}\,\nu)})=: 1+\lambda\,   (\vxiR\cdot\vxiR)\neq 0\label{nondeglam} 
\end{align}
Moreover, when the deforming one-form is timelike, we have to restrict $\lambda$ to be strictly less than 1 to remain in the framework of Lorentzian metrics. For $\lambda>1$, the metrics are Euclidean.\footnote{Such metrics are physically relevant in the scope of quantum three-dimensional gravity (see for instance Refs \cite{krasnov2000holography, krasnov2002analytic, krasnov2003holomorphic}), but will not be considered in this work.}
\strut\\
It is interesting to note that inserting the parametrization (\ref{cylUpm}, \ref{cylVpm}) into the expressions (\ref{IRLform1}--\ref{IRLform3}) of the invariant forms   so as to obtain their pull-backs on $\mathcal H$,
the timelike warped metric given by Eqs (\ref{glamKV}) and (\ref{tiwf}) can be written as:
\begin{align}
-\cosh^2(\upchi)\,d\uptau^2 +d\upchi^2+\sinh^2(\upchi)\,d\upvartheta^2+  \lambda \,\left(\cosh^2(\upchi)\,d\uptau+\sinh^2(\upchi)\,d\upvartheta\right)^2\qquad.\label{twarpmet}
\end{align}
\strut\\
{Here, we see that when $\lambda<0$ and $\upchi$ is large enough the closed curves of constant $\uptau$ and $\upchi$ become timelike.
For instance when $\lambda=-1$, Eq. (\ref{twarpmet}) corresponds to the famous G\"odel metric\footnote{As a cosmological solution it requires adding a flat factor to be a four-dimensional solution of Einstein's equations (see \cite{HawkingEllis} with a cosmological constant and a pressureless fluid, whose velocity is aligned with $\vec r_3$, as the energy-momentum source.)} (see Ref.\cite{Rooman:1998xf}), which is known to be an example of  a 
totally vicious 
space, as it has closed timelike curves throughout spacetime. Of course, taking a quotient does not resolve this issue.
\strut\\
If $\oneform{\xi}_{\text{\tiny {(R)}}}=\oneform{\rho}^{1}$ (spacelike warping) and $\lambda>-1$ the metric  signature is $(+,+,-)$; it becomes degenerate for $\lambda= -1$ and is of signature $(-,+,-)$ for $\lambda < -1$. In this case, to reobtain a usual Lorentzian metric with timelike vectors of negative norm, we have to replace the original quadratic form defining the metric by its opposite. 
As a consequence we obtain, for $\lambda<-1$ another branch of geometries where now the Gauss scalar curvature is $R=2(3-\lambda)$ instead of $R=2(\lambda - 3)$. \\
In what follows, we restrict ourselves to geometries whose light cones result from continuous deformations of those of the standard bi-invariant metric.  Therefore, instead of condition Eq. (\ref{nondeglam}), we impose:
\begin{align}
1+\lambda\,(\vxiR\cdot\vxiR)> 0\qquad .\label{contdef}
\end{align}
\strut\\
Finally, when the deforming one-form is lightlike, the parameter $\lambda$ is not restricted.\\
  On the other hand, for $\lambda\geq 0$, whatever is the warping, the surfaces of constant $\uptau$ are global spacelike sections of the manifolds. Thus, there cannot exist closed causal curves on these spaces. 
  \subsubsection{Hopf-type fibration}
In Ref. \cite{duff1999ads3} warped AdS$_3$ geometries are presented as kind of Hopf fibrations over the Lorentzian space AdS$_2$ or its Euclidean version: the Poincaré plane, with a warping of the fiber metric. We think that it is interesting to analyse this approach as it provides an alternative perspective on the geometries obtained in case of non-null warping:
this perspective favors a coset structure (see \cite{gilmore2008lie}, section {\bf 7.2}). \\ 
We remind the reader of the Iwasawa decomposition theorem (see for instance Refs \cite{Helgason}, chapter \RomanNumb{9} or \cite{Knapp}, section \RomanNumb{6}.4): a semisimple Lie group $G$ always is diffeomorphic to a product of a compact  Lie group $K$, an abelian Lie group $A$ and a nilpotent Lie group $N$. The factorization is not unique but requires the compact group $K$ to be the first or the third factor. In the particular case at hand, we obtain that the elements of the $SL(2,\,\mathbb {R})$   group can all be written as elements of the groups products:

\begin{align}
  &SL(2,\,\mathbb{R})\simeq\quad K\times N\times A  \qquad\text{or}\qquad  K\times A\times N  \qquad\text{or}\qquad A\times N\times K  \qquad\text{or}\qquad  N\times A\times K \label{KANIwa} 
  \end{align} 
  where
  \begin{subequations}
  \begin{align}
   &A:=\{A_a\}:=\{Exp[a\, X_1]\vert a\in\mathbb R\}\label{AIwa}  \\
   &K:=\{K_b\}:=\{Exp[b\, X_3]\vert b\in\mathbb [0,\,2\,\pi[\}\label{KIwa} \\
   &N:=\{N_x\}:=\{Exp[x\,\ft12( X_2+X_3)]\vert x\in\mathbb R\}\label{NIwa}
\end{align}   
\end{subequations}
with the $X_1,\,X_2,\,X_3$ the generators displayed in Eq. (\ref{sl2Rgen}).\\
In particular, let us also consider the left actions of the one-parameter subgroups $A$ (the abelian factor) and $K $ (the compact factor) on $\boldsymbol{z}$. They transform the matrix components (Eq. \eqref{SL2Relem}) as follows:
\begin{align}
 &A_{ \alpha}\boldsymbol{z}\ :  (U_\pm,V_\pm)\mapsto e^{\pm \alpha}(U_\pm,V_\pm)\qquad,\\
  &K_{ \alpha}\boldsymbol{z}\  :  (U_\pm+i\,V_\pm)\mapsto e^{i\, \alpha}(U_\pm+i\,V_\pm)  \qquad .
\end{align}
Before displaying explicit specific formulas, to go ahead let us agree to denote by $X_H$ a  specific Lie algebra generator.

\noindent The Lie algebra $sl(2,\,\mathbb{R})$ constitutes a three-dimensional vector space that the Cartan-Killing metric (see Ref.\cite{gilmore2008lie} section {\bf 4.8}) makes a $(2+1)$ Minkowski space $M^{(2,1)}$.  We parametrize it with   Lorentzian coordinates $\{x^1,\,x^2,\,x^3\}$ by expressing an arbitrary generator $X\in sl(2,\,\mathbb{R})$ in the fundamental representation of the Lorentz group, using the basis (\ref{sl2Rgen}), as $X=x^a\,X_a$   and writing its metric $\eta$ (proportional to the Cartan-Killing metric) as:
\begin{align}
    \eta(X_a,\,X_b)=\ft12 Tr(X_a\,X_b)=\eta_{ab}\qquad .\label{LieAlgmet}
\end{align}
\noindent Within this setting, we obtain a projection (${\proj}_H$) of $SL(2,\,\mathbb R)$ 
onto the adjoint orbit $M$ of the element $X_H$ in the Lie algebra $sl(2,\mathbb R)$:
\begin{align}
{\proj}_H: SL(2,\,\mathbb R)\to M \subset sl(2,\mathbb R) : z \mapsto z^{-1} X_H z  \qquad .
\end{align}
The stabilizer subgroup of $X_H$ in the coadjoint action is $\{\pm I\}\times H=:\pm H$ where $H$ is the connected one parameter subgroup of $SL(2,\,\mathbb R)$ generated by $X_H$. The coadjoint orbit $M$ therefore naturally identifies with the coset space ${\pm H}\backslash SL(2,\,\mathbb R)$ via the diffeomorphism:
\begin{align}
{\pm H}\backslash SL(2,\,\mathbb R)\to M : \pm Hz \mapsto z^{-1} X_H z \qquad .
\end{align}
The projection ${\proj}_H$ then defines a ${\pm H}$-principal bundle over the orbit $M$. We express it in coordinates as 
\begin{subequations}
\begin{align}
&{\proj}_H:{\boldsymbol z}\mapsto \{x^1,\,x^2,\,x^3\}\label{proj}\\
\intertext{where}
&x^a=\ft12\,Tr[{\boldsymbol z}^{-1}X_H{\boldsymbol z}X^a]\qquad .
\end{align}
\end{subequations}
 \noindent If $X_H$ is chosen to be a spacelike generator ($\eta(X_H,\,X_H)>0$) we obtain a one-sheet hyperboloid, an AdS$_2$ subspace,  in $M^{(2,1)}$. If $X_H$ is a timelike generator ($\eta(X_H,\,X_H)<0$) the image of the projection is a Poincaré plane: one of the sheets of a two-sheet hyperboloid (see Ref. \cite{gilmore2008lie}, section {\bf 7.2}). Finally, if $X_H$ is lightlike ($\eta(X_H,\,X_H)=0$) we obtain
an (open) nappe of the lightcone. \\
In the same way as a point of $\mathcal{H}$ is represented by a matrix (Eq. (\ref{SL2Relem})), a tangent vector $\vec v$ may be represented by a matrix:
\begin{align}
    {\boldsymbol{v}}:=\left(\begin{array}{cc}
      u_+   & v_+ \\
     -v_-    & u_-
    \end{array}\right)\qquad .\label{matvec}
\end{align}
At a given point ${\boldsymbol{z}}$, this vector  may be expressed as belonging to a right or a left  invariant field  generated by   Lie algebra elements $V_R$ or $V_L$:
\begin{align}
&{\boldsymbol{v}}=\left .\frac{d}{d\alpha}e^{\alpha\,V_R}\,\boldsymbol{z}\right\vert_{\alpha=0}= \left .\frac{d}{d\alpha}\boldsymbol{z}\,e^{\alpha\,V_L}\right\vert_{\alpha=0}
\end{align}
 where, using the previous matrix representation of $\vec v$ we obtain the expressions of the generators $V_R$ and $V_L$ in the fundamental $sl(2,\,\mathbb{R})$ representation:  
\begin{align}
&V_R={\boldsymbol{v}}{\boldsymbol{z}}^{-1}\quad,\quad V_L={\boldsymbol{z}}^{-1}{\boldsymbol{v}}\qquad .\label{RLvecrep}
\end{align}
The vanishing of the trace of these matrices reflects the condition (\ref{vHzero}): $\vec v(H)=0$.\\
The projections  at the point ${\boldsymbol{z}}$ of this vector in the (fundamental representation of the) Lie algebra is obtained by considering derivatives with respect to the parameter($\alpha$) of the right or left transformation defining them: 
\begin{align}
&\proj_{H\star}\left[ \vec v({\boldsymbol z})\right]=\left .\frac{d}{d\alpha}\left(\boldsymbol{z}^{-1}e^{-\alpha\,V_R}X_He^{\alpha\,V_R}\boldsymbol{z}\right)\,\right\vert_{\alpha=0}= {\boldsymbol z}^{-1}\left[X_H,\,V_R\right]\,{\boldsymbol z}=\left[{\boldsymbol z}^{-1}\,X_H\,{\boldsymbol z},\,V_L\right] \qquad .
\end{align}
From the expression   (\ref{LieAlgmet}) of the Lie algebra metric we see that scalar products of the projections of vectors belonging to  right invariant vector fields are constant  on the coset space, while those belonging to left invariant fields are constant along each fibre (but depend on the fibre). At each point $\boldsymbol{z}$ of AdS$_3$, as far as the fibre is not lightlike ($X_H$ not nilpotent) we may decompose the tangent space $T_{\boldsymbol{z}}$ into a one-dimensional subspace tangent to the fibre through ${\boldsymbol z}$: $\mathfrak{h}_{\boldsymbol z}$ and its $\uuline{g}\strut_{(0)}$ 
orthogonal subspace $\mathfrak{p}_{\boldsymbol z}$:
\begin{align}
T_{\boldsymbol{z}}=\mathfrak{h}_{\boldsymbol z}\oplus \mathfrak{p}_{\boldsymbol z}\qquad .
\end{align}
The $\uuline{g}\strut_{(0)}$ metric, restricted to ${\mathfrak{p}}_{\boldsymbol z}$ shares the same invariance as the pullback with respect to the projection $\proj$ of the Lie algebra. Thus they must be proportional, as they share the same orthogonal group. Indeed we obtain\footnote {An illustration of the identity valid for $2\times 2$ traceless matrices $X$ and $Y$ : $\Tr[(X\,Y)^2]-\Tr[X^2 Y^2]+\Tr[X^2]\Tr[Y^2]-(\Tr[X\,Y])^2=0$.} that on $\mathfrak{p}_{\boldsymbol z}\otimes \mathfrak{p}_{\boldsymbol z}$:
\begin{align}
    \proj_H^\star (\eta)=-4\,\left.\text{Tr}(X_H\,X_H)\,\uuline{g}\strut_{(0)}\right\vert_{\mathfrak{p}_{\boldsymbol z}\otimes \mathfrak{p}_{\boldsymbol z}}\label{degquadform}
\end{align}
 This (degenerate) quadratic form is right invariant and also invariant with respect to the left transformation subgroup $H$. \\
 Let us now consider the musical dual $\oneform{\xi}_{\text{\tiny {(R)}}} $ (with respect to $\uuline{g}\strut_{(0)}$) of the vector tangent to the orbits of the left action of the subgroup $H$. It  vanishes on the subspaces ${\mathfrak{p}}_{\boldsymbol{z}} $. It also is both right invariant and  invariant with respect to the $H$ subgroup of left transformations. We may use it to complete the quadratic form (\ref{degquadform}) into a non degenerate quadratic form $B_{\mu}$ : a metric, by adding a  term proportional to the tensor product $  \oneform{\xi}_{\text{\tiny {(R)}}}\otimes \oneform{\xi}_{\text{\tiny {(R)}}}$:
 \begin{align}
     B_\mu:= \proj_H^\star (\eta)+\mu  \,\oneform{\xi}_{\text{\tiny {(R)}}}\otimes \oneform{\xi}_{\text{\tiny {(R)}}}\qquad,\qquad (\mu \neq 0)
 \end{align}
 The metric so obtained constitutes a warped metric involving a non-lightlike one-form. \\
 Expressing this construction explicitly in coordinates, we obtain the following warped metrics:
 \begin{itemize}
\item{\em AdS$_2$ Hopf fibration}:  In case we consider an $A$-coset space, {\it i.e.} putting $X_H=X_1$ and  using the Iwasawa representation ($A\times N\times K$) the projection into the Lie algebra reads:
\begin{subequations}
\begin{align}
 &x^1=\ft12\,Tr[{\boldsymbol z}^{-1}X_1{\boldsymbol z}X_1]=(U_+ U_- -V_+V_-)\quad,\\
 &x^2=\ft12\,Tr[{\boldsymbol z}^{-1}X_1{\boldsymbol z}X_2]=(U_+V_-+U_-V_+)\quad,\\
 &x^3=-\ft12\,Tr[{\boldsymbol z}^{-1}X_1{\boldsymbol z}X_3]=-( U_+ V_--U_-V_+)\} \quad ,
\end{align}
\end{subequations}
that satisfy the equation $(x^1)^2+(x^2)^2-(x^3)^2=1$.\\
The coordinate system induced from the Iwasawa decomposition is:
\begin{subequations}
\begin{align}
&U_+=e^{  a}\left( \cos(b)-x\,\sin(b)\right)\qquad ,\\
&V_+=e^{ a}\left(x\, \cos(b)+\sin(b)\right)\qquad ,\\
&U_-=e^{ - a}\, \cos(b)\qquad ,\\
&V_-=e^{ -a} \sin(b)\qquad .
\end{align}
\end{subequations}
We obtain that the left action of $A$ reduces to an $a$-translation: $A_{\alpha}(\{a,\,x,\,b\})=\{a+\alpha,\,x,\,b\}$. \\
In terms of these coordinates, the pull-back ${\proj}^\star_A(\eta)$ of the Cartan-Killing metric (a metric of the AdS$_2$ space) is:
\begin{equation}
    {\proj}^\star_K(\eta)=4\,db\,\left((1+x^2)\,db+dx)\right)\qquad.
\end{equation}
The  warping right invariant one-form, the $\uuline{g}\strut_{(0)}$-musical dual of $\vec\partial_a$, the tangent vector field of the orbits of the $A$-subgroup left action, reads:
\begin{align}
\oneform{\xi}_{\text{\tiny {(R)}}}=\uform{\rho}^1=da-x\,db
\end{align}
leading to the warped metric expression:
\begin{align}
ds^2_{(\lambda)}=-(1+x^2)\,db^2-db\,dx+(1+\lambda)(da-x\,db)^2\qquad .\label{SRiemSum}
\end{align}
\item{\em Poincaré Hopf fibration}: In case the fibres are the orbits of the right action of the $K$ group. Adopting the parametrisation dictated by the decomposition ($K\times A\times N $):
\begin{subequations}
\begin{align}
&U_+=e^{   a}\, \cos(b)\qquad ,\\
&U_-=e^{ - a}\,\cos(b)-e^{  a}\,x\,\sin(b)\qquad ,\\
&V_+=e^{  a}\,x\, \cos(b)+e^{-a}\,\sin(b) \qquad ,\\
&V_-=e^{ a} \sin(b)\qquad ,
\end{align}
\end{subequations}
the $\uuline{g}\strut_{(0)}$ musical dual one-form to the tangent vector field of the orbits of the $K$-subgroup left transformations is:
\begin{align}
\oneform{\xi}_{\text{\tiny {(R)}}}=\uform{\rho}^3=-(db+\ft 12\,e^{2\,a}\,dx)\qquad .
\end{align}
 The coset space $K\setminus SL(2,\,\mathbb R)$ may be mapped on the lower sheet of the hyperboloid of equation: $-(x^3)^2+(x^1)^2+(x^2)^2=-1$ in  $M^{(2,1)}$, a Poincaré plane,  via the diffeomorphism ${\proj}_K:{\boldsymbol z}\mapsto \{x^1,\,x^2,\,x^3\}$ where:
 \begin{subequations}
\begin{align}
 &x^1=\ft12\,Tr[{\boldsymbol z}^{-1}X_3{\boldsymbol z}X_1]=(U_+ V_+- U_-V_-)\quad,\\
 &x^2=\ft12\,Tr[{\boldsymbol z}^{-1}X_3{\boldsymbol z}X_2]=-\ft12(U_+^2-U_-^2-V_+^2+V_-^2)\quad,\\
 &x^3=-\ft12\,Tr[{\boldsymbol z}^{-1}X_3{\boldsymbol z}X_3]=\ft12(U_+^2-U_-^2-V_+^2+V_-^2)\} \quad .
\end{align}
 \end{subequations}
The pull-back of the Cartan-Killing metric  now is:
\begin{align}
   {\proj}^\star_K(\eta)=4\,da^2+e^{4\,a}\,dx^2
\end{align}
leading to the warped metric:
\begin{align}
ds^2_{(\lambda)}= da^2+\ft 14 e^{4\,a}dx^2-(1-\lambda)(db-\sinh(2\,x)\,da)^2\qquad .\label{TRiemSum}
\end{align}
\end{itemize}
To summarise we have highlighted, on AdS$_3$ space, two trivial principal bundle structures whose base spaces are ``hyperbolic spheres'',  motivating the word ``Hopf-type''. \footnote{ These fibrations constitute the $SL(2,\,\mathbb{R})$ analogues of the standard Euclidean Hopf fibration: $$SU(2)\simeq S^3 \rightarrow U(1)\backslash SU(2)\simeq S^2\qquad .$$}
The writing of the metrics adapted to these fiber bundle descriptions of AdS$_3$ (Eqs \eqref{SRiemSum}, \eqref{TRiemSum}) exhibits  geometries given by Riemannian sums of a one-dimensional fiber of topology $\mathbb R$   with a two-dimensional AdS$_2$ space (Eq.\ (\ref{SRiemSum}))or a Poincaré plane (Eq.\ (\ref{TRiemSum})). 
}

{\subsection{Warped \texorpdfstring{AdS$_3$}{XXAdS3} geometries as solutions of gravity models}
\label{subsec:gravtheories}
Warped geometries occur as exact solutions in the framework of many three-dimensional gravity theories. For instance, in  Ref. \cite{Detournay:2005fz} they are presented as various deformations of string background field configurations of the $SL(2,\,\mathbb{R})$ Wess-Zumino-Witten model. Hereafter, we briefly consider three gravity theories for which a systematic, geometrical approach allows to clarify and extend already known solutions. \\
{We recall that since the one-from $\xiR$ is invariant, $\vxiR\cdot\vxiR$ is a constant. We denote it from now  by
\begin{align}
\epsR:=\vxiR\cdot\vxiR\qquad .
\end{align}
Accordingly, we have \begin{align}
\xi_{\text{\tiny(R)}\rho}\,g_{(\lambda)}^{\rho\sigma}\,\xi_{\text{\tiny(R)}\sigma}=\frac{\epsR}{1+\lambda\,\epsR}\qquad .\label{glambxixi}
\end{align}
 From the expression (\ref{glamKV}) of the metric we obtain as scalar curvature
\begin{align}
  &R=-2(3-\lambda\,\epsR)\label{Rscal}
\end{align}
and the two  identities, generalising similar ones occurring in Ref. \cite{banados2006three} 
\begin{align}
  &  R_{(\lambda)\,\mu\nu}=-2(1-\lambda\,\epsR)g_{(\lambda)\,\mu\nu}-4\,\lambda(1+\lambda\,\epsR)\,\xi_{\text{\tiny(R)}\mu}\,\xi_{\text{\tiny(R)}\nu}\label{Rddgddfd}\\
 &   \overset{\lambda}\nabla_\mu\,\xi_{\text{\tiny(R)}\nu}=\sqrt{1+\lambda\,\epsR}\,\eta_{\mu\nu\rho}\, g_{(\lambda)}^{\rho\sigma}\,\xi_{\text{\tiny(R)}\sigma}\label{covDfd}
\end{align}
 where $\eta_{\mu\nu\rho}:=\sqrt{-g}\,\epsilon_{\mu\nu\rho}$ is the Levi-Civita tensor, $R_{(\lambda)\,\mu\,\nu}$  the Ricci tensor obtained from the metric $g_{(\lambda)\,\mu\,\nu}$ (Eq. (\ref{glamKV})) and $\overset{\lambda}\nabla$ its associated Levi-Civita connection. In case we rescale the metric \eqref{glamKV} by a factor ${\ell}^2$ and $\oneform{\xiR}$ by ${\ell}$, the left hand-side of Eqs (\ref{Rscal}) and (\ref{Rddgddfd}) have to be multiplied by ${\ell}^2$ and the one of Eq. (\ref{covDfd}) by ${\ell}$. Notice that the determinant of the Ricci tensor allows to fix this factor. Indeed, taking into account a prefactor, we have
 \begin{align}
 {\ell}^6\, \det (R^\mu_\nu)=- 8(1-\lambda\,\epsR) \left(1-\lambda^2\,{\epsRsq}\right) \qquad.
 \label{DetRUd}
\end{align}
 These relations are particularly useful to establish the results that follow.
}
\subsubsection{Warped \texorpdfstring{AdS$_3$}{XXAdS3} geometries as solutions to topologically massive gravity}\label{SolTMG}
The condition (\ref{contdef})  allows us to interpret the metrics (\ref{glamKV}) as   solutions (see Ref.\ \cite{Anninos:2008fx} and references therein) of topologically massive gravity (see Refs \cite{Deser:1981wh,chow2010classification}):
\begin{align}
 &R_{\mu\,\nu} - \ft{1}{2} g_{\mu\, \nu} R +\Lambda\,g_{\mu\,\nu}-\ft{1}{\mu} \,C_{\mu\,\nu}=0\label {TMGeqs}\qquad ,
 \end{align}
{where the Cotton tensor is defined as:}
\begin{align}
  &C_{\mu\,\nu}:= {\sqrt{\vert g\vert}} \,\epsilon_{\mu\rho\sigma}\,\nabla^\rho\Big(R_\nu^\sigma-\ft 14\,\delta_\nu^\sigma\,R \Big)\qquad ,\label{CottonTens}
 \end{align}
for 
\begin{align}
    &\Lambda=-\frac{3-\lambda\, {\epsR}}3\qquad\mbox{\rm and}\qquad \mu=-3\,\sqrt {1+\lambda\,
     {\epsR}
    }
    \qquad .\label{LambmuTMG}
\end{align}
Note that these last equations show that for 
\begin{equation}
    3-\lambda\, {\epsR} \leq 0
\end{equation}
we obtain solutions of {topologicaly  massive} gravity
with zero or positive cosmological constant.}\par
 {In the same way we obtain solutions of new massive  gravity, advanced by Bergshoeff {\it et al.} \cite{Bergshoeff_2009a,Bergshoeff_2009b}, with field equations:
\begin{eqnarray}
\label{eomNMG} 
&R_{\mu\nu}-\frac{1}{2}g_{\mu\nu}R+\Lambda g_{\mu\nu}-\frac{1}{2\xi\,}\,K_{\mu\nu}=0\qquad,\label{NMGeqs}
\end{eqnarray}
where
\begin{eqnarray}
&K_{\mu\nu}=2\nabla^2R_{\mu\nu}-\frac{1}{2}\nabla_\mu\nabla_\nu R+\frac{9}{2}R\,R_{\mu\nu}-8R_\mu{}^\kappa R_{\nu\kappa}
+g_{\mu\nu}\left(3R_{\kappa\lambda}R^{\kappa\lambda}-\frac{1}{2}\nabla^2R-\frac{13}{8}R^2\right)\qquad. \nonumber
\end{eqnarray}
The metric (\ref{glamKV})  solves these equations for the values of the physical parameters given by:  
\begin{align}
    \Lambda=-\frac{\left(35+30\,\lambda\, {\epsR}-21\,\lambda^2\, {\epsRsq}\right)}{2\left(17+21\,\lambda\, {\epsR}\right)}\qquad\text{and}\qquad \xi=\frac{17+21\, \lambda\, {\epsR}}2
    \qquad.
\end{align}
In case of the so-called general massive gravity theory, defined as a combination of the topological and the new massive gravity theories, 
with field equations:
\begin{eqnarray}
R_{\mu\nu}-\frac{1}{2}g_{\mu\nu}R+\Lambda g_{\mu\nu}-\frac{1}{\mu}\,C_{\mu\nu}-\frac{1}{2\xi\,}\,K_{\mu\nu}=0\qquad\label{GMGeqs}
\end{eqnarray}
again the metric (\ref{glamKV}) constitutes a solution of the field equation for
\begin{subequations}
\begin{align}
    &\Lambda= -\frac{\left(35+30\,\lambda\, {\epsR}-21\,\lambda^2\, {\epsRsq}\right)}
    {2\left(17+21\,\lambda\, {\epsR}\right)}
    -\frac{\left(3-2\,\lambda\, {\epsR}-21\,\lambda^2\, {\epsRsq}\right)\,\sqrt{1+\lambda\, {\epsR}}}{2\,\mu\,\left(17+21\,\lambda\, {\epsR}\right)}\qquad,\\
    &  \xi=\frac{\left(17+21\, \lambda\, {\epsR}\right)\mu}{2\,\left(3\,\sqrt{1+\lambda\, {\epsR}}+\mu\right)}\qquad.
\end{align}
\end{subequations}
}
\subsubsection{Warped \texorpdfstring{AdS$_3$}{XXAdS3} geometries as solutions to Einstein-Cartan gravity}\label{SolECG}
In Refs \cite{Andrianopoli:2023dfm, PhysRevLett.133.031602} the authors consider the Einstein-Cartan gravity field equations:
\begin{align}
&\partial_\mu\,\beta_\nu-\partial_\nu\,\beta_\mu= 2\,\tau \,\sqrt{-g}\,\epsilon_{\mu\nu\rho}\,\beta^\rho\label{betaeq}\qquad,\qquad \\
&{R_{\mu\nu}-\ft{1}{2} g_{\mu\nu}\,R}=\tau ^2\,g_{\mu\nu}-\beta_\mu\,\beta_\nu-\beta_{(\mu;\nu)}+\beta^\rho_{;\rho}\,g_{\mu\nu}\qquad,
\label{Einseq}
\end{align}
where the metric is 3-dimensional Lorentzian, $\oneform{\beta}$ a one-form, and $\tau$ a constant.\\
Warped AdS$_3$ geometries provide solutions of them\footnote{These solutions appear in Refs\cite{Andrianopoli:2023dfm, PhysRevLett.133.031602};  expressing them as warped AdS$_3$ geometries clarifies their properties.} with the one-form $\oneform{\beta}$ being aligned with the one deforming the bi-invariant metric. A  {direct } calculation leads to  {the set of solutions}:
\begin{align}
{\ell}=\sqrt{1+\lambda\,\epsR}/\tau\qquad,\qquad q=2\,\lambda\,\sqrt{1+\lambda\,\epsR}
\end{align}
\begin{itemize}
\item Spacelike warping ($\epsR=+1$)
 \begin{subequations}
\begin{align}
&ds^2={\ell}^2\left(- \oneform{\rho}^0\otimes \oneform{\rho}^0+ \oneform{\rho}^1\otimes \oneform{\rho}^1+(1+\lambda )\,\oneform{\rho}^2\otimes \oneform{\rho}^2\right)\\
&\oneform{\beta}= q\,\oneform{\rho}^2
\end{align}
\end{subequations}
\item Timelike warping ($\epsR=-1$)
\begin{subequations}
\begin{align}
&ds^2={\ell}^2\left((\lambda-1) \oneform{\rho}^0\otimes \oneform{\rho}^0+ \oneform{\rho}^1\otimes \oneform{\rho}^1+ \oneform{\rho}^2\otimes \oneform{\rho}^2\right)\\
&\oneform{\beta}=q\,\oneform{\rho}^0
\end{align}
\end{subequations}
Notice that the condition $\lambda>0$ makes these spaces chronologically {safe} (no closed causal curves).
\item Lightlike warping ($\epsR=0$)
\begin{subequations}
\begin{align}
&ds^2= {\ell}^2\left(-\oneform{\rho}^0\otimes \oneform{\rho}^0+ \oneform{\rho}^1\otimes \oneform{\rho}^1+ \oneform{\rho}^2\otimes \oneform{\rho}^2+ \lambda  \,(\oneform{\rho}^0-\oneform{\rho}^2)\otimes (\oneform{\rho}^0-\oneform{\rho}^2)\right)\\
&\oneform{\beta}=q\,(\oneform{\rho}^0-\oneform{\rho}^2)
\end{align}
\end{subequations}
\end{itemize}
Here, $\lambda$ is the warping parameter used also in the rest of this work and $L$ is a free parameter.

{\subsubsection{Warped \texorpdfstring{AdS$_3$}{XXAdS3} geometries as Einstein-Maxwell-Chern-Simons gravity solutions}\label{SolEMCSG}
Warped   AdS$_3$ metrics also appear as solutions to
 the three-dimensional Einstein-Maxwell-Chern-Simons theory exposed in  Ref.\cite{banados2006three}. 
 It is a theory involving a 3-dimensional metric $g_{ab}$ and a potential $A_a$. The field equations are expressed in terms of the two-form Maxwell field $F=dA$ and its dual one-form $\mathcal F =\star F$; in components: $\mathcal F_a=\ft 12\,\sqrt{-g}\,\epsilon_{abc}F^{bc}$ where we choose $\epsilon_{012}=+1$. Expressed in coordinates, they read:
\begin{align}
&\partial_a\mathcal F_b-\partial_b\mathcal F_a-=2\,{\alpha_{M}}\,F_{ab}\qquad ,\label{BBCGMax}\\
&R_{ab}-\ft12 g_{ab}\,R= \Lambda\,g_{ab}+\ft12\left(\mathcal F_a\mathcal F_b-\ft12\,g_{ab}\,\mathcal F^c\mathcal F_c\right)\qquad ,\label{BBCGEin}
\end{align}
where $\Lambda$ denotes the cosmological constant and ${\alpha_{M}}$ the so-called topological electromagnetic mass {(a constant, not to be confused with the function introduced later, in Eq.\ (\ref{ClstpR}))}. These equations are very close to Eqs (\ref{betaeq}, \ref{Einseq}) and accordingly admit warped AdS$_3$ geometries with the potential $A$ aligned with the deforming one-form as solutions.\\
This leads to 
{the set of solutions for which:
\begin{align}
\label{233oneeq}
    q^2=-2\,\lambda /{\ell}^2\qquad,\qquad {\alpha_M}=-\sqrt{1+\lambda\, {\epsR}}/{\ell}<0\qquad,\qquad \Lambda=\left(1-\lambda\, {\epsR}\right)/{\ell}^{2}
\end{align}
\begin{itemize}
\item Spacelike warping (${\epsR}=+1$)
\begin{subequations}
\begin{align}
&ds^2={\ell}^2\,\left(- \oneform{\rho}^0\otimes \oneform{\rho}^0+ \oneform{\rho}^1\otimes \oneform{\rho}^1+(1+ \lambda) \,\oneform{\rho}^2\otimes \oneform{\rho}^2\right)\\
&A= q\,\oneform{\rho}^2
\end{align}
\end{subequations}
\item Timelike warping (${\epsR}=-1$)
\begin{subequations}
\begin{align}
&ds^2={\ell}^2\,\left((-1+ \lambda) \oneform{\rho}^0\otimes \oneform{\rho}^0+ \oneform{\rho}^1\otimes \oneform{\rho}^1+\,\oneform{\rho}^2\otimes \oneform{\rho}^2\right)\\
&A= q\,\oneform{\rho}^0
\end{align}
\end{subequations}
\item Lightlike warping (${\epsR}=0$)
\begin{subequations}
\label{233twoeq}
\begin{align}
&ds^2= {\ell}^2\,\left(- \oneform{\rho}^0\otimes \oneform{\rho}^0+ \oneform{\rho}^1\otimes \oneform{\rho}^1+\,\oneform{\rho}^2\otimes \oneform{\rho}^2+\lambda\,(\oneform{\rho}^0-\oneform{\rho}^2)\otimes (\oneform{\rho}^0-\oneform{\rho}^2)\right)\\
&A=  q\,(\oneform{\rho}^0-\oneform{\rho}^2)
\qquad.
\end{align}
\end{subequations}
\end{itemize}
}
All these geometries, as well as those obtained in sections \ref{SolTMG} and \ref{SolECG}, admit four Killing vectors: the three left invariant ones and the right one selected for the warping. They remain solutions that admit two Killing vectors once quotiented 
(see Sec.\ \ref{sec:warpedAdS3BH1}).\\ 
The three kinds of solutions \eqref{233oneeq}--\eqref{233twoeq} require that {${\alpha_{M}}<0$  (a consequence of our choice of the orientation and our assumption that $q$ is a constant)} in order for the Maxwell-Chern-Simons equation (\ref{BBCGMax}) to admit solutions in the framework of the ansatz used. {The condition to provide} (real) solutions of signature $\{+,\,+,\,-\}$ {makes $ {\alpha_{M}}$ real. For timelike warped geometries we need that $1>\lambda$ so } the cosmological constant is bounded {from below} by the mass parameter:{ $ 2/{\ell}^2>\Lambda>  {{\alpha_M}^2}$}.   Similarly the spacelike warped solution requires {$\lambda>-1$} for having {the metric signature $\{+,\,+,\,-\}$ and } the Maxwell-Chern-Simons potential real. In the case of solutions built on a lightlike warping ansatz, the cosmological constant has to be positive and is fixed by the mass parameter while the warping parameter always is arbitrary but negative: $\lambda=-q^2\,\Lambda$.\\
Just for completeness let us mention that by warping the round metric of $S^3\sim SU(2)$ we also may obtain solutions of Eqs (\ref{BBCGMax}, \ref{BBCGEin}) with negative cosmological constant: $\Lambda<- {{\alpha_M}^2}$. But in this case since the spacetime is compact and homogeneous, it is totally vicious (see Prop. 6.4.2, Ref.\cite{HawkingEllis}).
}

\section{Geodesics on Warped \texorpdfstring{AdS$_3$}{AdS3} Spaces}\label{ExpGeodWadSSol}
In this section, we will study geodesics on warped AdS$_3$ spaces.\footnote{Actually we will work on $SL(2,\,\mathbb R)$, but the raising to
its universal cover
is obvious.} Denoting by $\overset{\lambda}\nabla$ the Levi-Civita connection obtained from the warped metric for the value $\lambda$ of the deformation parameter, the geodesic equations read
\footnote{This equation can be reinterpreted as the equation of motion of a charged particle (of charge $q:=\lambda \,\xi_\alpha u^\alpha$), interacting with the right invariant ``Maxwell" field {obtained from the invariant Killing vector}:\,$F_{\mu\nu}:= \overset{0}{\nabla}_{\mu}\,\xi_{\nu}$,  satisfying the field equation $\overset{0}{\nabla}_\mu\,F^{\mu\alpha}=-2\,R^{\alpha\nu}\,\xi_\nu$.  }:
\begin{align}
    \overset{\lambda}{\nabla}_u u^\alpha=\overset{0}{\nabla}_u u^\alpha+\lambda\,(\xi_\beta u^\beta)\,\overset{0}{\nabla}_u \xi^\alpha=0\qquad ,\label{geodgenlambeq}
\end{align}
{
where in this section we will omit the index ``R'', to ease the notation.
}
This writing of the geodesic equations results from the particular properties that (right-)invariant Killing vector fields share on a group manifold. They are of  constant norm (denoted by $\epsilon$) and their orbits are geodesics for any deformation parameter $\lambda$:
\begin{align}
  \uuline{g}\strut_{(0)}(\vec \xi,\,\vec \xi)=\epsilon\qquad,\qquad \overset{\lambda}{\nabla}_\xi \xi^\alpha=0 \qquad.
\end{align}
It is worthwhile to notice that the warping we consider exhibits analogies
  to that introduced by Kerr and Schild \cite{1965cngg.conf..222K, Kerr:1965wfc}. Indeed, the Kerr metric can be obtained from the four-dimensional Minkowski metric using a lightlike affinely parametrised geodesic  vector and  a function as warping parameter. %{It is worthwhile to notice that the warping we consider is similar to that introduced by Kerr and Schild \cite{1965cngg.conf..222K,Kerr:1965wfc}.The Kerr metric can be obtained from the four-dimensional Minkowski metric using a lightlike affinely parametrised geodesic  vector and using a function as warping parameter.}
  \\
The  geodesic equations (\ref{geodgenlambeq}) may be integrated more or less easily in various coordinate systems (fibered, Schwarzschild etc.), but we find it simplest to use the  usual $M^{2, 2}$  coordinates and to start from a Lagrangian including a Lagrange multiplier $q$ to fix the motion on the AdS$_3$ hyperboloid. We will show that the geodesic equations then reduce to   elementary damped harmonic oscillator equations of motion: Eq.\ \eqref{geodesiceq}.

We consider affinely parametrized geodesics obtained from the Lagrangian
\begin{equation}\label{M22WLag}
L_\lambda = \frac{1}{2} \eta_{a b} \dot X^a \dot X^b 
+ \lambda \frac{1}{2} \left( \Xi_a \dot X^a \right)^2
+ q \frac{1}{2} \left({\eta_{a b}\,} X^a X^b +1\right)\qquad,
\end{equation}
where the metric of warped AdS$_3$ has been written as 
\begin{equation}
g_{a b} := \eta_{a b} + \lambda  \, \Xi_a  \,\Xi_b\,,  \qquad( \eta_{ab}) := 
-\frac 12\,\left(
\begin{array}{cccc}
 0 &1   & 0  &0 \\
 1 & 0  & 0  &0 \\
0  & 0  & 0  &1 \\
0  & 0  & 1  & 0   
\end{array}
\right)=:\frac 14\,(\eta^{ab})
\end{equation}
and its inverse metric is given by
\begin{equation}
    \label{inversemetric}
g^{ab} = \eta^{a b} - \frac{\lambda}{1+\epsilon\, \lambda\, } {\Xi^a\,} \Xi^b\qquad,
\end{equation}
where $\epsilon$ is the norm on $\mathcal H$ of the invariant Killing form used for the deformation
\begin{equation}
    {\eta_{a b}\,} {\Xi^a\,} \Xi^b = \epsilon \qquad,
\end{equation}
which we can always normalize to  $-1$, 0, or $+1$.\\
Any Killing form of AdS$_3$ may be written as 
\begin{equation}
    \Xi_a = K_{[a b]} X^b\qquad,
\end{equation}
where $K_{[a b]}$ denotes an antisymmetric tensor with constant components (the generators of $so(2,2)$ acting on usual coordinates of     $M^{2, 2}$). Left and right invariant form are obtained from self-dual and anti-self-dual such tensors. It holds that 
\begin{subequations}
    \label{iden-lag-emb}
\begin{align}
    \Xi_a X^a &= 0\qquad,\\
    \dot \Xi_a \dot X^a &= 0\qquad, 
\end{align}
\end{subequations}
From Eq.\ (\ref{KVLR}) we may distinguish six special matrix generators $\underset{p}{\mathcal{R}}\strut^a_{\ b}$  and $\underset{p}{\mathcal{L}}\strut^a_{\ b}$ such that the right and left invariant Killing vector components {read}:
\begin{equation}
    r_{(p)}^a=\underset{(p)}{\mathcal{R}}\strut^a_{\ b}\,X^b\qquad,\qquad  l_{(p)}^a=\underset{(p)}{\mathcal{L}}\strut^a_{\ b}\,X^b\qquad p=1,\,2,\,3\ \text{and }a=1,\dots,4\qquad .
\end{equation}
These generators satisfy the relations:
\begin{align}
  &   \underset{(p)}{\mathcal{R}}\strut^{(a}_{\ c}\, \underset{(q)}{\mathcal{R}}\strut^{b)}_{\ c}\,\eta^{cd}=-\eta^{(3)}_{pq}\,\eta^{ab}\qquad,\qquad   \underset{(p)}{\mathcal{L}}\strut^{(a}_{\ c}\, \underset{(q)}{\mathcal{L}}\strut^{b)}_{\ c}\,\eta^{cd}=-\eta^{(3)}_{pq}\,\eta^{ab}\qquad ,\label{normRL}\\
  &\underset{(p)}{\mathcal{R}}\strut^{a}_{\ c}\, \underset{(q)}{\mathcal{L}}\strut^{c}_{\ b}-\underset{(q)}{\mathcal{L}}\strut^{a}_{\ c}\, \underset{(p)}{\mathcal{R}}\strut^{c}_{\ b}=0\label{comRL1}
\end{align}
where $p,\,q=1,\,2,\,3$ and $a,\,b=1,\dots,4$.

Variation of the Lagrangian, Eq.\ (\ref{M22WLag}),  with respect to $q$ yields 
\begin{equation}
    \label{constraint1X}
    {\eta_{a b}\,} X^a X^b +1 = 0
\end{equation}
and as a consequence we have 
\begin{equation}
    \label{constraint2X}
    {\eta_{a b}\,} X^a \dot X^b = 0\qquad.
\end{equation}
Moreover, as  we restrict ourselves to   invariant Killing vectors $\Xi^a=\mathcal{R}\strut^a_{\ b}\,X^b$, we obtain from this last equation and Eq.\ (\ref{normRL}) that
\begin{equation}
    {\eta_{a b}\,} {\Xi^a\,} \dot \Xi^a =0\qquad, 
\end{equation}
{\it i.\ e.} on shell the invariant Killing vectors are of constant norm. 
\par\noindent
Variation with respect to $X^a$ yields 
\begin{equation}\label{VarXa1}
(\eta_{a b} + \lambda \Xi_a \Xi_b) {\ddot X}^b = q\, {\eta_{a b}\,} X^b - 2 \lambda (\Xi_b \dot X^b) \dot \Xi_a\qquad.
\end{equation}
Now and in the following, we define
\begin{align}
X_a &\equiv {\eta_{a b}\,} X^b \qquad ,\\
\Xi^a &\equiv \eta^{a b} \Xi_b\qquad , \\
k &\equiv \Xi_a \dot X^a\qquad . \label{defofk}
\end{align}
Thus, the equation of motion, Eq.\ (\ref{VarXa1}),  may be rewritten 
\begin{equation}\label{EqXa}
    (\eta_{a b} + \lambda \Xi_a \Xi_b) {\ddot X}^b = q\, X_a - 2\, \lambda\, k \dot \Xi_a\qquad.
\end{equation}
Multiplying with the inverse metric Eq.\ \eqref{inversemetric} we obtain 
\begin{equation}
    \label{geodnonsimp}
    \ddot X^a = \left(\delta^a_{~b} - \frac{\lambda}{1+\epsilon \lambda}{\Xi^a\,} \Xi_b \right)
    \left(q\, X^b - 2 \,\lambda\, k\, \dot \Xi^b\right)\qquad.
\end{equation}
Upon using \eqref{iden-lag-emb}, \eqref{geodnonsimp} reduces to
\begin{equation}
    \label{geodesiceq}
    {\ddot X}^a = - 2\, \lambda\, k \,\dot \Xi^a + q X^a\qquad.
\end{equation}
Hence, on shell we have 
\begin{equation}
\dot k = \Xi_a \ddot X^a = 0\qquad,
\end{equation}
confirming that $k$ is a constant of motion, as it must be, $\Xi$ being a Killing vector.
Contracting \eqref{geodesiceq} with $X_a$, we obtain
\begin{align}
&q\,X^a\,X_a=\ddot X^a\,X_a+2\,\lambda\,k\,\dot \Xi^a\,X_a
= \ddot X^a\,X_a- 2\, \lambda\, k^2\nonumber\\
&= \frac d{ds}(\dot X^a\,X_a ) -2\,L-\lambda\,k^2 = -2\,L-\lambda\,k^2\qquad.
\end{align}
Hence, as the Lagrangian is a constant of motion, $q$ also is constant on shell. We obtain:
\begin{equation}
    L=\frac12(q-\lambda\,k^2)\qquad.
\end{equation}
Let us now consider 
\begin{equation}
\dot X^a g_{ab} \dot X^b = \dot X^a \eta_{ab} \dot X^b + \lambda \dot X^a \Xi_a \Xi_b \dot X^b = \dot X^a \eta_{ab} \dot X^b + \lambda k^2\qquad.
\end{equation}
As
\begin{equation}
    \dot X^a \eta_{ab} \dot X^b= -2\, \lambda\, k^2 + q\qquad, \label{Xdotwitheta}
\end{equation}
it holds that 
\begin{equation}
    \dot X^a g_{ab} \dot X^b =q - \lambda k^2\qquad.
\end{equation}
Finally, as left and right invariant Killing vectors commute between themselves, we obtain, deforming the AdS$_3$ bi-invariant metric with a right invariant Killing vector, three extra constants of motion related to the left invariant Killing vector fields.
We obtain, as a consequence of Eq.\ (\ref{comRL}) the constants of motion
\begin{align}
     \underset{(p)}{\ell }:= l_{(p)}^a\,\eta_{ab}\dot X^b+\lambda\, k\,l_{(p)}^a\Xi_a\qquad;
\end{align}
that are related, on shell, to the previous ones by:
\begin{align}
    \underset{(p)}{\ell}\underset{(q)}{\ell}\,\eta^{(3){pq}}=q - \epsilon\,\lambda^2\,k^2 \qquad.
\end{align}
\subsection{Explicit solutions }
\label{spacelikedef}
We now consider more specifically a  spacelike  deformation with respect to $\vec r_{1}$, Eq. \eqref{KVLR}. 
In terms of null coordinates we have to consider
\begin{align}
    (K_{[ab]})&= 
   \frac12\, \left(
    \begin{array}{cccc}
     0 & +1  & 0 &0\\
    -1&0   & 0  &0 \\
    0 &0   & 0  &+1 \\
    0 &0   &-1 &0 
    \end{array}
    \right)\qquad, 
    \end{align}
such that  
\begin{equation}
     \Xi_a = K_{[a b]} X^b = (\oneform{\rho}^{1})_a =\ft 12 \{+U_-, -U_+,+V_-, -V_+\}\qquad,
\end{equation}
which yields the geodesic equations
\begin{subequations}
    \label{uxgeod}
\begin{align}
\ddot U_\pm \pm 2\, \lambda\, k \,\dot U_\pm - q\, U _\pm&= 0\qquad,\\
\ddot V_\pm \pm 2\, \lambda\, k\, \dot V_\pm - q\, V_\pm &=0\qquad.
\end{align}
\end{subequations}
Solving \eqref{uxgeod}, we find
\label{solution}
\begin{align}
   U_\pm &= \frac{e^{\mp p\,s}}2\left(\left(U_{\pm,0}+\frac{\dot U_{\pm,0}\pm p\,U_{\pm,0}}{\sqrt{p^2+q}} \right)e^{\sqrt{p^2+q}\,s}
     +\left(U_{\pm,0}-\frac{\dot U_{\pm,0}\pm p\,U_{\pm,0}}{\sqrt{p^2+q}}\right)e^{-\sqrt{p^2+q}\,s}\right)\qquad,
\end{align}
where we have defined $p = \lambda k$ and $U_\pm(0) = U_{\pm,0},\ \dot U_{\pm}(0) = \dot U_{\pm,0}$. 
The initial conditions need to be consistent with \eqref{constraint1X} and \eqref{constraint2X}. Hence, they must fulfill
\begin{subequations}
\begin{align}
   & U_{+,0}\,U_{-,0} +V_{+,0}\,V_{-,0} = +1\qquad, \\
   & U_{+,0}\, {\dot U}_{-,0} +U_{-,0} \,{\dot U}_{+,0}+V_{+,0}\, {\dot V}_{-,0} +V_{-,0} \,{\dot V}_{+,0} =0\qquad. \label{ud0is0}
\end{align}
\end{subequations}
Furthermore, due to \eqref{defofk} and \eqref{Xdotwitheta} we have 
\begin{align}
   &k  = \ft12(U_-\,\dot U_+-U_+\,\dot U_-+V_-\,\dot V_+-V_+\,\dot V_-) \qquad,\\
   &q - 2\, \lambda\, k^2  = - {\dot U_+}\, {\dot U_-} -{\dot V_+}\, {\dot V_-} \qquad.
\end{align}
The solution, Eq.\ \eqref{solution}, remains regular in the limit where $p \to 0$ and $q \to 0$.
The explicit form of the solutions shows the geodesic completeness of  spacelike  warped AdS$_3$. 

In case of a timelike warping, let's say with respect to the vector $\vec r_3$ we have to consider $\Xi_a=\ft12\{V_+,\,V_-,\,-U_+,\,-U_-\}$ and the equations of motion:
\begin{subequations}
\begin{align}
&\ddot U_+ - 2\, \lambda\, k \,\dot V_- - q\, U _+ = 0\qquad,\\
&\ddot V_- + 2\, \lambda\, k \,\dot U_+ - q\, V _- =0\qquad,\\
&\ddot U_- - 2\, \lambda\, k \,\dot V_+ - q\, U _- = 0\qquad,\\
&\ddot V_+ + 2\, \lambda\, k \,\dot U_- - q\, V _+ =0\qquad,
\end{align}
\end{subequations}
while for the lightlike warping, with respect to $\ft 1{\sqrt 2}(\vec r_2+\vec r_3)$, we obtain $\Xi_a=\ft 1{\sqrt 2}\,\{0,\,V_-,\,0,\,-U_-\}$ and the equations of motion:
\begin{subequations}
\begin{align}
&\ddot U_+ - 2\, \lambda\, k \,\dot V_- - q\, U _+ = 0\qquad,\\
&\ddot V_- - q\, V _- =0\qquad,\\
&\ddot V_+ + 2\, \lambda\, k \,\dot U_- - q\, V _+ =0\qquad,\\
&\ddot U_-  - q\, U _- = 0\qquad.
\end{align}
\end{subequations}
Lastly, let us emphasise that the warping of the metric qualitatively changes the behaviour of some timelike geodesics. While on undeformed AdS$_3$ ($\lambda=0$) all timelike geodesics are closed curves and thus each may be enclosed in a compact set, when $\lambda\neq 0$ this does not hold true anymore generically. From the previous equations we easily observe that for the case of spacelike warping there exist unbounded timelike geodesics {\it i.e.} that cannot be enclosed in a compact subset. We also have such unbounded geodesics in case of timelike or lightlike warpings if $\lambda>0$ while for $\lambda\leq 0$ they remain bounded (but not necessarily closed).

\section{Warped \texorpdfstring{AdS$_3$}{AdS3} Quotients}\label{sec:warpedAdS3BH1}

Interesting geometries can be obtained from warped AdS$_3$ spaces by taking quotients, similarly as to how the BTZ black holes were obtained from AdS$_3$ by taking quotients. This is done by specifying, in addition to the right invariant Killing vector $\vxiR$, a left invariant Killing vector field $\vxiL $. We assume them to be generically independent and define the identification vector 
\begin{align}
\label{identif}
\vec \partial=\LL\,\vec{\xi}_{\text{\tiny {(L)}}}+\LR\,\vec {\xi}_{\text{\tiny {(R)}}}\qquad .
\end{align}

The quotient structure is obtained by compactifying the orbits of this vector field {\it i.e.} considering the quotient manifold obtained through the action of the subgroups of left and right transformations (see Eq.\ \eqref{XRLxiRL})
\begin{align}
\bs{z}\sim Exp[2\,n\,\pi\, X_ L]\,\bs{z}\,Exp[-2\,n\,\pi \,X_ R]\qquad \text{with }n\in\mathbb Z\qquad,\label{Identif}
\end{align}
where $X_R$ ({\it resp.}  $X_L$ ) is the right ({\it resp.}  left) generator associated to $\LL\,\vxiL $ ({\it resp.}  $\LR\,\vxiR$).\footnote{Actually, any discrete subgroup  of $SL(2,\,\mathbb{R})$ isometrically acting (by
left-translations) on our warped AdS$_3$ spacetime could in principle replace the
$X_R$-commuting subgroup elements $Exp[2\,n\,\pi\, X_ L]$ we consider here. For instance, choosing the subgroup
isomorphic to the fundamental group of a compact Riemann surface would lead
to a compact quotient (and thus to a space admitting at least one closed causal
curve). Of course, the analysis, in those cases, is much more involved to carry out.  We thank an anonymous referee for inspiring us to consider this possible  generalisation.} \\
Before we discuss the nature of these spaces, let us introduce local coordinates adapted to this structure.  Hereafter, we assume the warping to be non-zero ($\lambda\,\vxiR\neq \vec0$). We may, {\it a priori}, distinguish between three kinds of structures: the self-dual quotient where $\LL=0$, the anti-self-dual quotient with $\LR=0$ (equivalent to the self-dual one) and the generic quotient ($\LL\,\LR\neq 0$). Moreover, each of these structures subdivides into subclasses according to the nature (timelike, spacelike or lightlike) of the right and left Killing vector fields involved.
\subsection{Preferred local coordinates on (warped) \texorpdfstring{AdS$_3$}{AdS3}}\label{AdcanlocCoord}
 Hereafter we build ``intrinsic''   
 coordinate systems, having in mind a generic identification vector.\\
Preferred coordinates are obtained as follows. We start from the two vector fields $\vxiR$ and $\vxiL $ that we write as:
\begin{align}
\vxiR=x^a\,\vec r_{(a)}\qquad ,\qquad \vxiL =y^a\,\vec l_{(a)}
\end{align}
and the vectors normal to surfaces $H=\mathrm{const.}$, see Eq.\ (\ref{nH}).\\
Then we introduce a (non-trivial) one-form $\oneform\sigma$ such that 
\begin{align}
\oneform\sigma(\vxiR)=\oneform\sigma(\vxiL )=\oneform\sigma(\vnH )=0\qquad.\label{defsigma}
\end{align}
Defining
\begin{align}
  \kappa_{a\,b}=\vec r_{(a)}\cdot\vec l_{(b)} \qquad,\qquad  \kappa^a_{\phantom{a}b}:=H\,\oneform{\rho}^a(\vec l_{(b)}) = (\eta^{(3)})^{a\,c}\,\kappa_{c b}\qquad,\label{kappaab}
\end{align}

we express $\oneform\sigma$ in terms of right invariant one-forms: $\oneform\sigma=\sigma_a\,\oneform\rho^{(a)}$ where (up to a factor):
\begin{align}
\sigma_a=\epsilon_{a\,b\,c}\,x^b\,\kappa^c_{\hphantom {c}p}\,y^p\qquad .
\end{align}
From this one-form we may build a vector $\vec\sigma=\sigma^a\,\vec r_{(a)}$ orthogonal to the three previous ones $\vxiR$, $\vxiL $ and $\vnH $.\\
Let us now define several scalars on $M^{2,2}$:
\begin{align}
&\vxiR\cdot\vxiR=\epsR\,H\qquad, \qquad (\epsR:=x^a\,\eta_{a\,b}\,x^b)\\
&\vxiL \cdot\vxiL =\epsL\,H\qquad, \qquad (\epsL:=y^a\,\eta_{a\,b}\,y^b)\\
&\vxiR\cdot\vxiL =\kappa =x^a\,\eta_{a\,b}\,\kappa^b_{\hphantom{b}c}\,y^c\qquad ,\label{defkappa}\\
&\vec{\sigma}\cdot\vec{\sigma}=H\,\oneform{\sigma}(\vec \sigma)=H\,\left(\kappa^2-\epsR\,\epsL\,H^2\right) \qquad .\label{normsig}
\end{align}
Hereafter, we  denote by $\Delta$:
\begin{align}
\Delta:= \kappa^2-\epsR\,\epsL\,H^2 \qquad . \label{defDelta}
\end{align}
It is only when $\Delta\neq 0$ that $\vxiR$ and $\vxiL $ are linearly independent.\newline
Without loss of generality, we also assume that the vectors $\vxiR$ and $\vxiL $ are normalised such that:
\begin{align}
 \epsR=\pm1,\ 0\qquad\text{and}\qquad \epsL=\pm1, \ 0\qquad.
 \end{align}
 
 These conventions fix for each choice of {the}  vectors $\vxiL $ and $\vxiR$ the range of variation of $\kappa$ that soon will be interpreted as a ``radial'' coordinate.\\
The four vector fields: $\vnH ,\,\vec \sigma,\,\vxiR,\,\vxiL $ define, at each point where $\Delta\neq 0$, a frame of the tangent space. By duality, we obtain two fields of one-forms $\{ \oneform{p}^R,\,\oneform{p}^L\}$ that  with $\{-\ft1{4\,H}\,dH,\,\oneform{\sigma}\}$ provide frames  of the cotangent spaces where $H\neq 0$. Thus, the one-forms $ \oneform{p}^R$ and $\oneform{p}^L$ are such that their contractions are:
\begin{align}
&\oneform{p}^R(\vxiR)=1\qquad,\qquad \oneform{p}^L(\vxiL )=1 \qquad,\label{pRLxiRL}\\
&\oneform{p}^R(\vxiL )=\oneform{p}^R(\vec\sigma)=\oneform{p}^R(\vnH )=0\label{pRxiLsign}\qquad,\\
&\oneform{p}^L(\vxiR)=\oneform{p}^L(\vec\sigma)=\oneform{p}^L(\vnH )=0\qquad .\label{pLxiRsign}
\end{align}
Let us notice that:
\begin{align}
\oneform\sigma=\ft12\,H\,d(H^{-1}\,\kappa)\qquad,
\label{Clstsig}
\end{align}
while the components of $\oneform{p}^R=p^R_a\,\oneform\rho^{(a)}$ and $\oneform{p}^L=p^L_a\,\oneform\rho^{(a)}$  are:\footnote{Warning: notice that both are expressed with respect to right invariant one-forms but are not invariant one-form themselves.}
\begin{align}
&p^R_a=\eta^{(3)}_{a\,b}\left(\kappa\,\kappa^b _{\hphantom{b} p}\,y^p-H^2\,\epsL\,x^b\right)/\Delta\qquad ,\\
&p^L_a=\eta^{(3)}_{a\,b}\left(\kappa\,x^b-\epsR\,\kappa^b _{\hphantom{b} p}\,y^p\right) \,H/\Delta\qquad .
\end{align}
{These one-forms are closed and thus locally exact (see Appendix \ref{proofdpLR}), which allows to write them as differentials of functions $\alpha$ and $\beta$}:
\begin{align}
&d  \oneform p^R =0 \qquad{i.\,e.}\qquad  \oneform p^R= d\alpha\qquad,\label{ClstpR}\\
&d  \oneform p^L =0 \qquad {i.\,e.}\qquad  \oneform p^L=d\beta\qquad .\label{ClstpL}
\end{align}
Now we have all the ingredients needed to rewrite the metric of AdS$_3$ in a ``preferred'' system of coordinates adapted to the choice of the generators we made. It will be expressed in terms of the pullback  $h_\star$ of the embedding of AdS$_3$ on $\mathcal H\subset M^{2,\,2}$. To avoid an inflation of notation, we denote the pullback of a function with the same letter as the function itself and write:
\begin{align}
&h_\star(\kappa)=\kappa \qquad ,\\
&h_\star(\Delta)=\Delta=\kappa^2-\epsR\,\epsL\qquad .
\end{align}
Eqs \eqref{Clstsig}, \eqref{ClstpR} and \eqref{ClstpL} are of paramount importance. They show that the pullbacks of the forms $\oneform\sigma,\, \oneform p^L,\, \oneform p^R$ are (locally) exact on $\mathcal H$ (exact on $\mathcal H$ where $\Delta\neq 0$).
 They define local coordinates: $\{\kappa,\,\alpha,\,\beta\}$. We obtain for the one-forms:
\begin{align}
&h_\star(\oneform\sigma)=1/2\,d\kappa\qquad ,\\
&h_\star(\oneform p^R)= d\alpha\qquad ,\\
&h_\star(\oneform p^L)= d\beta\qquad .
\end{align}
The metric components of ${ \uuline{g}}\strut_{(0)}$ are easily obtained by comparing the scalar products of the restrictions on $\mathcal H$ 
of the vectors $\vec\sigma$, $\vxiR$ and $\vxiL $ amongst themselves. Comparing their values, calculated using the expression of the metric in terms of the right invariant one-form, Eq.\ \eqref{glamKV}, and in terms of $(\alpha,\,\beta,\,\kappa)$ coordinates, we obtain:
\begin{align}
{ \uuline{g}}\strut_{(0)}=\epsR\,d\alpha^2+2\,\kappa\,d\alpha\,d\beta+\epsL\,d\beta^2+\frac1{4\,(\kappa^2-\epsR\,\epsL)}\,d\kappa^2\qquad .\label{metcan}
\end{align}
Let us notice that when both $\vxiR$ and $\vxiL $ are timelike ($\epsR=\epsL=-1$), we obtain that $\vert \kappa\vert \geq 1$, a restriction which ensures that the metric signature remains Lorentzian.

To obtain the warped metric, we notice that $\oneform \xiR =x_a\oneform\rho ^a$ (defined on $\mathcal H$), that  is the musical 
dual  one-form  of  the vector $\vxiR$, can be written: 
\begin{align}
\oneform \xiR=h_\star \left( \epsR\,\oneform p^R+\kappa\,\oneform p^L\right)=\epsR\,d\alpha+\kappa\,d\beta \qquad.\label{fxir}
\end{align}
Thus, the expression of the warped metric is obtained by adding to 
Eq.\ (\ref{metcan}) the $SL(2,\mathbb R)$ symmetry breaking term $ \lambda\,\oneform\xiR\otimes\oneform\xiR=\lambda\,(\epsR\,d\alpha+\kappa\,d\beta)\otimes (\epsR\,d\alpha+\kappa\,d\beta)$, leading to:
\begin{align} { \uuline{g} }\strut_{(\lambda)}=\epsR\,(1+\epsR\,\lambda)\,d\alpha^2+2\,\,(1+\epsR\,\lambda)\,\kappa\,d\alpha\,d\beta+(\epsL+  \lambda\,\kappa^2)\,d\beta^2+\frac1{4\,(\kappa^2-\epsR\,\epsL)}\,d\kappa^2\qquad .\label{wmetcan} \end{align}
Equations (\ref{ClstpR}), (\ref{ClstpL}) become singular when $\Delta$ vanishes. This corresponds to a singularity of the $g_{(\lambda)\,\kappa\kappa}$ component of the metrics (\ref{metcan}) and (\ref{wmetcan}). However, the determinant of these metrics remains regular, indicating that the surfaces of constant {$\kappa$} corresponding to zeros of $\Delta$ are coordinate singularities. They may define horizons or potential topological singularities once some identification are performed. When $\epsR=\epsL$ they are located where $\xiR=\pm \xiL$. If $\epsR=0$ and $\epsL=1$ or $\epsL=0$ and $\epsR=1$, they are located where $\xiR\cdot\xiL=0$. If $\epsR=0$ and $\epsL=-1$ or $\epsL=0$ and $\epsR=-1$, they are never reached, as they are pushed to infinity.

The hyperboloid $\mathcal H$ is invariant with respect to the symmetry that maps a point on its antipodal one:
\begin{align}
z\mapsto A(\bs{z}):=-\bs{z}\qquad,\qquad \{U_+,\,U_-,\,V_+,\,V_-\}\mapsto  \{-U_+,-U_-,-V_+,-V_-\}\qquad.
\end{align}
The left and right invariant Killing vectors being also invariant, the coordinates $\alpha$, $\beta$ and $\kappa$ are  even functions of the null coordinates.  They define local coordinate charts. In section [\ref{sec: proj}] we show how they are glued together, such that they cover all the space.
\vspace{1mm}\\
Moreover, elementary hyperbolic geometry shows that for $\epsR=-1$ and $\epsL =-1$, $\kappa$ is of constant sign and of absolute value never less than 1 ($\vert\kappa\vert\in\mathbb R_{\geq1}$). 
If both  $\epsR=0$ and $\epsL =0$, $\kappa$ is still of constant sign. When $\epsR=1$ or $\epsL=1$, there is no restriction: $\kappa\in\mathbb R$.\\
Notice also that $\alpha$ is an even function of $\vxiL $ but  $\kappa$ and $\beta$ are odd functions. As a consequence, $\Delta$, considered as function of $\kappa$, has two single roots: $\kappa=\pm 1$ if and only if $\epsR=\epsL=+1$, a double root $\kappa=0$ when $\epsR\,\epsL=0$ but only one single root $\kappa=+1$ or $\kappa=-1$ when $\epsR=\epsL=-1$.

\noindent To close this section, we observe that, as $\oneform \xiR\wedge d\oneform \xiR=\epsR\,d\alpha\wedge d\kappa\wedge d\beta$, Eq.\ \eqref{fxir} tells us that, in the present three-dimensional case, the one-form $\oneform \xiR$ is a contact form unless $\vxiR$ is lightlike ($\epsR=0$, in which case it is integrable). In particular (even when $\vxiR$ is lightlike), the vector field
$\vxiR$ generates a one-dimensional Lie group which is the structure group of the bundle with total space 
    AdS$_3$ and symplectic base equivariantly symplectomorphic to the coadjoint orbit of the element $\oneform \xiR|_e$ in the dual $sl(2,\mathbb{R})^\star$ of the Lie algebra $sl(2,\mathbb{R})$. Note that, when contact, the vector field $\vxiR$ is (proportional to) the Reeb field of the contact structure (see for instance Ref.\cite{geiges2004contactgeometry}) {\it i.e.} the vector field ${\bf r}$ determined by the conditions $i_{{\bf r}}{\rm d}\oneform \xiR=0$ and $\oneform \xiR({\bf r})=1$.

\subsubsection{Locally warped  geometries as  gravity model solutions}
\label{locwarp}
The geometry studied by Anninos {\it et al.} in Ref. \cite{Anninos:2010pm} (also considered by \cite{Clement:1994sb}) constitutes a very interesting extension of the warped geometries considered up to here. Let us briefly discuss them in the framework of our preferred coordinate system. It consists of a lightlike warping ($\epsR=0$)  but modulated,  by a function $\lambda(\kappa)$, depending only on $\kappa$ (instead of a constant)
so that  {$\vxiR$ and $\vxiL$} remain Killing vectors.  This ansatz provides new solutions of various  massive gravity theories. They  are particularly easy to obtain using the preferred coordinate system. For instance, as obtained in Ref.\cite{Anninos:2010pm}, for 
\begin{align}
    \lambda(\kappa)=\lambda_0\,\kappa^{\mu/2-3/2}+c_1/\kappa+c_2/\kappa^2,\qquad{ \Lambda=-1,}
    \end{align}
$\lambda_0,\  c_1, \ c_2$  being constants, {the metric ( \ref{wmetcan}), with $\epsilon_{\alpha,\beta,\kappa}=1$, is a solution} of topologically massive gravity  ( \ref{TMGeqs}).  Using the identities Eqs (\ref{glambxixi}) -- (\ref{DetRUd}) we find that it is only for  $\mu=-1$ or $+1$ that the geometry is locally  AdS$_3$ and for $\mu=3$ that it is a warped one in the sense of Eq.(\ref{glamKV}). The computation of the Killing vectors confirms this:  if $\mu \neq -1,\ 1 ,\ 3$ there are only two (local) Killing vectors, but four when $\mu=3$ and six when $\mu=-1$ or $+1$. 
\\
\par In the same way we obtain solutions of the equations   of new and general massive  gravity theories advanced by Bergshoeff {\it et al.} \cite{Bergshoeff_2009a,Bergshoeff_2009b}. The field equations of new massive gravity are displayed in Eq. (\ref{NMGeqs}).
The metric {( \ref{wmetcan})} with $\epsR=0$ and the constant $\lambda$ replaced by
\begin{align} 
\lambda(\kappa)=&\kappa^{-3/2}\left(\lambda_+\,\kappa^{\sqrt{(1+ {2\,} \xi)/8}}+\lambda_-\,\kappa^{-\sqrt{(1+ {2\,} \xi)/8}} \right) 
 +c_1/\kappa+c_2/\kappa^2\ ,
\end{align}
$\lambda_+,\ \lambda_-,\  c_1, \ c_2$  being constants, solves them  {when the cosmological constant takes the value $\Lambda=-(1+1/4\,\xi)$}.\\
\par  In case of the so-called general massive gravity theory, defined as a combination of the topological and the new massive gravity theories 
{\it i.e.} with field equations (\ref{GMGeqs})
the metric {( \ref{wmetcan})} with $\epsR=0$, where now the constant $\lambda$ is replaced by the function:
\begin{align}{
    \lambda(\kappa)=}&{ \kappa^{-(3/2+\xi/4\,\mu)}\left(\lambda_+\,\kappa^{\sqrt{(2\,\mu^2+4\,\mu^2\,\xi+\xi^2)  }/4\,\mu}+\lambda_-\,\kappa^{-\sqrt{(2\,\mu^2+4\,\mu^2\,\xi+\xi^2)  }/4\,\mu}\right)}\nonumber\\
    &+c_1/\kappa+c_2/\kappa ^2 
\end{align}
($\lambda_+,\ \lambda_-,\  c_1, \ c_2$ being four arbitrary constants), solves them .
\\
\par In the framework of the Einstein-Maxwell-Chern-Simons theory we also obtain locally lightlike ($\epsR=0$) warped configurations that solve the field equations (\ref{BBCGMax}) and (\ref{BBCGEin}). {Putting $\epsilon_{\alpha,\kappa,\beta}=1$, they} read as :
\begin{subequations}
\begin{align}
{
    q=q_0 \kappa^{-(1+ {\alpha_M}\,{\ell})}+c_3/\kappa\ ,\quad \lambda(\kappa)= {-}\frac{q_0^2\,{\alpha_M}\,{\ell}^3}{2(1+2\,{\alpha_M}\,{\ell})}\kappa^{-2(1+\mathbf{\alpha} \,{\ell})}+c_1/\kappa+c_2/\kappa^2\ ,\quad \Lambda=1/{\ell}^2\ ,
    }
\end{align}
$q_0$, $c_1$, $c_2$ and $c_3$ being constants. \\
{In the special case where the topological mass ${\alpha_M}=-1/2\ell$, the solution involves a logarithmic term}:
\begin{align}
{
q=q_0/\sqrt{ \kappa}+c_3/\kappa\ ,\quad \lambda(\kappa)={-}\frac{q_0^2\,{\alpha_M}\,{\ell}^2}{4\,\kappa}\log(\kappa)+c_1/\kappa+c_2/\kappa^2\ ,\quad \Lambda=1/{\ell}^2\ .}
\end{align}
\end{subequations}
 \par 
Notice that by allowing for $q$ to be a function, we are no more restricted to impose ${\alpha_M}<0$.
\\
When $\epsR=0$ and we restrict the local warping function to $\lambda(\kappa)=c_1/\kappa+c_2/\kappa^2$, the modified metric remains locally AdS$_3$ but with different values of the parameters $\LL,\ \LR$ {(See next section)} and with $\epsL$ replaced by $\text{sign}(\epsL+c_1)$ . When $\epsR\neq0$ there is no non-zero residual local warping that preserves the local AdS$_3$ character of the metric.   We deserve a detailed study of these geometries for future works.
To be complete, let us mention that the same considerations apply to the metric (\ref{wmetcanPhisd}) but lead to a Gödel-like spacetime (totally vicious spacetime). In this case, the solutions are the same except that the parameter $\mu$ has to be replaced by $-\mu$ if we adopt the orientation $\epsilon_{\tau\kappa\varphi}=1$ in both cases. This sign flip is an immediate consequence of the change of coordinates, equations (\ref{tauab}, \ref{phiab}) and (\ref{tauabsd}, \ref{phiabsd}), which link the metrics (\ref{wmetcanPhi}) and (\ref{wmetcanPhisd}) to the metric (\ref{wmetcan}).
\subsection{Preferred local coordinates on (warped) \texorpdfstring{AdS$_3$}{AdS3} quotients} 

The (non-zero) identification vector used to quotient the warped  geometry is given by a linear combination of a  left invariant generator $\vxiL ${ --- that can be chosen among the left invariant generators similarly as in Eqs (\ref{spwf}--\ref{liwf}) –––} and the vector $\vxiR$ defining the warping
\begin{align}
\vec \partial=\LL\,\vxiL +\LR\,\vxiR=\LL\,\partial_{\beta}+\LR\,\partial_{\alpha}\neq \vec 0\qquad.\label{vecpart}
\end{align}
This vector is defined up to a sign. We always may assume that $\LR\geq 0$  and $\LL\geq 0$ ($\LL> 0$ if $\LR=0$) acting with an isometry in case of spacelike warping and distinguishing between future and past oriented in case of timelike or lightlike warping (see Table \ref{TableItoIII}). Let us remind that the $sl(2,\,\mathbb R)$ Lie algebra --- and as a consequence, the AdS$_3$ tangent spaces --- has a three-dimensional Minkowski space geometrical structure. Their generators --- and as a consequence, the invariant vector fields they define --- live on various orbits: the origin, one-sheeted spacelike hyperboloids, future or past (half) null cones, future or past sheets of two-sheeted timelike hyperboloids. Thus, we have to consider the following {thirteen}   possible cases depending on the nature (spacelike,  timelike or lightlike) of the vectors $\vxiR$ and $\vxiL $.  
\begin{table}[ht]
\begin{center}
   \begin{tabular}{ |l | c |c| c | }
     \hline
      {\bf\ Type} & $\vxiR$ & $\vxiL $ & $\kappa \ (= \vxiR\cdot\vxiL )\  \text{range} $\\\hline
     $\ \bs{I }_a$ &spacelike &  timelike&$\kappa\in \mathbb R $\\\hline
     $\ \bs {  I }_b$ &spacelike & spacelike&$\kappa\in \mathbb R $\\\hline
     $\ \bs {\rmII}_a$ &spacelike & lightlike&$\kappa\in \mathbb R $\\ \hline
     $\ \bs { I }_c^+$ & $f-$timelike  & $f-$timelike &$\kappa\leq -1 $\\\hline
      $\ \bs { I }_c^-$ & $f-$timelike  &$p-$timelike &$\kappa\geq 1 $\\\hline
       \rule[0cm]{0cm}{4mm} {${\tilde{\bs {{ I }}}}_a$} & timelike  & spacelike &$\kappa\in\mathbb R   $\\\hline 
    $\ \bs {\rmII }_b^+$ & $f-$timelike  & $f-$lightlike &$\kappa< 0   
    $\\\hline 
        $\ \bs {\rmII }_b^-$ &$f-$timelike  & $p-$lightlike &$\kappa > 0$\\\hline 
        \rule[0cm]{0cm}{4mm}\  {${ {\tilde {\bs{\rmII}} }}_b^+$} & $f-$lightlike  & $f-$timelike &$\kappa< 0  $\\\hline
        \rule[0cm]{0cm}{4mm}   {${{\tilde{\bs{\rmII}}}}_b^-$} & $f-$lightlike  & $p-$timelike &$\kappa> 0   $\\\hline
        \rule[0cm]{0cm}{4mm}\  {${{\tilde {\bs{\rmII} }}}_a$} & lightlike  & spacelike &$\kappa\in\mathbb R   $\\\hline
        $\ \bs {\rmIII }^+$ & $f-$lightlike  &$f-$lightlike &$\kappa\leq 0   $\\ \hline
         $\ \bs {\rmIII }^-$ & $f-$lightlike & $p-$lightlike &$\kappa\geq 0$\\ \hline  
         \end{tabular}
         \caption{$f-$: future pointing, $p-$: past pointing.\label{TableItoIII}}
\end{center}
 \end{table} \\
  Moreover, acting with appropriate $O(2,\,2)$ transformations we may assume prescribed values of the vector fields $\vxiR$ and  $\vxiL $ at the origin (see section \ref{Expexp}). 
  In section \ref{sec: proj}, we will analyse the causal structure of the metric \eqref{wmetcanPhi},  using projection diagrams. For this, we recall the ranges of parameters in \eqref{wmetcanPhi}. A priori, $\epsR, \epsL \in \{-1, 0, 1\}$, $\LL, \LR \in \mathbb{R}$, $\tau \in \mathbb{R}$ and $\kappa \in \mathbb{R}$ up to additional restrictions coming either from horizons, closed timelike curves or topological singularities. Since the overall sign in front of the identification vector \eqref{vecpart} does not matter, we may assume $\LR \geq 0$ without loss of generality. Moreover, as we distinguish between future and past-oriented generators (when they are timelike or lightlike),  the relevant range of parameters we have to take into account is: $\epsR, \epsL \in \{-1, 0, 1\}$ (and
  $\LR\geq 0$  and $\LL\geq 0$; $\LL> 0$ if $\LR=0$).

  The orbits of the identification vector are closed. Accordingly, for the resulting spacetime to be non-totally vicious, see Ref. \cite{minguzzi2008causal}, there must be a non-pathological chronological  region, a domain where the identification vector is spacelike with respect to the physical metric {\it i.e.} where
\begin{align}
{ \uuline{g} }\strut_{(\lambda)}(\vec\partial,\,\vec\partial)={ \uuline{g}}\strut_{(0)}(\vec\partial,\,\vec\partial)+\lambda \,\left(\oneform{\xi}(\vec\partial)\right)^2
\end{align}
is positive.

A necessary condition characterizing  chronologically {safe} regions, the regions without closed causal curves, is:
\begin{align}
&{ \uuline{g} }\strut_{(\lambda)}(\vec\partial,\,\vec\partial)=\epsL\,\LL^2 +\epsR \LR^2 +2\,\LL\,\LR\,\kappa+\lambda \left(\epsR\,\LR + \LL\,\kappa\right)^2 >0\qquad .\label{ChronoCrit}
\end{align}
We insist that {\it a priori} this condition is only necessary but not sufficient. 
 \\
If $\LL=0$ this requires $\epsR=+1$: among the self-dual quotient space geometries only the spacelike ones could be non-totally vicious.\\
When $\LL\neq 0$ and
\begin{align}
(1+\lambda\, \epsR)\,\LR^2 -\epsL\,\lambda\,\LL^2<0\qquad ,
\end{align}
and if moreover $\lambda<0$ the space admits closed causal curves through any point. \\ Anyway, to avoid closed causal curves at infinity, we recover a condition already imposed:
$\lambda>0$.
Taking into account the condition Eq. (\ref{contdef}), this restricts $ 1>\lambda>0 $  in case of timelike warping and $ \lambda >0 $ when considering spacelike or lightlike warping.\\

On the other hand, if
\begin{align}
(1+\lambda\, \epsR)\,\LR^2 -\epsL\,\lambda\,\LL^2>0
\label{Dd2}
\end{align}
there are two values of $\kappa$ , depending on the ratio 
$B=L_R/L_L$, that we denote $\kappa_{\text{\tiny CTC}}^+$ and $\kappa_{\text{\tiny CTC}}^-$ 
\begin{align}
    \kappa^\pm_{\text{\tiny CTC}}=\frac{-(1+\epsR\,\lambda)\,B\pm\sqrt{(1+\epsR\,\lambda)\,B^2-\epsL\,\lambda}}{\lambda}\label{kappaCTCpm}
\end{align}
for which ${ \uuline{g} }\strut_{(\lambda)}(\vec\partial,\,\vec\partial)$ vanishes.\\ 
Defining $r=C( \kappa -\kappa_{\text{\tiny CTC}}^+)$ or $r=C( \kappa -\kappa_{\text{\tiny CTC}}^-)$, we obtain an expression of the metric such that the chronological singularity is located on $r=0$ and the horizons or topological singularities are either located at $r_\pm=C(\kappa_{\text{\tiny CTC}}^+\pm 1)$ or $r_\pm=C(\kappa_{\text{\tiny CTC}}^-\pm 1)$. This clarifies the origin of the isometry, noted in \cite{Jugeau:2010nq}, between metrics with two different values  of $r_+$ and $ r_-$,
resulting in the same two invariants $\LR$ and $\LL$ (or $T_L$ and $T_R$).

 For the following considerations, it is more convenient to use an angular coordinate $\varphi$ that explicitly reflects the $2\pi$-periodicity implied by the identifications, along with another coordinate that ranges over~$\mathbb{R}$:
  \begin{subequations}
  \label{tauphitoalphabeta1}
  \begin{align}
  &\tau=\alpha-\frac{\LR}{\LL}\,\beta\qquad ,\label{tauab}\\
  &\varphi =\beta/\LL\qquad .\label{phiab}
  \end{align}
  \end{subequations}
 Using these coordinates, the metric is given by: 
\begin{align} 
{ \uuline{g} }\strut_{(\lambda)}=~& \epsR\,(1+\epsR\,\lambda)\,d\tau^2
+2\, (1+\epsR\,\lambda)\,(\LL\,\kappa +\epsR\,\LR)\,d\tau\,d\varphi\nonumber\\
&+\left(\epsL\,\LL^2 +\epsR \LR^2 +2\,\LL\,\LR\,\kappa+\lambda \left(\epsR\,\LR + \LL\,\kappa\right)^2\right)\,d\varphi^2\nonumber\\
&+\frac1{4\,(\kappa^2-\epsR\,\epsL)}\,d\kappa^2
\label{wmetcanPhi} \end{align}
and the identification of points simply reads $(\tau, \kappa, \varphi) \sim (\tau, \kappa, \varphi + 2 \pi)$. Notice that when   $ \epsL=0$    we may  reabsorb $\LL\neq 0$ into $\kappa$ by rescaling the lightlike vector $\vxiL $.\\
 On the other hand, to avoid closed causal curves at infinity ($\vert\kappa\vert \mapsto \infty$) we have to impose the condition:
 \begin{equation}
     \lambda>0\qquad . \label{lambpos}
 \end{equation}
  Moreover, if $\LR=0$, the vector $\xiR$ only plays a r\^ole in the construction of the coordinate system. In this case, it has no intrinsic meaning and all three coordinate systems leading to the metrics (\ref{wmetcanPhi}) with $\epsR=\pm 1$ or $0$ are related by coordinate transformations, implicitly given by the equations of subsection \ref{Expexp}.

The coordinate change, Eqs (\ref{tauab}-- \ref{phiab}) is not adapted for self-dual configurations, those with $\LL=0$ (and $\LR>0$). In these cases we define
  \begin{subequations}
  \label{tauphitoalphabeta2}
 \begin{align}
  &\tau=\beta \qquad ,\label{tauabsd}\\
  &\varphi =\alpha/{\LR}\qquad, \label{phiabsd}
  \end{align}
  \end{subequations}
  which leads to the local expression of the metric:
  \begin{align} 
{ \uuline{g} }\strut_{(\lambda)}=& ( \epsL+\lambda\,\kappa^2)\,d\tau^2
+2\, \kappa\, (1 +\lambda\,\epsR)\,\LR\,d\tau\,d\varphi
+\epsR\,(1+\lambda\,\epsR)\,\LR^2\,d\varphi^2
+\frac1{4\,(\kappa^2-\epsR\,\epsL)}\,d\kappa^2\qquad .\label{wmetcanPhisd} \end{align}
For the quotient space with this warped metric to be not totally vicious we have to require  $\epsR=1$. Then, we may assume condition (\ref{contdef}) instead of  (\ref{lambpos}). 
\subsection{Photon rings %and turning points
}
%I changed title here
In Ref.\cite{DKLSW2024} the importance of photon rings, in the context of $\bs{I}_b$ geometries (Eq. (\ref{wmetcanPhi})) with $\epsR=\epsL=+1$), is emphasised. Here we want to briefly describe them, using the preferred coordinates, as they are particularly simple in this framework (and in our opinion). \newline
These geometries present chronological singularities when the ratio 
\begin{equation}
\label{ratioB}
B= \frac{\LR}{\LL}\qquad, 
\end{equation}
of the weight of the right and left invariant Killing vectors defining the identification vector $\vec\partial$ satisfies (see Eq. (\ref{Dd2})):
\begin{align}
B^2 \geq {\frac \lambda{(1+\lambda)}}\qquad .
\end{align}
In this case, the chronological singularities are located at 
\begin{align}
    { \kappa}_{\text{\tiny CTC}}^\pm =\frac{-(1+\lambda) B \pm\sqrt{(1+\lambda)B^2-\lambda}}{\lambda}<0\qquad .
    \label{kappactcphoton}
\end{align}
The qualitative analysis of null geodesics in the framework of black holes is easily obtained from the expression of the metric (\ref{wmetcanPhi}) with $\epsR=\epsL=+1$. Let us define 
\begin{align}
 b=L_{(\text{\tiny L})}\,p_\tau/p_\varphi   \qquad,
\end{align} the (rescaled by $\LL$) ratio of the conserved momenta associated to the $\tau$ and $\varphi$ coordinates, themselves associated to the two commuting Killing vectors of the metric. 
From the zero-length condition of the null geodesic and their two constants of motion, we infer that:
\begin{align}
    \dot\kappa^2=4\Big((b(\kappa+ B)-1)^2-b^2\frac{\kappa^2-1}{1+\lambda}\Big)=:U[\lambda,B,b,\kappa]\qquad .
\end{align}
where, for future use, we defined a function $U[\lambda,B,b,\kappa]$ that has to be non-negative along the null geodesic.\newline

Photon rings are null geodesics invariant with respect to the two parameter isometry group generated by $\vxiL $ ans $\vxiR$.  They are obtained by solving the equations
\begin{align}
    U[\lambda, B, b ,\kappa]=0\qquad,\qquad \partial_\kappa U[\lambda, B,b,\kappa]=0\qquad
\end{align}
whose solutions are 
\begin{align}
 &b_\pm=   \frac{(1+\lambda)\,B\pm\sqrt{\lambda(1+\lambda)}}
 {(1+\lambda)\, B^2-\lambda}\qquad ,\\
&\tilde \kappa_\pm  = \pm \sqrt{ (1+\lambda)/\lambda}\qquad .
\end{align}
One is given by a positive value of the $\kappa$ coordinate, the other by a negative value. They are outside the region $-1\leq\kappa\leq 1$, bounded by the horizons (between which $\kappa$ is a timelike coordinate). When they exist they correspond to an inner photon ring (located between the singularity and the inner horizon) and an outer photon ring (located between the outer horizon and infinity). \newline
 When $B^2<\lambda/(1+\lambda)$ there is no chronological singularity and the geometries present two photon rings. When $B^2=\lambda/(1+\lambda)$ the metric presents  a  double chronological singularity, coinciding with the inner photon ring $\tilde{\kappa}^-$.  As a consequence the relevant geometries possess only one photon ring: 
 \begin{align}
  {\tilde\kappa}^+=\sqrt{ (1+\lambda)/\lambda}\qquad .
 \end{align}
 This endures as far as  $\sqrt{\lambda/(1+\lambda)}< B <(2+\lambda)/\sqrt{\lambda(1+\lambda)}$ in which cases we obtain that 
 \begin{align}
 \kappa_{\text{\tiny CTC}}^-< {\tilde \kappa}^-<\kappa^+_{\text{\tiny CTC}}   <0<\tilde \kappa^+\qquad,
 \end{align}
while for $(2+\lambda)/\sqrt{\lambda(1+\lambda)}< B$ the two photon rings are greater than $\kappa_{\text{\tiny CTC}}^+$.

\subsection{Decrypting the coordinate systems}
\label{subsec:decrypt}
In the following sections we provide, for various choices of the vector fields $\vxiR$ and $\vxiL $, explicit expressions of the preferred coordinates $\kappa$, $\alpha$ and $\beta$ in terms of null coordinates, as well as the inverse expressions: the parametrisation of the null coordinates defining the embedding of $\mathcal H$ in $M^{2,2}$ leading to the metrics expressed in local coordinates, Eqs (\ref{metcan}), (\ref{wmetcan}).\\
The first ones are obtained from Eq.\ (\ref{defkappa}) and the integration of Eqs (\ref{ClstpR}), (\ref{ClstpL}). The inverse relations are obtained by comparing the scalar products of  right and left Killing vector fields (see Eqs (\ref{kappaab})), expressed in local coordinates (see hereafter) with the same ones expressed in null coordinates by noticing that: 
\begin{subequations}
\begin{align}
    &U_\pm^2=\frac12\,\left((\kappa_{2\,2}-\kappa_{3\,3})\pm(\kappa_{2\,3}-\kappa_{3\,2})\right)\label{Upm2}\\
    &V_\pm^2=-\frac12\,\left((\kappa_{2\,2}+\kappa_{3\,3})\mp(\kappa_{2\,3}+\kappa_{3\,2})\right)\label{Vpm2}\\
    &U_+\,U_-=\frac12\,\left(1+\kappa_{1\,1}\right)\qquad,\qquad V_+\,V_-=\frac12\,\left(1-\kappa_{1\,1}\right)\\
    &U_+\,V_+=\frac12\,\left(\kappa_{3\,1}-\kappa_{2\,1}\right)\qquad,\qquad U_-\,V_+=\frac12\,\left(\kappa_{1\,2}-\kappa_{1\,3}\right)\\
    &U_+\,V_-=\frac12\,\left(\kappa_{1\,2}+\kappa_{1\,3}\right)\qquad,\qquad U_-\,V_-=-\frac12\,\left(\kappa_{3\,1}+\kappa_{2\,1}\right)\label{UVpm}
\end{align}
\end{subequations}
These equations provide the key to relate the various coordinate systems introduced in the literature to describe warped AdS$_3$ spaces .
The expressions of the Killing vector fields provide a straightforward way to obtain the equations of the embedding leading to a given local AdS$_3$ metric $g_{\mu\nu}$ in a coordinate system $x^\mu$.

The algorithm consists of the following steps. 
\begin{enumerate}
\item From the warped AdS$_3$ metric $g_{ \mu\nu}$ compute the 4 Killing vectors.
\item Identify among these four vectors one, $ \xi^\mu$, that commutes with all the others and its dual 1-form: $\xi_\nu=g_{\mu\nu}\xi^\mu$
\item Consider the metric $\gamma$ with components $\gamma_{\mu\nu}:=g_{ \mu\nu}-\lambda \,\xi_\mu\,\xi_\nu$ and compute the value of $\lambda$ such that the metric $\gamma_{\mu\nu}$ becomes an Einstein metric: $R(\gamma)^\mu_\nu\propto \delta ^\mu_\nu$
\item Rescale the metric $\gamma_{\mu\nu}$ by a factor such that $R(\gamma)^\mu_\nu=-2\, \delta ^\mu_\nu$
\item The rescaled metric is locally an AdS$_3$ metric. Compute (locally) its six Killing vectors (we already know four of them).
\item Split them into sets of invariant vector fields $\{\vec l_{a}\}$ and $\{\vec r_{a}\}$, normalised according to Eqs (\ref{norm}]), and such that they obey the commutation relations Eqs (\ref{comRL}) and the positivity constraints Eqs (\ref{Upm2}, \ref{Vpm2}) imply ($ - \kappa_{3\,3}\geq \text{Max.}\{\vert \kappa_{3\,2}\vert,\, \vert \kappa_{2\,3}\vert,\, \vert \kappa_{2\,2}\vert\}$)  .
\item Compare the  scalar products $\kappa_{a\,b}:=\vec r_{a} \cdot\vec l_{b} $ expressed in local coordinates with their expressions in null coordinates and deduce  parametrisations of the hyperboloid, Eq.(\ref{AdS30hyp}),  in terms of local coordinates.
\item Doing this for various coordinate systems provides implicit relations between them.
\end{enumerate}
In Appendix \ref{loctonullcoord} we illustrate the procedure with the example of the warped AdS$_3$ black hole written in the coordinates used in Eq.\  (4.1) Ref.\ \cite{Anninos:2008fx}.
\subsubsection{Killing vectors }\label{sectKV}
In Appendix \ref{Hparam} we describe an algebraic method to compute the parametrisation of the null coordinates of the embedding of $\mathcal H$ in     $M^{2, 2}$ from the expression in local coordinates of the bi-invariant metic. The starting point is the expressions of the Killing vectors.\\
As the right and left sector are similar, we restrict our considerations to the left one. The right one is obtained by permuting $\epsL$ with $\epsR$ and $\beta$ with $\alpha$. 

Two cases have to be considered, depending on whether  $\epsL $ is zero or not. From the metric Eq. (\ref{metcan}) elementary quadratures lead to:
\begin{itemize}
\item If  $\epsL\neq 0$\\
Let $s_\kappa=\text{sign}( \kappa^2-\epsR\,\epsL)$. (Let us remind that in general $s_\kappa=+1$ unless $\epsR=\epsL=+1$ in which case  $\kappa$ runs from $-\infty$ to $+\infty$.)\\
 We obtain the left invariant vectors:
\begin{align}
&\vec{\mu}_1=\varepsilon_1\,  \partial_\beta\quad ,\label{mu1}\\
&\vec \mu_2=\varepsilon_2\,\left(\frac{\cosh(2\,\sqrt{\epsL}\,\beta)}{ \sqrt{\vert\kappa^2-\epsR\,\epsL\vert}}(\partial_\alpha-\epsL\,\kappa\,\partial_\beta)+2\,s_\kappa\,\sqrt{\vert \kappa^2-\epsR\,\epsL\vert}\,\frac{\sinh(2\,\sqrt{\epsL}\,\beta)}{\sqrt{\epsL}}\,\partial_\kappa\right)\label{m2}
\quad,\\
&\vec \mu_3=\varepsilon_1\,\varepsilon_2\,\left(\frac{\sinh(2\,\sqrt{\epsL}\,\beta)}{ \sqrt{\epsL}\sqrt{\vert\kappa^2-\epsR\,\epsL\vert}}( \epsL\,\partial_\alpha-\kappa\,\partial_\beta)+2\,s_\kappa\,\sqrt{\vert \kappa^2-\epsR\,\epsL\vert}\,\cosh(2\,\sqrt{\epsL}\,\beta)\,\partial_\kappa \right)\quad ,\label{mu3}\\
&\varepsilon_1=\pm 1\qquad,\qquad \varepsilon_2=\pm 1
\end{align}
They are orthogonal between themselves and of norms:
\begin{align}
\vec \mu_1\cdot\vec \mu_1=\epsL\qquad,\qquad\vec \mu_2\cdot\vec \mu_2=-s_\kappa\,\epsL\qquad,\qquad\vec \mu_3\cdot\vec \mu_3=s_\kappa \qquad.
\end{align}
The correspondence between them and the ``standard" ones satisfying the commutation relations Eqs (\ref{comRL}) depends on the values of  $\epsL$ and $s_\kappa$. 
The Lie brackets of the vector fields Eqs (\ref{mu1}--\ref{mu3}) are given by:
\begin{align}
[\vec \mu_1,\,\vec \mu_2]= 2\, \vec \mu_3\quad,\quad[\vec \mu_2,\,\vec \mu_3]= 2\,s_\kappa\,\epsL\,\vec \mu_1\quad,\quad[\vec \mu_3,\,\vec \mu_1]=-2\,\epsL\,\vec \mu_2\quad.
\end{align}
Accordingly we obtain (up to  $SO(2,1)$ transformations): 
\begin{align}
&\text{If }\epsL=+1\text{ and }s_\kappa=-1\ :\ \vec l_1= \vec \mu_1\quad,\quad \vec l_2= \pm\vec \mu_2\quad,\quad \vec l_3= \pm\vec \mu_3\\
&\text{If }\epsL=+1\text{ and }s_\kappa=+1\ :\ \vec l_1= \vec \mu_1\quad,\quad \vec l_2= \pm\vec \mu_3\quad,\quad \vec l_3= \pm\vec \mu_2\\
&\text{If }\epsL=-1\text{ and thus }s_\kappa=+1\ :\ \vec l_1= \vec \mu_3\quad,\quad \vec l_2=\pm \vec \mu_2\quad,\quad \vec l_3= \pm\vec \mu_1
\end{align}

 The right invariant Killing vectors are given by a similar correspondence, but with opposite signs. The two possibilities of choosing the signs of $\vec l_2$ and $\vec l_3$ allow to always satisfy the positivity constraints  implied by  Eqs.\ (\ref{Upm2}), (\ref{Vpm2}).

\item {If $\epsL= 0$}\\
The left invariant Killing vectors reduce to:
 \begin{align}
&\vec{\mu}_1= \partial_\beta\quad ,\label{mu10}\\
&\vec \mu_2=2\,(\kappa\,\partial_\kappa-\beta\,\partial_\beta)\quad ,\label{mu20}\\
&\vec \mu_3=\frac 1\kappa\,\partial_\alpha+4\, \kappa\,\beta\,\partial_\kappa-\left(2\,\beta^2+\frac{\epsR}{2\,\kappa^2}\right)\partial_\beta\quad .\label{mu30}
\end{align}
Two  are lightlike. Their non-zero scalar products are:
\begin{align}
\vec \mu_1\cdot\vec \mu_3=1\qquad,\qquad\vec \mu_2\cdot\vec \mu_2=1\qquad,
\end{align}
and the linear combinations
\begin{align}
&\vec{l}_1= (\vec \mu_3+\vec\mu_1)/\sqrt{2}\quad ,\quad \vec{l}_2= \vec \mu_2\quad ,\quad \vec{l}_3= (\vec \mu_3-\vec\mu_1)/\sqrt{2}
\end{align}
satisfy the commutation relations Eqs (\ref{comRL}).

\end{itemize}
\subsubsection{Interlude: Killing spinors}
Having obtained the expressions of the Killing vectors, a natural question consists in looking 
for Killing spinors. 
As it is well known, their existence often signals that the underlying geometry 
is a solution, in the sector of vanishing fermionic fields, of the field equations of some supergravity models. {Indeed supersymmetric warped AdS$_3$ solutions were found in some 3-dimensional supergravities; see for instance \cite{PhysRevD.75.125015, Deger_2014, OColgain:2015jlg}. Here, however, we restrict ourselves to the discussion of the existence of Killing spinors.}

Only lightlike warped geometries possess such spinor fields when $\lambda\neq 0$. They have $\lambda $-independent scalar curvature $R=-6$ and admit a Killing spinor $\Sigma$ -- a solution to the equations:
\begin{align}
    \nabla_\mu\,\Sigma =\pm \frac 12\,\gamma_\mu\,\Sigma\qquad,\label{KSEq}
\end{align}
the sign depends on the representation of the Dirac matrices $\gamma_\mu$ used.\\
With respect to the frame:
\begin{align*}
& \oneform{f}^{(1)}=\frac 1{2\,\kappa}   \,d\kappa\qquad,\nonumber\\
 & \oneform{f}^{(2)}=\kappa\,d\alpha+\frac 12\,(\epsL+1+\lambda\,\kappa^2)   \,d\beta\qquad,\nonumber\\
 &\oneform{f}^{(3)}=\kappa\,d\alpha+\frac 12\,(\epsL-1+\lambda\,\kappa^2)   \,d\beta\qquad,\nonumber\\
\end{align*}
this spinor reads
\begin{align}
    \Sigma=\sqrt{\kappa}\,\frac 12\,(Id-\gamma^{1}) \,S\qquad,
\end{align}
where $S$ is an arbitrary constant spinor. Its norm is zero and the Killing vector it defines is $\partial_\alpha$, the (lightlike) one involved  in the warping of the AdS$_3$ geometry:
\begin{align}
 & \overline{\Sigma}\,\Sigma=0\qquad,\\
 & \overline{\Sigma}\gamma^\mu\,\Sigma=\{1,\,0,\,0\}\qquad.
\end{align}
This spinor is invariant with respect to the identification along the orbits of $\vec\partial$. Thus, it remains well-defined in the framework of the (extremal) black holes these identifications define.\footnote{Scrupulous readers may wonder if this Killing spinor is globally well defined. The resolution of the Killing spinor equation, Eq. (\ref{KSEq}),  using a global coordinate system yields a positive answer to this question.}\\
The existence of this Killing spinor opens up the way to interesting questions.    For instance, to examine the supersymmetric properties of warped AdS$_3$ black holes in the so-called quadratic ensemble \cite{ADMW}. These questions will not be considered here.

\subsubsection{Singularities, horizons and all that ...}\label{Singhoretc}
We now will examine in detail the various possibilities that the choices of $\vxiR$, $\vxiL $ and $\vec \partial$ define under the assumptions:
\begin{align}
    &\LR\geq 0\quad,\quad \lambda>0\quad,\quad 1+\epsR\,\lambda >0\qquad .
\end{align}
Singularities are of two types: chronological and topological.
Chronological singularities are defined as the boundaries of the regions where causal closed lines exist, topological singularities as fixed points of the identification relation Eq.\ (\ref{Identif}) such that the resulting space is no more a differentiable manifold (see for instance Ref.\cite{HawkingEllis}, section {\bf 5.8}).\\
Let us first consider chronological singularities. When they exist, these boundaries are given by the two roots (Eq.\ \eqref{kappaCTCpm})
of the quadratic equation defined by the condition Eq.\ \eqref{ChronoCrit}.
Moreover, to have a black hole geometry, the corresponding singularities have to be hidden by horizons. Horizons are null surfaces of constant $\kappa$ corresponding to roots of $\Delta$: surfaces where $\oneform\sigma $
becomes lightlike (see Eqs.\ (\ref{defDelta}), (\ref{defsigma})). \\

To continue the discussion let us recall that:
 \begin{itemize}
 \item[---] the vectors $\vxiR$ and $\vxiL $ are each defined up to an $SO^+ (2,1)$ 
 transformation. \item[---] their scalar product defines, Eq. (\ref{defkappa}),  the coordinate $\kappa$
 \item[---] the identification vector $\vec \partial$ is defined up to a sign.
 \end{itemize}
 The identification of the Killing vector fields expressed in null coordinates with those expressed in local coordinates is restricted by several considerations. The vectors $\vxiR$ and $\vxiL $ are given by $\partial_\alpha$ and $\partial_\beta$. Moreover, the vector $\vec r_{2,3}$ and $\vec l_{2,3}$ must be chosen so that their scalar products satisfy the positivity conditions implied by Eqs (\ref{AdS30hyp}), (\ref{Upm2}), (\ref{Vpm2}).\\
 There are several ways to break the $SL(2,\mathbb{R})_R$ remaining isometry group of the metric by choosing the vector $\vxiL $.  This leads to the list in Table \ref{TableItoIII} of warped geometries. The equations giving the action on null coordinates of the discrete isometry group generated by $Exp[2\,\pi\,n\,\vec\partial]$  ($n\in \mathbb Z$) and used to define quotients are immediate from their expressions in terms of the local coordinates $\alpha$, $\beta$ and $\kappa$ that transform as:
 \begin{align}
  & \alpha\mapsto \alpha+2\,\pi\,\LR\qquad,\qquad \beta\mapsto \beta +2\,\pi\,\LL\qquad,\qquad  \kappa\mapsto \kappa\qquad .\label{disctrans}
 \end{align}

 On the other hand, the fixed points are the ones such that 
 \begin{align}
  Exp[\LL\,X_L]\,\boldsymbol{z}\,Exp[\LR\,X_R]=\boldsymbol{z}\qquad\text{\em i.e.}\qquad \boldsymbol{z}^{-1}\,Exp[\LL\,X_L]\,\boldsymbol{z}= Exp[-\LR\,X_R]\qquad.
 \end{align}
This means that the two generators $\LL \,X_L$ and $-\LR \,X_R$  belong to the same adjoint orbit in the $sl(2,\,\mathbb R)$ Lie algebra. 

\noindent When the fixed points set is non-empty, it has the following structure.

1. If $X_L$ is timelike then its stabilizer subgroup in the adjoint action is (conjugated to) $K=SO(2)$. In this case, according to the Iwasawa decomposition, there exists a unique $z_0$ in $AN$ such that $-X_R=\Ad_{z_0}(X_L)$. Hence the fixed points consist in the lateral class $z_0\,K$ i.e. a circle.

2. If $X_L$ is spacelike, then its stabilizer subgroup in the adjoint action is (conjugated to) $\pm A$ and there exists a unique (up to sign) $z_0\in KN$ such that $-X_R=\Ad_{z_0}(X_L)$. Hence, the fixed points consist of the set $\pm z_0\,A$ i.e. two lines.

3. If $X_L$ is lightlike, then its stabilizer subgroup in the adjoint action is (conjugated to) $\pm N$ and there exists a unique (up to sign) $z_0\in KA$ such that $-X_R=\Ad_{z_0}(X_L)$. Hence the fixed points consist of the set $\pm z_0\,N$ i.e. two lines again.

\subsubsection{Explicit expressions}\label{Expexp}
 Let us now examine the various cases in Table \ref{TableItoIII} in detail. In particular, we recall that the $sl(2,\,{\mathbb R})$ Lie algebra has the structure of a $(2+1)$-dimensional Minkowski space with each generator belonging to one of the various co-adjoint orbits: spacelike hyperboloid, future or past sheet of a timelike hyperboloid, future or past sheet of the null cone and the origin (zero vector). Thus, elementary hyperbolic geometry provides a check of the range of the $\kappa$ coordinate defined by the product (\ref{defkappa}) of the right and left Killing vectors involved. 
 \begin{itemize}
\item{Spacelike warping: } $\vxiR=\vec r_1$, $ \epsR=+1$; $\lambda>0$
\begin{itemize}
\item $\bs{I}_a$:
  \begin{equationInItem}
    \begin{alignat}{3}
      \qquad\vxiL  &= \vec l_3 & &\qquad,\qquad &
      \epsL &= -1\qquad, \nonumber \\
      \kappa &= U_+\,V_- - U_-\,V_+\in \mathbb R & &\qquad,\qquad &
      \Delta &= (U_+^2+V_+^2)(U_-^2+V_-^2)\geq 1\qquad, \\
      \alpha &=\frac14\ln\left(\frac{U^2_+\,+V_+^2}{U_-^2+V_-^2}\right) & &\qquad,\qquad &
      \beta &= \frac12 \arctan\left(\frac{U_+\,V_-+U_-\,V_+}{U_+\,U_--V_+\,V_-}\right)
      \label{abIa}
    \end{alignat}
  \end{equationInItem}
    and with $\kappa=\sinh(2\,\rho)$
  \begin{align}
      &U_\pm=e^{\pm\,\alpha}\,\left(\cosh(\rho)\,\sin(\beta)\mp \sinh(\rho)\,\cos(\beta)\right)\qquad,\\
      &V_\pm=-e^{\pm\,\alpha}\,\left(\cosh(\rho)\,\cos(\beta)\pm \sinh(\rho)\,\sin(\beta)\right)\qquad.
  \end{align}
  These equations, combined with the expressions of $\kappa$ and $\alpha$ provide  an unambiguous value of $\beta$, on the contrary of the one given by the second of Eq.\ (\ref{abIa}).
  \\
  In this case condition Eq.\ (\ref{Dd2}) always is fulfilled. The boundaries defining the chronological singularities are such that
  \begin{align}
    \kappa_{\text{\tiny CTC}}^-<-B\quad,\quad -B<\kappa_{\text{\tiny CTC}}^+\leq\frac{(1-B^2)}{2\,B}\qquad ,
  \end{align}
  where $B$ is given by Eq.\ \eqref{ratioB}.
\item $\bs{I}_b$:
\begin{equationInItem}
\begin{alignat}{3}
    \vxiL &=\vec l_1 & &\qquad,\qquad &
    \epsL &= +1\qquad, \nonumber \\
    \kappa &= U_+\,U_- -V_+\,V_-\in \mathbb R & & \qquad,\qquad &
    \Delta &= -4\,U_+\,U_-\,V_+\,V_-\qquad,\label{SIbhor} \\
    \alpha &=\frac14\ln\left\vert\frac{U_+\,V_+}{U_-   \,V_-}\right\vert & &\qquad,\qquad &
    \beta &= \frac14\ln\left\vert\frac{U_+\,V_-}{U_-   \,V_+}\right\vert\qquad,
\end{alignat}
\end{equationInItem}
  Here we have to distinguish between several sectors:\\
       For $\kappa=\cosh(2\,\rho)\geq 1$, $\rho\in \mathbb R$
  \begin{align}
      &U_\pm=\hphantom{\pm }\sigma_U\,e^{\pm(\alpha+\beta)}\,\cosh(\rho) \qquad,\\
      &V_\pm=\pm \sigma_V\,e^{\pm(\alpha-\beta)}\,\sinh(\rho)\qquad ,
  \end{align}
    for $\kappa=-\cosh(2\,\rho)\leq -1$, $\rho\in \mathbb R$
  \begin{align}
      &U_\pm=\pm\sigma_U\,e^{\pm(\alpha+\beta)}\,\sinh(\rho) \qquad ,\\
      &V_\pm= \hphantom{\pm } \sigma_V\,e^{\pm(\alpha-\beta)}\,\cosh(\rho)\qquad .
  \end{align}
  In all these four expressions $\sigma_U$ and $\sigma_V$ are equal to $+1$ or $-1$. \\
  For $\kappa=\cos(2\,\rho)\leq 1$, $0\leq \rho< 2\,\pi $
  \begin{align}
      &U_\pm=  e^{\pm(\alpha+\beta)} \cos(\rho)\qquad,\\
      &V_\pm=  e^{\pm(\alpha-\beta)}\ \sin(\rho)\qquad.
  \end{align}
  If $B^2<\lambda/(1+\epsR\,\lambda)$, then all the space is chronologically safe. Otherwise, the chronological singularities are located in the region $\kappa< -1$. Moreover, notice that the horizon $\kappa=+1$ ({\it resp.} $\kappa=-1$) corresponds to the null surfaces $V_+=0$ or $V_-=0$ ({\it resp.} $U_+=0$ or $U_-=0$). When $L_{\text{\tiny R}}=L_{\text{\tiny L}}$, the intersection line ($U_+=0,\ U_-=0$) is a line of fixed points with respect to the identifications (\ref{disctrans}) and thus constitutes a topological singularity. 
\item $\bs{\rmII}_a$: 
\begin{equationInItem}
\begin{alignat}{3}
    \vxiL  &= (\vec l_2 -\vec l_3)/\sqrt{2} & &\qquad,\qquad &
    \epsL &= 0\qquad, \nonumber \\
    \kappa &= \sqrt{2}\,U_-\, V_+\in \mathbb R & &\qquad,\qquad &
    \Delta &= 2\,U_-^2\,V_+^2\qquad,\label{SIIahor} \\
    \alpha &= \frac12\,\ln\left\vert\frac{V_+}{U_-}\right\vert& & \qquad,\qquad &
    \beta &=  \frac{U_+\,U_--V_+\,V_-}{2\,\sqrt{2}\,U_-  \,V_+}\qquad,
\end{alignat}
\end{equationInItem}
%\end{itemize}
    and with $\kappa=\sigma_\kappa\,e^{2\,\rho}\in \mathbb R$ where $\sigma_\kappa=\pm 1$ and $s=\pm1$:
  \begin{align}
      &U_+= s\,e^{(\alpha-\rho)}\frac{1+2\,\sigma_\kappa\,\beta\,e^{2\,\rho}}{2^{3/4}}\qquad &,\qquad 
      &U_-= s\,\frac{e^{-(\alpha-\rho)}}{2^{1/4}}\qquad ,\\
       &V_+= s\,\sigma_\kappa\,\frac{e^{(\alpha+\rho)}}{2^{1/4}}\qquad &,\qquad 
       &V_-= s\,e^{-(\alpha+\rho)}\frac{1-2\,\sigma_\kappa\,\beta\,e^{2\,\rho}}{2^{3/4}}\qquad.
  \end{align}
  In this case the singularities are located in the region $\kappa<-B/2\leq 0${; the horizons are the surfaces $U_-=0$ or $V_+=0$ (corresponding to limit values of the coordinates: $\rho=-\infty,\ \alpha=+\infty,\ \beta=\pm \infty$)}.
  \end{itemize}
    \item{Timelike warping: } 
    $\vxiR= \vec r_3$, $\epsR=-1$; $1>\lambda>0$
    \begin{itemize}
    \item $\bs{I}_c^-$:
      \begin{equationInItem}
        \begin{align}
          \vxiL &=-\vec l_3 \qquad,\qquad \epsL=-1\qquad, \\
          \kappa&=+\frac12\,(U_+^2+U_-^2+ V_+^2+V_-^2)\geq 1\qquad,\nonumber\\
          \Delta&=\frac14\,\left((U_+ +U_-)^2+(V_++V_-)^2\right)\,\left((U_+ - U_-)^2+(V_+-V_-)^2\right)\qquad, \\
          \alpha&=\frac12\, \arctan\left[\frac{U_+^2 + V_+^2 -U_-^2-V_-^2}{2\,(U_-\,V_+ - U_+ \,V_-)}\right]\qquad,\qquad\\
          \beta&=  \frac12\, \arctan\left[\frac{2\,(U_+\,V_+ - U_- \,V_-)}{V_+^2 - U_+^2 +U_-^2-V_-^2}\right]\qquad,
        \end{align}
      \end{equationInItem}
  and with $\kappa=\cosh(\rho)$ and $\rho\geq 0$
  \begin{align}
      &U_\pm= + \cosh(\rho)\,\sin(\alpha-\beta)\pm \sinh(\rho)\,\sin(\alpha+\beta)\qquad, \\
      &V_\pm=-\cosh(\rho)\,\cos(\alpha-\beta)\pm \sinh(\rho)\,\cos(\alpha+\beta)\qquad.
  \end{align}
  
  \item $\bs{I}_c^+$:
    \begin{equationInItem}
      \begin{align}
        \vxiL &=\vec l_3 \qquad,\qquad \epsL=-1 \\
        \kappa&=-\frac12\,(U_+^2+U_-^2+ V_+^2+V_-^2)\leq -1\qquad,\nonumber\\
        \Delta&=\frac14\,\left((U_+ +U_-)^2+(V_++V_-)^2\right)\,\left((U_+ - U_-)^2+(V_+-V_-)^2\right)\qquad,
        \\
        \alpha&=+ \frac12\, \arctan\left[\frac{U_+^2 + V_+^2 -U_-^2-V_-^2}{2\,(U_-\,V_+ - U_+ \,V_-)}\right]\qquad,\\
        \beta &=+ \frac12\, \arctan\left[\frac{2\,(U_+\,V_+ - U_- \,V_-)}{U_+^2 - V_+^2 -U_-^2+V_-^2}\right]
      \end{align}
    \end{equationInItem}
  and with $\kappa=-\cosh(\rho)$
  \begin{align}
      &U_\pm= \left(\cosh(\rho)\,\sin(\alpha+\beta)\mp \sinh(\rho)\,\sin(\alpha-\beta)\right)\qquad,\\
      &V_\pm=\left(\cosh(\rho)\,\cos(\alpha+\beta)\pm \sinh(\rho)\,\cos(\alpha-\beta)\right)\qquad.
  \end{align}

  In both  cases   $\bs{I}_c^\mp$  condition (\ref{Dd2}) is satisfied. Thus, there are closed causal curves located in the region where $\kappa$ is between $\kappa_{\text{\tiny CTC}}^-<-1$ and $\kappa_{\text{\tiny CTC}}^+>1$. Moreover the line $\kappa=1$    constitutes a line of fixed points, a topological singularity line in case $\bs{I}_c^-$ when identifications are done with $\LL=\LR$.
  \item $\bs{\rmII}_b^-$:
    \begin{equationInItem}
      \begin{alignat}{3}
        \vxiL  &=(\vec l_2 -\vec l_3)/\sqrt{2} & &\qquad,& \epsL &=0\qquad, \nonumber \\
        \kappa &= \frac{U_-^2 + V_+^2}{\sqrt{2}} & &\qquad,\qquad & \Delta &=\frac{(U_-^2 + V_+^2)^2}{2}\qquad, \\
        \alpha &= \arctan \left(\frac{V_+}{U_-}\right)) & &\qquad, \qquad &
        \beta &=  \frac{U_+\,V_+ -U_-\,V_-}{\sqrt{2} \,(U_-^2+  \,V_+^2)}
      \end{alignat}
    \end{equationInItem}
  and with $\kappa=e^{2\,\rho}$ and $0\leq \alpha<2\,\pi$:
  \begin{align}
      &U_+=  2^{-1/4}\,(\cos(\alpha)\,e^{-\rho}+2\,\beta\,\sin(\alpha)\,e^{\rho})\quad&,\quad &U_-= 2^{1/4}\cos(\alpha)\,e^{\rho}\qquad,\\
      & V_-=  2^{-1/4}\,(\sin(\alpha)\,e^{-\rho}-2\,\beta\,\cos(\alpha)\,e^{\rho})\quad&,\quad &V_+=  2^{1/4}\,\sin(\alpha)\,e^{\rho}\qquad.
  \end{align}
   \item $\bs{\rmII}_b^+$:
     \begin{equationInItem}
       \begin{alignat}{3}
         \vxiL  &=(\vec l_2 +\vec l_3)/\sqrt{2} & &,\qquad & \epsL &=0\qquad, \nonumber \\
         \kappa &= -\frac{U_+^2 + V_-^2}{\sqrt{2}} & &,\qquad & \Delta &=\frac{(U_+^2 + V_-^2)^2}{2}\qquad, \\
         \alpha &= \arctan \left(\frac{V_-}{U_+}\right)  & &\qquad, &
         \beta &=  \frac{U_+\,V_+ -U_-\,V_-}{\sqrt{2} \,(U_+^2 +  V_-^2)}\qquad,
       \end{alignat}
     \end{equationInItem}
    and with $\kappa=-e^{2\,\rho}$ and $0\leq \alpha<2\,\pi$:
  \begin{align}
      &U_+=  2^{1/4}\,\cos(\alpha)\,e^{\rho}\quad,&\quad &U_-= 2^{-1/4}(\cos(\alpha)\,e^{-\rho}-2\,\beta\,\sin(\alpha)\,e^{\rho})\qquad,\\
      & V_-=  2^{1/4}\,\sin(\alpha)\,e^{\rho}\quad,&\quad &V_+=  2^{-1/4}(\sin(\alpha)\,e^{-\rho}+2\,\beta\,\cos(\alpha)\,e^{\rho})\qquad.
  \end{align}
    In both cases, the singularities are located in regions $\kappa<B/2$.
       \end{itemize}
   \item{Lightlike warping: } 
$
\vxiR=(\vec r_2+\vec r_3)/\sqrt{2}\qquad,\qquad \epsR=0
$
\begin{itemize}
\item $\bs{\rmIII^+}$:
  \begin{equationInItem}
    \begin{alignat}{3}
      \vxiL  &= (\vec l_2 +\vec l_3)/\sqrt{2} & &\qquad,\qquad & \epsL &= 0\qquad, \\
      \kappa &= -V_-^2 & &\qquad,\qquad & \Delta &= V_-^4\qquad,\label{LIIIpbhor} \\
      \alpha &= -\frac{U_+}{\sqrt{2}\, V_-} & &\qquad,\qquad & \beta &=  -\frac{U_-}{\sqrt{2}\,V_-}\qquad,
    \end{alignat}
  \end{equationInItem}
  and with $\kappa= -e^{2\,\rho}$ and $s=\pm 1$:
  \begin{align}
      &U_+= -s\,\sqrt{2}\,\alpha\,e^{\rho}\quad&,\quad &U_-=-s\,\sqrt{2}\,\beta \,e^{\rho}\quad,\\
      &V_+= s\,(e^{-\rho}-2\,\alpha\,\beta\,e^{\rho})\quad&,\quad &V_-=s \,e^{\rho}\qquad.
  \end{align}
  From Eq.\ (\ref{kappaCTCpm}) we directly read that $\kappa_{\text{\tiny CTC}}^+=0$.  Thus, the safe region on this local patch is given by $\kappa<\kappa_{\text{\tiny CTC}}^-=-2\,B/\lambda\leq 0$.
\item $\bs{\rmIII^-}$:
  \begin{equationInItem}
    \begin{alignat}{3}
      \vxiL  &= (\vec l_2 -\vec l_3)/\sqrt{2} & &\qquad,\qquad & \epsL &= 0\qquad, \\
      &\kappa= U_-^2 & &,\qquad\qquad & \Delta &= U_-^4\qquad,\label{LIIImhor} \\
      \alpha &= \frac{V_+}{\sqrt{2}\, U_-} & &,\qquad\qquad &
      \beta &= -\frac{V_-}{\sqrt{2}\,U_-}
    \end{alignat}
  \end{equationInItem}
  and with $\kappa= e^{2\,\rho}$ and $s=\pm 1$:
  \begin{align}
      &U_+= s\,(e^{-\rho}+2\,\alpha\,\beta\,e^{\rho})\quad,&\quad &U_-=s\,e^{\rho}\quad ,\\
      &V_+= s\,\sqrt{2}\,\alpha\, e^{\rho}\quad,&\quad &V_-=-s\,\sqrt{2}\,\beta\,e^{\rho}\quad .
  \end{align}
   There is no singularity on this local patch (as in this case $\kappa_{\text{\tiny CTC}}^+=0$) {but there is one on its boundary. The surface $U_-=0$ constitutes a lightlike chronological singularity, containing a topological singularity on the line $V_+=V_-=1$ when identifications are done with $\LL=\LR$.}
   \end{itemize}
\end{itemize}
{
The parametisations obtained for the choices of the vectors $\xiR$ and $\xiL$ corresponding to the four types $\tilde{\bs{I}}_a$, $\tilde{\rmII}_a$, $\tilde{\rmII}_b^+$ and $\tilde{\rmII}_b^-$ are similar to those obtained for the types $ \bs{I}_a$, $ \rmII_a$, $ \rmII_b^+$ and $ \rmII_b^-$ except that the functions $\alpha$ and $\beta$ have to be interchanged, after having performed a discrete $O(2,\,2)$ transformation.\footnote{ These transformations are for type $ \bs{I}_a:\ V_+\leftrightarrow V_-$, for $ \rmII_a:\ V_+\leftrightarrow -V_-$, for $ \rmII_b^+:\ U_+\leftrightarrow U_-$ and for $ \rmII_b^-:\ V_+\leftrightarrow V_-$.} Indeed, they correspond to the exchange of $\epsL$ with $\epsR$, in accordance with
the writing of the metric  Eq.\ (\ref{metcan}). But once identifications are performed their causal structures may be different. This is why they have to be taken into account.}\newline
Let us also emphasize that from the explicit solutions of the geodesic equations obtained in section \ref{ExpGeodWadSSol}, together with the previous expressions of the $\alpha,\ \beta$ and $\kappa$ coordinates in terms of the null coordinates provide solutions of the  geodesic equations in the adapted coordinates, and more generally using the algorithm described in Appendix \ref{loctonullcoord} in any coordinate system.

%%%%%%%%%%%%%%%%%%%%%%%%%%%%%%%%%%%%%%%%%%%%%%%%%%%%%%%%

%%%%%%%%%%%%%%%%%%%%%%%%%%%%%%%%%%%%%%%%%%%%%%%%%%%%%%%%%%%%%%%%%%%%%%%%%%%%%%%%%%%

 \section{Projection diagrams}
\label{sec: proj}
In this section we discuss the causal structure of warped geometries, using \emph{projection diagrams}, which were introduced in 
\cite{Chrusciel:2012gz}. 
In subsection  \ref{subsec:summary}, we give a summary of the results obtained.
In subsection \ref{subsec:proj-def}, we recall the definition of a projection diagram given in \cite{Chrusciel:2012gz} and provide some comments regarding this definition, in line with what is said there. Subsections \ref{subsec:generic} - \ref{subsec:lambda0} provide our results. Subsection \ref{ProjMICS} provides an alternative method of visualizing the causal structure of warped spaces.
We warn that while projection diagrams have many similarities to the commonly used Carter-Penrose diagrams, there are also important differences. Carter-Penrose diagrams are obtained by performing a conformal completion of the spacetime and then taking a slice, while projection diagrams are obtained by projecting to a subset of two-dimensional Minkowski space: $M^{1,1}$, in a certain way and then performing a conformal completion. 
The causal structure of the warped black holes of \cite{Anninos:2008fx}, type $\bs{I}_b$ and $\bs{\rmII}_a$ in our context, has been discussed in \cite{Jugeau:2010nq}.  Our results agree with their analysis, for the case $\bs{I}_b$ compare %left and top right diagram in 
Fig. \ref{FIG2A} and Fig.\ \ref{FIG2B} to Fig.\ 7 and Fig.\ 8 in  \cite{Jugeau:2010nq}; for the case $\bs{\rmII}_a$, compare Fig. \ref{FIG4} to Fig.\ 9 in  \cite{Jugeau:2010nq}.
{
The geometry $\tilde{\rmII}_a$    appears in Eq. (2.4) of Ref. \cite{Anninos:2010pm}  when $z=2$. Using their notations, in which $\alpha$ and $\beta$ are constants (not to be confused with the functions introduced in Eqs (\ref{ClstpR}),(\ref{ClstpL})), their geometry corresponds to the local metric  (\ref{wmetcanPhi}) with $\epsR=0$, $\epsL=1$, $\kappa = r/\alpha$, $\LL=\alpha$, $\LR=\beta/2$ and $\lambda=1$. The paper \cite{Anninos:2010pm} claims that the geometry describes a black hole when $\beta \geq 2 \alpha$, where the causal singularity is hidden behind an horizon, in accordance with Eq. (\ref{ChronoCrit}) and  Fig.\ \ref{FIG4}. For $z=0$ or $z= 1$ their metric is locally isometric to AdS$_3$, for all other value of $z$ it is an AdS$_3$ space warped by a lightlike Killing vector. Moreover, unless $z=2$ (or $0, 1$), the warping factor is not constant, but local (see Section \ref{locwarp}).}
\par 
\subsection{Summary of the causal structure}
\label{subsec:summary}
\begin{figure}[H] 
  \centering
    \begin{subfigure}{0.3\textwidth}
      \centering
      \includegraphics{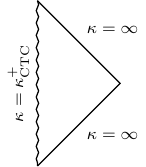}
     \caption{\label{FIG1} }
      %\label{NOHa}
  \end{subfigure}
  \hspace{0.5cm}
  \begin{subfigure}{0.3\textwidth}
      \centering
      \includegraphics{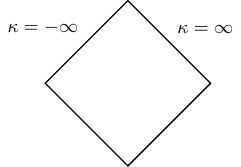}
      \caption{\label{FIG5A}}
      %\label{NOHd}
  \end{subfigure}
  \hspace{0.5cm}
  \begin{subfigure}{0.3\textwidth}
      \centering
      \includegraphics{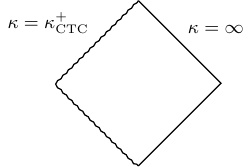}
      \caption{\label{FIG5B}}
      %\label{NOHe}
  \end{subfigure}
    \vspace{6mm}\strut\\
  \centering
    \begin{subfigure}{0.3\textwidth}
      \centering
      \includegraphics{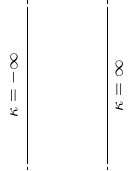}
     \caption{\label{FIG6B} }
      %\label{NOHa}
  \end{subfigure}
  \hspace{0.5cm}
  \begin{subfigure}{0.3\textwidth}
      \centering
      \includegraphics{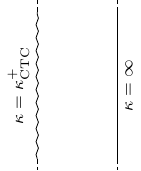}
     \caption{\label{FIG6} }
      %\label{NOHb}
  \end{subfigure}
  \hspace{0.5cm}
  \begin{subfigure}{0.3\textwidth}
      \centering
      \includegraphics{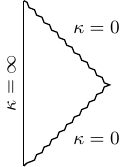}
     \caption{\label{FIG6A} }
      %\label{NOHc}
  \end{subfigure}
 \caption[] {\label{NOH}{\bf Configurations without a horizon}, encountered in types
 \hspace{-100.5mm}\tabular[t]{@{}l@{}} \\(a):  $\bs{I}_a$, $\rmII_a$, $\bs{I}_b$, $\bs{I}_c^\pm, \rmII_b^\pm$, $\tilde{\bs{I}}_a$, $\tilde\rmII^\pm_b$, ${\rmIII}^+$, ${\tilde \rmII}_a$, each only for $\lambda \neq 0$ 
 \\(b): $\tilde{\bs{I}}_a$,  only for $\lambda \neq 0$ 
 \\ (c): $\rmII_a$, $\rmII_b^\pm$, ${\rmIII}^\pm$, each only for $\lambda \neq 0$ 
 \\ (d): self-dual case ($\LL = 0$), $\epsR = 1$, $\epsL = -1$
 \\ 
 (e): $\bs{I}_a$, $\bs{I}_c^-$, $\rmII_b^-$ each only for $\lambda = 0$
 \\ (f): ${\rmIII}^-$ only for $\lambda = 0$
\endtabular}
\end{figure}

\begin{figure}[H]
  \centering
  \begin{subfigure}{0.49\textwidth}
      \centering
      \includegraphics{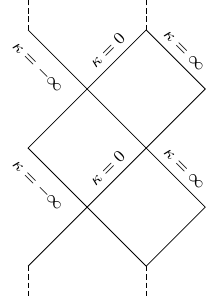}
     \caption{\label{FIG8} }
      %\label{ONEHa}
  \end{subfigure}
  \begin{subfigure}{0.49\textwidth}
    \centering
    \includegraphics{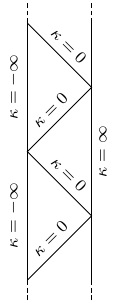}
    \caption{\label{FIG7}}
    %\label{ONEHc}
\end{subfigure}
 \caption[] {\label{ONEH}\tabular[t]{@{}l@{}}{\bf Configurations with one horizon, no singularity}, encountered in types
  \\ (a): ${\tilde \rmII}_a$, only for $\lambda \neq 0$ 
  \\ (b): self-dual case ($\LL = 0$), $\epsR = 1$, $\epsL =0$ 
 \endtabular}
\end{figure}

\begin{figure}[H]
  \begin{subfigure}{0.49\textwidth}
    \centering
    \includegraphics{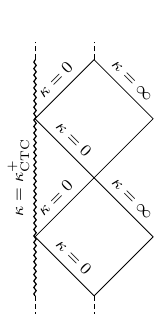}
   \caption{\label{FIG4} }
    %\label{ONEHb}
\end{subfigure}
  \begin{subfigure}{0.49\textwidth}
      \centering
      \includegraphics{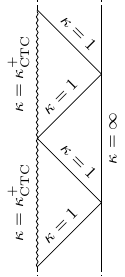}
      \caption{\label{FIG3B}}
     % \label{ONEHd}
  \end{subfigure}
    \caption[] {\label{ONEH2}\tabular[t]{@{}l@{}}{\bf Configurations with one horizon}, encountered in types\\ (a): $\rmII_a$, ${\tilde \rmII}_a$ for $\lambda \neq 0$
  \\ (b): $\rmII_a$ for $\lambda = 0$
 \endtabular}
    \end{figure}

\begin{figure}[H]
  \centering
  \begin{subfigure}{0.49\textwidth}
      \centering
      \includegraphics{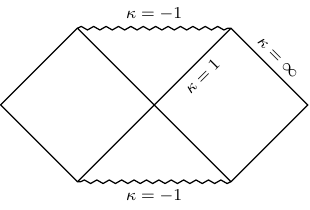}
     \caption{\label{FIG2B} }
      % \label{ONEHSLa}
  \end{subfigure}
  \begin{subfigure}{0.49\textwidth}
      \centering
      \includegraphics{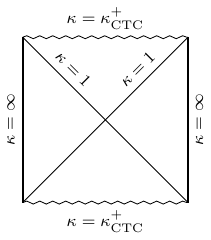}
     \caption{\label{FIG9}}
      %\label{ONEHSLb}
  \end{subfigure}
    \caption[] {\label{ONEHSL}\tabular[t]{@{}l@{}}{\bf Configurations with one horizon} \\ (a): $\bs{I}_b$ for $\lambda \neq 0$
  \\ (b): $\bs{I}_b$ for $\lambda = 0$
 \endtabular}
     \end{figure}

\begin{figure}[H]
  \centering
  \begin{subfigure}{0.47\textwidth}
      \centering
      \includegraphics[scale = 0.88]{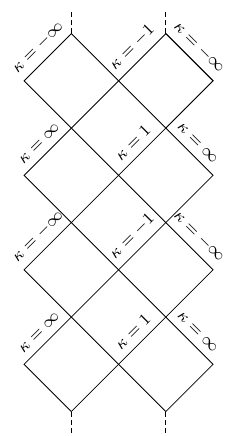}
     \caption{\label{FIG2C} }
      %\label{TWOHa} 
  \end{subfigure}
  \begin{subfigure}{0.43\textwidth}
      \centering
      \includegraphics[scale = 0.88]{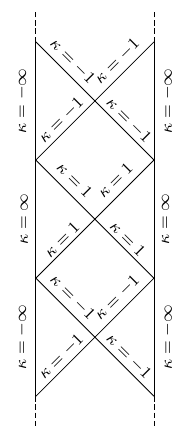}
     \caption{\label{FIG3C} }
      %\label{TWOHb}
  \end{subfigure}
    \caption[] {\label{TWOH2}\tabular[t]{@{}l@{}}{\bf Configurations with two horizons, no singularity}, encountered in types 
    \\ (a): $\bs{I}_b$ for $\lambda \neq 0$ 
    \\ (b): self-dual case ($\LL = 0$), $\epsR =1$, $\epsL =1$
    \endtabular}
  \end{figure}

\begin{figure}[H]
  \begin{subfigure}{0.51\textwidth}
      \centering
      \includegraphics[scale = 0.88]{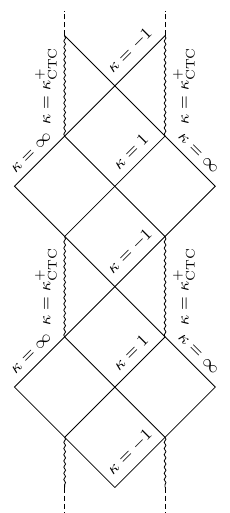}
      \caption{\label{FIG2A} }
      %\label{TWOHc}
  \end{subfigure}
  \hspace{1cm}
  \begin{subfigure}{0.3\textwidth}
      \centering
      \includegraphics[scale = 0.88]{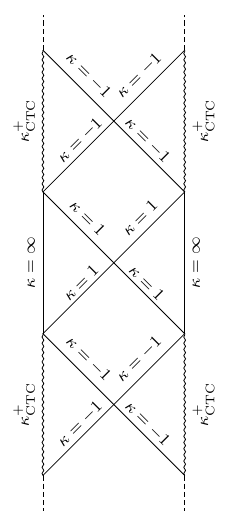}
      \caption{\label{FIG3A}}
      % \label{TWOHd}
  \end{subfigure}
    \caption[] {\label{TWOH22}\tabular[t]{@{}l@{}}{\bf Configurations with two horizons, encountered in types} 
    \\ (a): $\bs{I}_b$ for $\lambda \neq 0$
    \\ (b): $\bs{I}_b$  for $\lambda = 0$
    \endtabular}
     \end{figure}

Here, before going into the details, we summarise the results obtained on the  various causal structures when the warping parameter $\lambda$ is nonzero. 
For the metric \eqref{wmetcanPhi} which has $\LL \neq 0$,  we find the following causal structure, classified according to Table \ref{TableItoIII}: 
\newline
\begin{itemize}
    \item Spacelike warping ($\epsR=+1$): \newline
    The case ${\bs I}_a $ has a naked singularity with projection diagram given by Fig.\ \ref{FIG1} \\
The case  ${\bs I}_b $ ($\epsL=+1$) is the standard warped version of the BTZ black hole whose causal structure has also been discussed in \cite{Jugeau:2010nq}.  The spacetime possesses closed timelike curves when the ratio $B = \LR/\LL$, defined in \eqref{ratioB}, 
satisfies
$B^2\geq \lambda/(1+\lambda)$. In addition, for $B=1\ (i.e.\ \LR = \LL)$, the locus $\kappa=-1$ contains a line of fixed points, leading to a topological singularity, as pointed out in Sec. \ref{Expexp}. Extending from $\kappa = \infty$ inward, the corresponding projection diagrams in the different regimes are given in Fig.\ \ref{FIG2A} and Fig.\ \ref{FIG2B}. Extending from $\kappa = -\infty$ inward, the projection diagram is given by Fig.\ \ref{FIG1}. For $B^2< \lambda/(1+\lambda)$ the spacetime is devoid of closed timelike curves and the projection diagram given by \ref{FIG2C}.
\\
  In case $\bs{\rmII}_a$ ($\epsL=0$) 
  with $\LR \neq 0$
  we obtain extremal black hole configuration if we extend from $\kappa = \infty$ inward., see Fig.\   \ref{FIG4}. Extending from $\kappa = -\infty$ yields Fig.\ \ref{FIG1}. When $\LR = 0$, the projection diagram is given by Fig.\ \ref{FIG5B}. 
    \item Lightlike warping ($\epsR=0$): 
The cases $\bs{\rmIII^+}, \bs{\rmIII^-}$, $\tilde\rmII^+_b$, $\tilde\rmII^-_b$ have naked singularities. 
For $\tilde{\bs{\rmII}}_a$, if $B^2 \geq \lambda$ and if we extend from $\kappa = \infty$ inward, we obtain Fig.\ \ref{FIG4}, while if we extend from $\kappa =- \infty$ inward, we obtain Fig.\ \ref{FIG1}.
For $B^2 < \lambda$, the metric is devoid of CTCs and the projection diagram given by Fig.\ \ref{FIG8}.
 \
    \item Timelike warping ($\epsR=-1$): 
    The cases ${\bs I}^+_c, {\bs I}^-_c$, $\bs{\rmII}_b^+, \bs{\rmII}_b^-$ have naked singularites. 
 The case ${\tilde{\bs {{ I }}}}_a$ has naked singularities when $B^2 \geq {\lambda}/{(1+\lambda)} \geq 0$ and in that regime the projection diagram is given by Fig.\ \ref{FIG1}.
    When $B^2 < {\lambda}/{(1-\lambda)} \geq 0$ the spacetime has no CTCs and the projection diagram is given by Fig.\ \ref{FIG5A}.
\end{itemize}
For the self-dual configurations ($\LL = 0$) we find the following: 
\begin{itemize}
\item Spacelike warping ($\epsR=+1$):
For $\epsR=+1$ there are never any closed timelike curves.
For 
$\epsL=-1$ there are no horizons. 
For $\epsL= 0$ and $\epsL= 1$ the causal structure is given by 
Fig. \ref{FIG3B}
and Fig. \ref{FIG7}, respectively. 
In these cases, the horizons don't appear to be black hole horizons (compare to discussion in Sec. 2.3 of \cite{Detournay:2019xgl}). 
\item Null or timelike warping ($\epsR \in \{0, -1\}$): The spacetime has closed null curves through every point.
\end{itemize}
To summarize, the cases ${\bs I}_b $, $\bs{\rmII}_a$ and $\tilde{\bs{\rmII}}_a$  have the causal structure of a black hole.

\subsection{Definition}
\label{subsec:proj-def}
In \cite{Chrusciel:2012gz}, a new class of two-dimensional diagrams, 
the so-called \emph{projection diagrams}, were introduced as a tool to visualize the global structure of spacetimes.
These diagrams can be used to depict non-spherically symmetric or non-block diagonal metrics with two-dimensional diagrams, using 
a two-dimensional auxiliary metric constructed out of the spacetime. 
For the convenience of the reader, we now recall its definition
 \cite{Chrusciel:2012gz}:\par
\begin{em}
    \label{defpro}
  Let $(M, g)$ be a smooth spacetime and let $M^{1, n}$ denote $(n+1)$-dimensional Minkowski spacetime. 
  A projection diagram is a pair
  $(\pi, U)$, where 
  \begin{equation}
      \pi: M \rightarrow W\,, \qquad W \subset M^{1,1}\nonumber 
  \end{equation}
  is a continuous map, differentiable on an open dense set. 
  $U \subset M$ is an open set, assumed to be nonempty, on which $\pi$ is a smooth submersion, such that
  \begin{enumerate}
      \item every smooth timelike curve $\sigma \subset \pi(U)$ is the projection of a smooth timelike curve $\omega$ in $(U, g)$: $\sigma = \pi \circ \omega$;
      \item the image $\pi\circ\omega$ of every smooth timelike curve $\omega\subset U$ is a timelike curve in $M^{1,1}$.
  \end{enumerate}
\end{em}
\par\noindent
    The two requirements on timelike curves ensure that causal relations on $\pi(U)$ reflect --
  as accurately as possible -- causal relations on $U$. 
  By continuity, it follows that images of causal curves in $U$ are causal in $\pi(U)$.
  Many spacelike curves in the original spacetime are mapped to either null or timelike 
  curves in the diagram. 

  The map $\pi$ in the definition of the projection diagram is used to systematically construct an auxiliary 
  two-dimensional metric out of spacetime. 
  One then performs a conformal completion, which is always possible in two dimensions, and draws a diagram of the 
  two-dimensional auxiliary metric.
  This auxiliary metric shares important causal features with the original metric
  due to the requirements on $\pi$.
  
  We give some comments on the 
  definition.
  While it is assumed for simplicity that $(M, g)$, $\pi|_U$ and the causal curves in the definition are smooth, this is unnecessary for most purposes. 
  In the definition, it is assumed that the map $\pi$ maps $U$ to a subset of $M^{1,1}$ but this can be modified. 
  In some applications it might be more natural to consider different two-dimensional manifolds as the target space; an example of such a spacetime
  is already
  given in \cite{Chrusciel:2012gz} and requires only minimal modification of the definition. For additional explanations 
  with regards to the definition consult Ref. \cite[section 3.1]{Chrusciel:2012gz}.

Importantly, the requirements in the definition ensure that if there exists a black hole region in the diagram,
there exists a black hole region in the original spacetime. Indeed, if the diagram has a black hole region $B$, then
$B = W - J^-(\scri^+(W))$ is nonempty. Here, $J^-(\scri^+)$ denotes the causal past of 
$\scri^+$, future null infinity. 
This implies, that 
all curves leaving the black hole region $B$ and hitting $\scri^+(W)$ at late times are spacelike.
We now define a new set $\tilde B$ for which it holds that $B = \pi(\tilde B)$. Assume that there exist causal curves, 
moving forward in time, 
that leave $\tilde B$ and that go to future null infinity; then this must also be the case in $B = \pi(\tilde B)$ via the axioms of the 
projection diagram. This immediately leads to a contradiction.  Therefore, if a black hole region exists in the diagram,
a black hole region must exist in the original spacetime.

\subsubsection{Conformal diagrams of two-dimensional, static metrics}
In what follows we will need to construct conformal diagrams for metrics of the form 
\begin{equation}
\label{walker2d}
    g_{(2)} = - \hat F(\kappa) d\tau^2 + \frac{d\kappa^2}{\hat F(\kappa)}\,,
\end{equation}
where $\tau \in \mathbb{R}$ and $\hat F(\kappa)$ is an analytic function on an interval.
The geometry of the  spacetime  and its potential extendability will depend
upon the sign of $\hat F$, the zeros of $\hat F$ and their order.
    A systematic study of conformal diagrams 
 for two-dimensional metrics has been carried out by Walker \cite{Walker:1970}, which was reviewed in \cite{Chrusciel:2012gz}. For convenience of the reader, we recall the details needed for the considerations in this work here and refer for an  in-depth discussion of this topic to \cite{Walker:1970, Chrusciel:2012gz}, as well as \cite{chrusciel2020geometry}.
 
 Following \cite{Walker:1970}, we consider separately maximal intervals (``blocks'') on which ${\hat F(\kappa)}$ is finite and does not change sign. These intervals define the ranges of $\kappa$ and lead to connected Lorentzian manifolds 
 on which $g_{(2)}$ is defined everywhere. 
 The conditions under which such manifolds can be patched together were discussed in \cite{Walker:1970}.
 Following \cite{Walker:1970, Chrusciel:2012gz}, we first bring the metric \eqref{walker2d} into a  manifestly conformally flat form by choosing a value $\kappa^*$ 
 such that ${\hat F}(\kappa^*) \neq 0$ and 
introduce a new coordinate
 \begin{equation}
 \label{xkappa}
  x(\kappa) = \int_{\kappa^*}^\kappa \frac{ds}{{\hat F}(s)}\qquad,
 \end{equation}
which yields
  \begin{equation}
    g_{(2)}  =  {\hat F(\kappa)} (- d\tau^2 + dx^2)\qquad.  \end{equation}
 When $x(\kappa)$ ranges over $\mathbb{R}$ on the block, we obtain that the block is conformal to a diamond (the conformal diagram of the usual 2d Minkowski space). When $x(\kappa)$ is only diverging at one end, we obtain that  the block is conformal to 
 a triangle. When $x(\kappa)$ is finite at both ends, we obtain a strip.

 It was shown in \cite{Walker:1970} that 
 four blocks can be glued together across a boundary where
 \begin{equation}
    {\hat F}(\kappa_0) = 0\qquad, \qquad {\hat F}'(\kappa_0) \neq 0\qquad,
 \end{equation}
 such that the Kruskal extension so obtained is real analytic.
 If the function ${\hat F}(\kappa)$ has a higher-order zero at $\kappa = \kappa_0$
 \begin{equation}
    {\hat F}(\kappa_0) = 0\qquad, \qquad {\hat F}'(\kappa_0) = 0\qquad,
\end{equation}
 one may only glue together two blocks, as was shown in \cite{Walker:1970} by performing an Eddington-Finkelstein extension. 
 
\subsubsection{Extendability of spacetime}
In the last subsection, we discussed under which conditions a two-dimensional spacetime of the form \eqref{walker2d} can be extended across a surface $\kappa = \kappa_0$, on which $F(\kappa_0) = 0$.
 As shown in \cite{Chrusciel:2012gz}, since $\kappa$ is a real analytic function in terms of the Kruskal coordinates,
 also the  spacetime  metric $g_{\mu \nu}$ extends smoothly across $\kappa = \kappa_0$.
The quotienting procedure may give rise to closed timelike curves or topological singularities and we will cut our spacetime off in such a way that the resulting spacetime is devoid of closed timelike curves/topological singularities.
In the regions where closed timelike curves appear, causality is not represented in any useful way in the projection diagram. For this reason these regions are removed from the diagram.

\subsection{The generic cases: $\lambda>0,\ \LR\geq 0,\,\LL>0$}
\label{subsec:generic}
We now discuss the causal structure of  \eqref{wmetcanPhi} and \eqref{wmetcanPhisd} using projection diagrams.
We start with \eqref{wmetcanPhi}
where $\epsR, \epsL =\{\pm1,\ 0\}$, $\LR \geq 0$ and $\LL >0$.
In addition, we split the cases according to the classification in Table \ref{TableItoIII}.
 The case $\LL = 0$ is covered by the metric \eqref{wmetcanPhisd}, which we will cover later (see Sec. \ref{selfdualcausal}).
We recall that to avoid a change of metric signature we impose, see Eq.\ (\ref{contdef}),
\begin{equation}
   (1+ \epsR \lambda)  > 0\qquad.
   \label{restict1}
\end{equation}
to stay in the domain where the signature of the metric is given by $(-, +, +)$.

For our considerations it is useful to put the metric, Eq. (\ref{wmetcanPhi}), into the form
\begin{equation}
   { \uuline{g} }\strut_{(\lambda)}=-N^2(\kappa) F(\kappa) d\tau^2
    + \frac{d{\kappa}^2}{F(\kappa)}+R^2({\kappa})(d\varphi+N^\varphi(\kappa) d\tau)^2
    \label{anninos3}
\end{equation}
with 
\begin{align}
    N^2(\kappa) &= \frac{\LL^2 (1+\lambda \, \epsR )}{4 R^2(\kappa)} \qquad, \\
   F(\kappa) &= 4 (\kappa^2 - \epsL \epsR)\qquad, \\
   N^\varphi(\kappa) &= \frac{(1+\lambda\,\epsR) (\kappa \,
   \LL + \epsR\,\LR)}{R(\kappa)}\qquad, \\
   R^2(\kappa) &= \LL^2 \left(\kappa ^2 \lambda +\epsL\right)+2 \,\kappa 
   \LL \LR (1+\lambda\,\epsR)+ \LR^2
  \epsR (1+\lambda\,\epsR)\qquad.
  \label{R2kappa}
\end{align}
The spacetime is defined as the region where $R^2(\kappa)$  is non-negative.  The region $R^2(\kappa)>0$ is where no closed timelike curves occur. We will refer to the locus $\kappa = \kCTCpm$ for which 
\begin{equation}
\label{defkappaCTC}
    R^2(\kCTCpm) = 0\qquad,
\end{equation}
as the chronological (or topological) singularity that defines the boundary of the manifolds we consider.\newline
For\footnote{ Let us also remind that for $\lambda < 0$ there are closed timelike curves at large values of $\kappa$ which is why we focus on $\lambda \geq 0$ in this work (see Eq.\ \eqref{lambpos}). } $\lambda \neq 0$,   there are two roots of this equation, given by Eq. \ref{kappaCTCpm}.\newline
For $\lambda=0$ but $\LL\,\LR\neq 0$ one root ($\kCTCm$) is pushed to infinity while the other reduces to
\begin{align}
\kCTCp=-(\epsL\,B+\epsR\,B^{-1})\qquad \label{kappaCTClineq}, 
\end{align}
where $B$ is given by \eqref{ratioB}.
If $\lambda $ and  the product $\LL\,\LR$ vanish, $R(\kappa)$ reduces to the constant $\epsL\,\LL^2+\epsR\,\LR^2$ which fixes $\epsR$ or $\epsL$ to be equal to one.\\
We then write the metric as 
\begin{equation}
    { \uuline{g} }\strut_{(\lambda)} = N(\kappa) \left(- \hat F d\tau^2 + \frac{d\kappa^2}{\hat F} \right) +R({\kappa})^2(d\varphi+N^\varphi(\kappa) d\tau)^2
    \label{anninos4}
\end{equation}
with 
\begin{equation}
    \label{hatFgen}
\hat F = N(\kappa) F(\kappa) = \frac{\sqrt{1+\epsR\, \lambda}}{\lambda}\,
\frac{2\, (\kappa^2 - \epsL \epsR)}{\sqrt{(\kappa-\kCTCp)(\kappa-\kCTCm)}}
\qquad.
\end{equation}

To discuss the causal structure of the warped AdS$_3$ quotients, we construct projection diagrams. The projection $\pi$ is given by $(\tau, \kappa, \varphi) \mapsto (\tau, \kappa)$ and maps $(M, g)$ 
to a subset of $M^{1,1}$ with local metric
\begin{equation}
    \label{2daux}
\dsig =  N(\kappa) \left(- \hat F d\tau^2 + \frac{d\kappa^2}{\hat F} \right)\qquad,
\end{equation}
where $N(\kappa)$ is strictly positive in the domain of interest.
To build the various projection diagrams we have to look at the domains of definition of the $\kappa$ variable ($R(\kappa)\geq 0$) and, the  behaviour of $x(\kappa)$ at infinity and, when relevant, the behaviour of $x(\kappa)$ near $\kappa=0,\,\pm 1$. Table \ref{TableGenwAdS3} summarizes our results.

%TABLE GENWarpAdS
\begin{table}[h]
\begin{adjustbox}{width=\columnwidth,center}
%\begin{center}
%\hspace{-2cm}
\begin{tabular}{|c|c|c|c|c|c|c|c|c|c|c|}
\hline
\multicolumn{11}{|c|}{ \rule[6mm]{0mm}{0mm}\ \rule[-3mm]{0mm}{0mm}Generic cases\hspace{15mm}$x(\kappa_0)\propto \sqrt{\lambda/(1-\lambda \epsR)}\,\int^{\kappa_0}\sqrt{(\kappa-\kCTCp)(\kappa-\kCTCm)}/(\kappa^2-\epsR\epsL)\,d\kappa\qquad,\qquad x(\pm \infty)=div.(0)$} \\
\hline Type&$\{\epsR,\ \epsL\}$& $B\geq 0,\ \lambda>0$&$\kCTCp$&$\kCTCm$&$x(\kCTCp)$&$x(\kCTCm)$&$x(+1)$&$x(-1)$&$x(0)$&Fig. \\
\hhline{|=|=|=|=|=|=|=|=|=|=|=|}
%Ia
$\rule[0.5mm]{0cm}{4mm}\rule[-2.5mm]{0cm}{4mm}\ \bs{I}_a,\ \kappa\in \mathbb R$&$\{+1,\ -1\}$& &$\in \mathbb R$&$\in \mathbb R^-_0$& & & & & &$\kappa\geq\kCTCp$ or $\kappa\leq\kCTCm<0$, Fig. \ref{FIG1}\\ 
\hhline{|=|=|=|=|=|=|=|=|=|=|=|}
%IIa
\rule[0.5mm]{0cm}{4mm}\rule[-2.5mm]{0cm}{4mm}${\rmII}_a\ \kappa\in \mathbb R$&$\{+1,\ 0\}$& $B>0
$ &$<0$  &$<\kCTCp<0$   & &$f.$ & & & &$\kappa\leq\kCTCm$ Fig. \ref{FIG1} \\
\hline\rule[0.5mm]{0cm}{4mm}\rule[-2.5mm]{0cm}{4mm} &$\{+1,\ 0\}$& $B>0$ &$<0$  &$<\kCTCp<0$   &$f.$ & & & & $div.(1)$ &$\kappa\geq\kCTCp$ Fig. \ref{FIG4} \\
\hline\rule[0.5mm]{0cm}{4mm}\rule[-2.5mm]{0cm}{4mm} &$\{+1,\ 0\}$& $B=0$ &$=0$  &$=0$   &$div.(0)$ &$div.(0)$ & & & $div.(0)$ &$\kappa\geq 0$ or $\kappa\leq 0$ Fig. \ref{FIG5B} \\
\hhline{|=|=|=|=|=|=|=|=|=|=|=|}
%Ib
$\rule[0.5mm]{0cm}{4mm}\rule[-2.5mm]{0cm}{4mm}\ \bs{I}_b,\ \kappa\in \mathbb R$&$\{+1,\ +1\}$&$B^2<{\lambda}/{(1+\lambda)}$ & $\in \mathbb C$ &$\in \mathbb C$   & & &$div.(0)$ &$div.(0)$ & &Fig. \ref{FIG2C}\\
\hline
$\rule[0.5mm]{0cm}{4mm}\rule[-2.5mm]{0cm}{4mm} $& &$B^2\geq{\lambda}/{(1+\lambda),\ \neq 1}$  & $\leq -1/B<-1$ & & $f.$  &  &$div.(0)$ &$div.(0)$ & &$\kappa\geq\kCTCp$  Fig. \ref{FIG2A}\\
\hline
$ \rule[0.5mm]{0cm}{4mm}\rule[-2.5mm]{0cm}{4mm}$& &$B =1$ & $=-1$ &$=-(\lambda+2)/\lambda<-1$ & $f.$  &  &$div.(0) $ &$f.$ & &$\kappa\geq -1$ Fig. \ref{FIG2B}
\\
\hline
$\rule[0.5mm]{0cm}{4mm}\rule[-2.5mm]{0cm}{4mm} $& &$B^2\geq {\lambda}/{(1+\lambda)}$ &    &$\leq-1/B<-1$ &    & $f.$& & & & $\kappa\leq\kCTCm$ Fig. \ref{FIG1} \\
\hhline{|=|=|=|=|=|=|=|=|=|=|=|}
%Ic+
$\rule[0.5mm]{0cm}{4mm}\rule[-2.5mm]{0cm}{4mm}\ \bs{I}_c^+,\ \kappa\leq -1$&$\{-1,\ -1\}$&$  $ & $\geq 1$ &$<-1$   & & & &  & &$\kCTCm>\kappa  $ Fig. \ref{FIG1} \\
\hhline{|=|=|=|=|=|=|=|=|=|=|=|}
 %Ic-
\rule[0.5mm]{0cm}{4mm}\rule[-2.5mm]{0cm}{4mm}$\bs{I}_c^-,\ \kappa\geq 1 $&$\{-1,\ -1\}$&  &  $\geq 1$ & $\leq -1$  & & &$f.$ & & &Fig. \ref{FIG1} \\
\hhline{|=|=|=|=|=|=|=|=|=|=|=|}
%IIb+
\rule[0.5mm]{0cm}{4mm}\rule[-2.5mm]{0cm}{4mm}$ {\rmII}_b^+,\ \kappa \leq 0 $  &$\{-1,\ 0\}$& $B>0,\ 1>\lambda>0 $ &$>0$  &$<0$   & & & & & &Fig. \ref{FIG1} \\ \hline 
\rule[0.5mm]{0cm}{4mm}\rule[-2.5mm]{0cm}{4mm}   & &$B=0,\ 1>\lambda>0 $ &$=0$  &$=0$   & $div.(0)$& $div.(0)$ & & &$div.(0)$ &Fig. \ref{FIG5B} \\
\hhline{|=|=|=|=|=|=|=|=|=|=|=|}
%IIb-
\rule[0.5mm]{0cm}{4mm}\rule[-2.5mm]{0cm}{4mm}${\rmII}_b^-,\ \kappa\geq 0$&$\{-1,\ 0\}$&$B>0, \ 1>\lambda>0 $ &$>0$  & $<0$  & & & & & &Fig. \ref{FIG1}\\
\hline\rule[0.5mm]{0cm}{4mm}\rule[-2.5mm]{0cm}{4mm} & &$B=0,\ 1>\lambda>0 $ &$=0$  & =0  &$div.(0)$ & $div.(0)$& & &$div.(0)$ &Fig. \ref{FIG5B}\\
\hhline{|=|=|=|=|=|=|=|=|=|=|=|}
%tIa
$\rule[0.5mm]{0cm}{4mm}\rule[-2.5mm]{0cm}{4mm}\tilde{\bs{I}}_a,\ \kappa\in \mathbb R$&$\{-1,\ +1\}$&$B^2<{\lambda}/{(1-\lambda)}$ &$\in \mathbb C$ & $\in \mathbb C$ & & & & & &Fig. \ref{FIG5A}\\
 \hline
  &&\rule[0.5mm]{0cm}{4mm}\rule[-2.5mm]{0cm}{4mm}$B^2\geq{\lambda}/{(1-\lambda)}$ & $\geq -1/B$ &$\leq -1/B $&$f.$ & $f.$  & & & &Fig. \ref{FIG1} \\
\hhline{|=|=|=|=|=|=|=|=|=|=|=|}
%tIIb+
\rule[0.5mm]{0cm}{4mm}\rule[-2.5mm]{0cm}{4mm}$ \tilde{\rmII}_b^+,\ \kappa \leq 0 
$&$\{0,\ -1\}$&$\lambda>0 $ &$(\sqrt{B^2+\lambda}-B)/\lambda  $ & $-(\sqrt{B^2+\lambda}+B)/\lambda  $ & & & & & &Fig. \ref{FIG1}\\
\hhline{|=|=|=|=|=|=|=|=|=|=|=|}
%tIIb-
\rule[0.5mm]{0cm}{4mm}\rule[-2.5mm]{0cm}{4mm}$ \tilde{\rmII}_b^-,\ \kappa \geq 0 $&$\{0,\ -1\}$&$\lambda>0 $ & $(\sqrt{B^2+\lambda}-B)/\lambda $  & $ -(\sqrt{B^2+\lambda}+B)/\lambda $  & & & & & &Fig. \ref{FIG1}\\
\hhline{|=|=|=|=|=|=|=|=|=|=|=|}
%IIIp
\rule[0.5mm]{0cm}{4mm}\rule[-2.5mm]{0cm}{4mm}${\rmIII}^+,\ \kappa\leq 0$  &$\{0,\ 0\}$&$B>0  $ & $=0$ &$=-2\,B/\lambda $   & & & & &  &$\kappa\leq\kCTCm$, Fig. \ref{FIG1}\\ \hline
\rule[0.5mm]{0cm}{4mm}\rule[-2.5mm]{0cm}{4mm}   & &$B=0  $ & $=0$ &$=0$   & $div.(0)$&$div.(0)$ & & &$div.(0)$  &$\kappa\leq\kCTCm$, Fig. \ref{FIG5B}\\
\hhline{|=|=|=|=|=|=|=|=|=|=|=|}
%IIIm
\rule[0.5mm]{0cm}{4mm}\rule[-2.5mm]{0cm}{4mm}${\rmIII}^-,\ \kappa\geq 0$&$\{0,\ 0\}$ & & $=0$    &$=-2\,B/\lambda<0 $  & & & & &$div.(1/2)$  &$\kappa\geq 0$ Fig. \ref{FIG5B}\\
\hhline{|=|=|=|=|=|=|=|=|=|=|=|}
%tIIa
\rule[0.5mm]{0cm}{4mm}\rule[-2.5mm]{0cm}{4mm}$\tilde{\rmII}_a,\ \kappa \in \mathbb R$ &$\{0,\ +1\}$& $\lambda>B^2$ &$\in\mathbb C$  & $\in\mathbb C$  & & & & &$div.(1)$ &Fig. \ref{FIG8}  \\
\hline
\rule[0.5mm]{0cm}{4mm}\rule[-2.5mm]{0cm}{4mm} &$\{0,\ +1\}$&$\lambda\leq B^2 $ & &$ \leq -1/\sqrt{\lambda}  $  &  &$f.$ & & & &$\kappa\leq \kCTCm$ Fig. \ref{FIG1}\\
\hline
\rule[0.5mm]{0cm}{4mm}\rule[-2.5mm]{0cm}{4mm} &$\{0,\ +1\}$&$\lambda\leq B^2 $ &$\leq -1/\sqrt{\lambda} $ &   &$f.$ &  & & &$div.(1)$ &$\kappa\geq\kCTCp$ Fig. \ref{FIG4}\\
\hhline{|=|=|=|=|=|=|=|=|=|=|=|}

\end{tabular}
%\end{center}
\end{adjustbox}
\caption{ \label{TableGenwAdS3} A summary of all the elements needed to fix the causal structure of the maximal  extension of the various types of warped geometries displayed in Table \ref{TableItoIII} 
with metric \eqref{wmetcanPhi}.
($B$ is given by \eqref{ratioB}, $f.$ = finite, $div.(0)$= logarithmic divergence, $div.(n) $=divergent (as $1/\epsilon^n$))}
\end{table}

\subsection{The warped self--dual cases: $\lambda>0,\ \LR\geq 0,\,\LL=0$}
\label{selfdualcausal}
If $\epsR\leq 0$ the full spacetime contains closed timelike curves. We thus consider $\epsR=+1$ where no closed timelike curves occur and $\kappa$ belongs to $\mathbb R$. The metric (\ref{wmetcanPhisd}) can be rewritten as:
\begin{align}
ds^2=-(\kappa^2-\epsL)d\tau^2+\frac{d\kappa^2}{4(\kappa^2-\epsL)}+(1+\lambda)\,\LR^2(d\varphi+\frac{\kappa}{\LR}\,d\tau)^2\qquad 
\end{align}
  such that
\begin{align}
\hat F(\kappa)=2\,(\kappa^2-\epsL)\qquad.
\end{align}
The integrals defining $x(\pm \infty)$ are finite and thus are vertical lines in the diagram.
\begin{table}[h!]
%\begin{adjustbox}{width=\columnwidth,center}
\begin{center}
\hspace{-2cm}\begin{tabular}{ |c|c|c|c|c|}
\hline
\multicolumn{5}{|c|}{\footnotesize{$x(\kappa_0)\propto\int^{\kappa_0}1/(\kappa^2-\epsL )\,d\kappa\ ,\  \epsR=+1\ ,\  \kappa\in\mathbb R\ ,\  x(\pm \infty)=f.$}}\\
\hline  $  \epsL$&$x(+1)$&$x(-1)$&$x(0)$&Fig.\\
\hhline{|=|=|=|=|=|}
%epsL=-1
\rule[0.5mm]{0cm}{4mm}\rule[-2.5mm]{0cm}{4mm} $-1$  & &  & & Fig.\ \ref{FIG6B}\\ 
\hhline{|=|=|=|=|=|}
%epsL=0
\rule[0.5mm]{0cm}{4mm}\rule[-2.5mm]{0cm}{4mm} $  \phantom{-}0$  & &  &$div.(1)$ &  Fig. \ref{FIG7}\\ 
\hhline{|=|=|=|=|=|}
%epsL=-1
\rule[0.5mm]{0cm}{4mm}\rule[-2.5mm]{0cm}{4mm} $+1$  &$div.(0)$ &$div.(0)$  & &Fig. \ref{FIG3C}\\ 
\hhline{|=|=|=|=|=|}
\end{tabular}
\end{center}
%\end{adjustbox}
\caption{ \label{TableSDmetwAdS3} A summary of all the elements needed to fix the causal structure of the maximal  extensions of self-dual ($\LL = 0$) warped geometries with $\epsR = 1$ which are described by \eqref{wmetcanPhisd} ($f.$ = finite, $div.(0)$= logarithmic divergence, $div.(n) $=divergent (as $1/\epsilon^n$)) }
\end{table}
As $\LL=0$ the geometry of the warped AdS$_3$ quotient does not depend on the nature of the vector $\xiL$, but  the local coordinate expressions of the metric and the corresponding domains of the various coordinate patches do. The three expressions of the metric (\ref{wmetcanPhisd}) for $\epsL=\pm 1,\,0$  are local descriptions of the complete space depicted by the strip, with the various coordinate patches covering different domains of the full space (See section \ref{Expexp}).

Note that the function $\hat F(\kappa)$ is independent from $\lambda$. Thus, the discussion of the causal structure of warped geometries applies {\it mutatis mutandis} to the unwarped situation (self--dual BTZ black hole geometries).

\subsection{The special case of no warping $\lambda = 0$}
\label{subsec:lambda0}
To be complete and allow the reader to compare the projection diagrams with previous work on the causal structure of BTZ black holes \cite{Banados:1992gq}, we suppress the warping by setting $\lambda=0$.
Accordingly, the root $\kCTCm$ of Eq. (\ref{defkappaCTC}) is pushed to minus infinity; the chronological singularity rests at $\kCTCp=-(\epsL\,B +\epsR\,B^{-1}) $ and $x(\infty)$ is finite. As now the left and right vectors are on the same footing, we have fewer cases to consider: the exchange of $\xiR$ with $\xiL$ here reduces to the replacement of $B$ by $B^{-1}$ (for $B \neq 0$). For $B = 0$, the causal structure is the same as in the unwarped case.
 
\begin{table} [h!]
\begin{adjustbox}{width=\columnwidth,center}
%\begin{center}
%\hspace{-2cm}
\begin{tabular}{|c|c|c|c|c|c|c|c|c|c|}
\hline
\multicolumn{9}{|c|}{\rule[6mm]{0mm}{0mm}\ \rule[-3mm]{0mm}{0mm} $x(\kappa_0)\propto \sqrt{B }\,\int^{\kappa_0}\sqrt{(\kappa-\kCTCp)}/(\kappa^2-\epsR\epsL)\,d\kappa$\qquad,\qquad $x( + \infty)=f.$}\\
\hline Type&$\{\epsR,\ \epsL\}$& $B>0$ &  $\kCTCp$&$x(\kCTCp)$& $x(+1)$&$x(-1)$&$x(0)$&Fig.\\
\hhline{|=|=|=|=|=|=|=|=|=|=|}
%Ia
$\rule[0.5mm]{0cm}{4mm}\rule[-2.5mm]{0cm}{4mm}\ \bs{I}_a,\ \kappa\in \mathbb R$&$\{+1,\ -1\}$&  &$1/2(B^{-1}-B)\in \mathbb R$&$f.$& & &  &$\kappa\geq \kCTCp$,  Fig. \ref{FIG6}\\ 
\hhline{|=|=|=|=|=|=|=|=|=|=|}
%IIa
\rule[0.5mm]{0cm}{4mm}\rule[-2.5mm]{0cm}{4mm}${\rmII}_a\ \kappa\in \mathbb R$&$\{+1,\ 0\}$&  &$-B/2<0$  &$f.$   & &  & $div.(1)$  &$\kappa\geq -B/2$, Fig. \ref{FIG3B} \\
\hhline{|=|=|=|=|=|=|=|=|=|=|}
%Ib
$\rule[0.5mm]{0cm}{4mm}\rule[-2.5mm]{0cm}{4mm}\ \bs{I}_b,\ \kappa\in \mathbb R$&$\{+1,\ +1\}$&  $B=1$   &$=-1$ & $f.$ &$div.(0)$ &$f.$ &  &$\kappa\geq-1$, Fig. \ref{FIG9}\\
\hline
$\rule[0.5mm]{0cm}{4mm}\rule[-2.5mm]{0cm}{4mm} $& &$B\neq 1$  & $ <-1$ &    $f.$&$div.(0)$ &$div.(0)$ & &$\kappa\geq\kCTCp$, Fig. \ref{FIG3A} \\
\hhline{|=|=|=|=|=|=|=|=|=|=|}
%Ic-
$\rule[0.5mm]{0cm}{4mm}\rule[-2.5mm]{0cm}{4mm}\ \bs{I}_c^-,\ \kappa\geq 1$&$\{-1,\ -1\}$& & $\geq 1$ &$f.$   &&  & &$\kappa\geq \kCTCp>1  $, Fig. \ref{FIG6} \\
\hhline{|=|=|=|=|=|=|=|=|=|=|}
%IIb-
\rule[0.5mm]{0cm}{4mm}\rule[-2.5mm]{0cm}{4mm}${\rmII}_b^-,\ \kappa\geq 0$&$\{-1,\ 0\}$&  &$B/2>0$  & $f.$  & & & &$\kappa\geq B/2$, Fig. \ref{FIG6}\\
\hhline{|=|=|=|=|=|=|=|=|=|=|}
%III-
\rule[0.5mm]{0cm}{4mm}\rule[-2.5mm]{0cm}{4mm}$ {\rmIII}^-,\ \kappa\geq 0$&$\{0,\ 0\}$&      &$= 0 $&  &   & &$div.(1/2)$  &$\kappa\geq 0$ Fig. \ref{FIG6A}\\
\hhline{|=|=|=|=|=|=|=|=|=|=|}

\end{tabular}
%\end{center}
\end{adjustbox}
\caption{ \label{TableGenBTZ} A summary of all the elements needed to fix the causal structure of the maximal  extension of generic BTZ geometries (see Ref. \cite{Banados:1992gq})
which here are given by \eqref{wmetcanPhi} with $\lambda = 0$.
 ($B$ is given by \eqref{ratioB}, $f.$ = finite, $div.(0)$= logarithmic divergence, $div.(n) $=divergent (as $1/\epsilon^n$)) 
}
\end{table}

\subsection{An alternative projection using a modified Iwasawa coordinate system }\label{ProjMICS}
In the previous sections, we displayed the causal structure of the different Lorentzian spaces defined by the various ways of warping the bi-invariant $SL(2,\,\mathbb{R})$-metric and taking quotients of the $SL(2,\,\mathbb{R})$ manifold. Here we would like to present an alternative description of the causal structure obtained directly from the quotient structure. The idea has been described in detail in Ref.\cite{Bieliavsky:2003de}, section {\bf 4}. We limit ourselves to the black hole configurations.  The idea consists of the following.  
Once the identifications leading to black hole structure are fixed, we build a global coordinate system, adapted to the so-called \emph{bi-action}: the combined right action generated by the left invariant component 
$\LL\,\xiL$  of the  vector $\vec\partial$ defining the identification and the left action generated by $\LR \xiR$, as given in Eq.(\ref{Identif}). Assuming $L_{\text{\tiny R}}>L_{\text{\tiny L}}$, it consists of the parametrisation of the matrix $\mathbf{z}$, Eq.(\ref{SL2Rmat}), obtained by writing it as\footnote{In case where $\LR <\LL $ a global parametrisation is obtained by writing $\mathbf{z}= :a(\LL\,v)\,n(u)\,k(t^\prime)\,
     a(\LR\,v)$} :
\begin{align}
    \mathbf{z}&=\left(\begin{array}{cc}
      e^{\LR\,v}   &0  \\
        0 & e^{-\LR\,v}
    \end{array}\right)
     \left(\begin{array}{cc}
      1   &u  \\
        0 & 1
    \end{array}\right)
     \left(\begin{array}{cc}
      \cos(  t^\prime)   &\sin(t^\prime)  \\
        -\sin(t^\prime) & \cos(t^\prime)
    \end{array}\right)
    \left(\begin{array}{cc}
      e^{\LL\,v}   &0  \\
        0 & e^{-\LL\,v}
    \end{array}\right)\\
    &=:a(\LR\,v)\,n(u)\,k(t)\,
     a(\LL \,v)\qquad .
     \end{align}

     This decomposition dictated by the quotient furnishes a  fibre structure --   a foliation parametrised by $v$ -- of the space. Consider a sheet ($v$ constant) of this foliation. To each of its point corresponds   a circle of the warped AdS$_3$ black hole: the orbit of the one-parameter subgroup generated by the $e^{v\,\vec\partial}$ translations. 
Plugging this parametrisation in the metric Eq. (\ref{glamKV}) we obtain:
{\small \begin{align}
{ \uuline{g} }\strut_{(\lambda)}=&-{dt^\prime}^2-dt^\prime\,du-2\,\LR\,u\,dt^\prime\,dv-\LL\,\sin(2\,t^\prime)\,du\,dv
 +\Big[\LR^2+\LL^2+2\,\LR\,\LL\,\Big(\cos(2\,t^\prime)-u\,\sin(2\,t^\prime)\Big)\Big]dv^2\nonumber\\
&+\lambda\,\Big[
\Big(\LR+ \LL\big(\cos(2\,t^\prime)-u\,\sin(2\,t^\prime)\big)\Big)\,dv-u\,dt^\prime\Big]^2
\qquad.
\label{MICSds2}
\end{align}
}
In these coordinates the identification yielding the black holes structure, is simply: 
\begin{align}
(t^\prime,\,u,\,v)\mapsto (t^\prime,\,u,\,v+2\,\pi\,m)\ ,\quad m\in \mathbb Z\qquad .
\end{align}

Then  project on the sheet, parallelly to the $\vec\partial$ direction,  the local light cone. So causal curves with respect to the warped metric project inside the light cone projection.

The  space-time  causal structure may be described by considering the two fields of directions resulting from the projection on a surface of constant $v$-coordinate of the light-cone, parallel to the $\partial_v$ direction. To lighten the notation let us write the metric Eq.\ (\ref{MICSds2}): $ds^2=g_{ab}dx^adx^b$, with $\{x^a\}:=\{t,\,u,\,v\}$ and introduce a 2D metric $d\sigma^2=\gamma_{a\,b}dx^a\,dx^b$. The field directions we are looking for correspond to the two null directions of this 2D metric\footnote{Incidentally let us notice that the curvature of this metric blows up on the chronological singularity and the metric signature changes, becoming Euclidean. }, defined by:
\begin{align}
    d\sigma^2:=(g_{a\,v}\,g_{b\,v}-g_{a\,b}\,g_{v\,v})\,dx^a\,dx^b=:\gamma_{a\,b}dx^a\,dx^b\qquad,\qquad \{x^a\}:=\{t,\,u\}\label{dsigma2}
\end{align}
{\it i. e.}, using the variable $p=\arctan (u)$ instead of $u$:
\begin{align}
    \frac{dt}{dp}:=-\frac{\gamma_{t\,u}\pm\sqrt{\gamma^2_{t\,u}-\gamma_{t\,t}\,\gamma_{u\,u}}}{\cos^2(p)\,\gamma_{t\,t}}\qquad.
    \label{dtdp}
\end{align}
Notice that these fields are only defined on the chronologically safe region. We easily verify that: $\gamma^2_{t\,u}-\gamma_{t\,t}\,\gamma_{u\,u}=-\det(g_{a\,b})\,g_{u\,u}$. We also easily check that the curves defined by the intersections of the horizons, Eqs.\ \eqref{Hm}, \eqref{Hp},  with the surfaces of constant $v$-coordinate constitute solutions of Eq.(\ref{dtdp}).
\begin{figure}[ht]
    \centering
        \includegraphics[scale=0.35
]{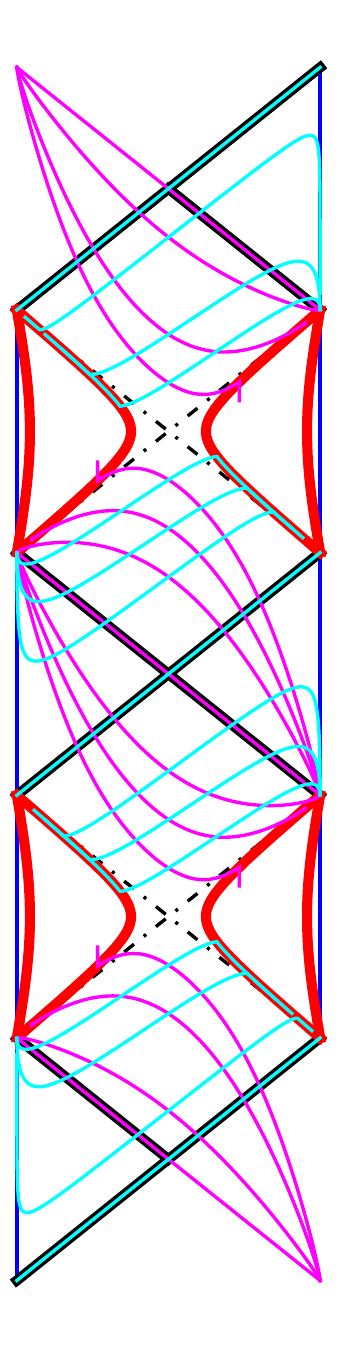}
    \caption{Universal features of spacelike warped AdS space. The time axis $t$ is vertical, while the $p$ coordinate axis is inclined at 45 degrees. Several curves are drawn: chronological singularity boundaries (in red), inner horizons (in dot-dashed black), outer horizon (in full black ) In cyan and magenta, integrals of Eq.\ (\ref{dtdp}) issued from the $t$-axis, at values of $t$ respectively equal to: $-4\,\pi/10,\,-3\,\pi/10,\,-2\,\pi/10,\,-\pi/10,\,+\pi/10,\,+2\,\pi/10,\,+3\,\pi/10,\,+4\,\pi/10$. The integral curves issued from $t=-\pi/2,\, 0$ or $\pi/2$ are the outer horizons.
    }
    \end{figure}

We also find it illuminating to draw several sections of the space, represented by the cylinder (see Appendix \ref{CovSLR} ) given by the product of a unit disk labeled by   $\tanh(\upchi)\,e^{i\,\upvartheta} $  and a time axis $\uptau $, using the    $\uptau,\,\upvartheta,\, \upchi$ coordinates introduced in Eqs. \eqref{cylUpm}, \eqref{cylVpm}.
Surfaces of constant $\kappa$ are $\phi$ invariant. They read:
\begin{align}
    \kappa =&\,U_+\,U_--V_+\,V_-\\=&\cosh^2(\upchi)\,\cos(2\,\uptau)-\sinh^2(\upchi)\,\cos(2\,\upvartheta)\label{cstkappatauchi}\\=&\cos(2\,t^\prime)-u\,\sin(2\,t^\prime)\qquad.\label{cstkappaut}
\end{align}
In particular, we reobtain results displayed in Ref. \cite{Bieliavsky:2003de}, that the horizons are the surfaces:
\begin{align}
&\mathcal H^-_1:\qquad t=\pi/2+m\,\pi&\qquad &\mathcal H^-_2:\qquad u=\cot(t)\label{Hm}\\
&\mathcal H^+_1:\qquad t=m\,\pi&\qquad &\mathcal H^+_2:\qquad u=-\tan(t)\label{Hp}
\end{align}
 as follows immediately from Eq. \eqref{cstkappaut} for $\kappa=\pm 1$.

More concretely let us introduce on a temporal section of fixed value of $\uptau$ Cartesian coordinates: $x=\tanh(\upchi)\,\cos(\upvartheta)$ and $y=\tanh(\upchi)\,\sin(\upvartheta)$, ($x^2+y^2\leq 1)$. The quadratic equation defining the intersection of the surfaces of fixed value of $\kappa$ with the temporal section  is obtained directly from Eq.\ (\ref{cstkappatauchi}) and reads:
\begin{align}
     \tanh^2(\upchi)=\frac{\kappa-1-\sin^2(\uptau)}{\kappa-1-\sin^2(\upvartheta)}
    \qquad\text{or}\qquad
     (\kappa-1)\,x^2+(\kappa-2)\,y^2=(\kappa-1)-\sin^2(\uptau)\qquad.
\end{align}

They are given by the intersection of conical curves with the unit disk
\begin{itemize}
    \item{$\kappa=2+k^2>2$}: arc of an ellipse,
    $$(1+k^2) x^2+k^2 y^2=k^2+ \cos^2(\uptau)$$
\item{$\kappa=2$}: two straight line segments
$$x=\pm \cos(\uptau)$$
\item{$2>\kappa=1+k^2>1$} arc of a hyperbola$$k^2\,x^2-(1-k^2)\, y^2=k^2- \sin^2(\uptau)$$
If $k^2= \sin^2(\uptau)$ the hyperbola degenerates into two intersecting straight line segments.
\item{$\kappa=1$}: two straight line segment (external horizon)
$$ y^2= \sin^2(\uptau)$$
\item{$\kappa=1-k^2<1 $}: arc of an ellipse pieces$$ k^2\, x^2+(1+k^2)\, y^2=k^2+ \sin^2(\uptau)$$
(For $\uptau$ a multiple of $\pi/2$ degeneracy's occur) 
\end{itemize}
The equations providing the intersections of constant $v$ surfaces with the same temporal sections are:
\begin{align}
    &x=\frac{\big(\sinh[(\LR+ \LL)\,v]\,\cos^2(t^\prime)
    +e^{2\,\LL\,v}\,\cosh[(\LR-\LL)\,v]\,\sin^2(t^\prime)\big)\cos(\uptau)}
    {\cosh[(\LR+\LL)\,v]+e^{-(\LR-\LL)\,v}\,\sinh[2\,\LL \,v]\,\sin^2(\uptau)}\nonumber\\
    &\hphantom{x=}-\frac{\ft 12\,e^{-(\LR-\LL)\,v}\,\sin(2\,t^\prime)\sin(\uptau)}
    {\cosh[(\LR+\LL)\,v]+e^{-(\LR-\LL)\,v}\,\sinh[2\,\LL \,v]\,\sin^2(\uptau)}\\
    &y=\frac{\big(\cosh[(\LR+\LL)\,v]\,\cos^2(t^\prime)
    +e^{2\,\LL\,v}\,\sinh[(\LR-\LL)\,v]\,\sin^2(t^\prime)\big)\sin(\uptau)}
    {\cosh[(\LR+\LL)\,v]+e^{-(\LR-\LL)\,v}\,\sinh[2\,\LL \,v]\,\sin^2(\uptau)}\nonumber\\
    &\hphantom{y=}-\frac{\ft 12\,e^{-(\LR-\LL)\,v}\,\sin(2\,t^\prime)\cos(\uptau)}
    {\cosh[(\LR+\LL)\,v]+e^{-(\LR-\LL)\,v}\,\sinh[2\,\LL \,v]\,\sin^2(\uptau)}
\end{align}
By extracting from them the expressions of $\cos(2\,t^\prime)$ and $\sin(2\,t^\prime)$ we easily see that these curves also are
ellipses, tangent to the circle at infinity ($x^2+y^2=1$) at the point $x_c=\cos(\uptau)$, $y_c=\sin(\uptau)$, corresponding to the asymptotic region ($\kappa=\pm \infty$) of the warped AdS$_3$ space. 
\begin{figure}[h]
\begin{center}
 \includegraphics[width=4 cm]{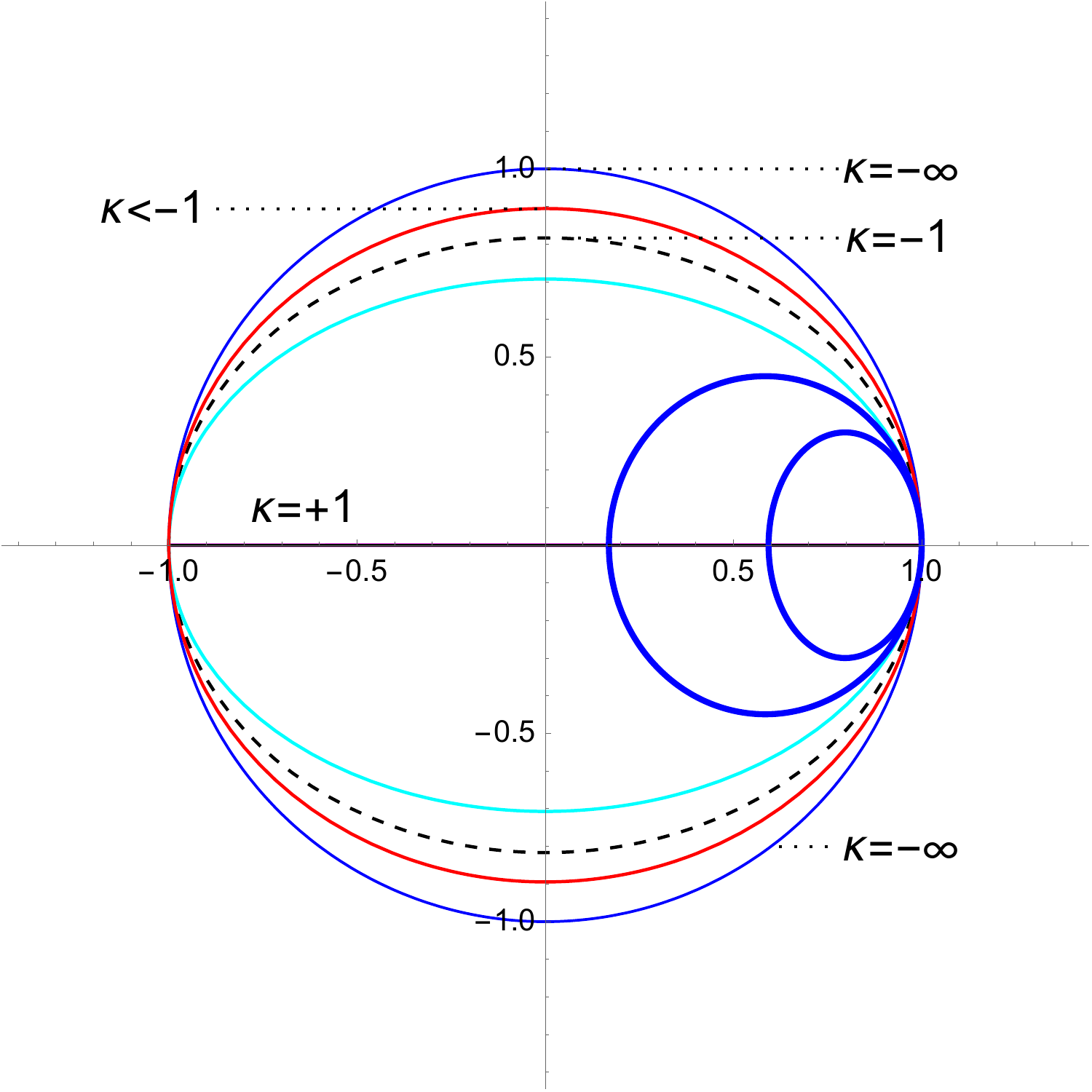}
 \includegraphics[width=4 cm]{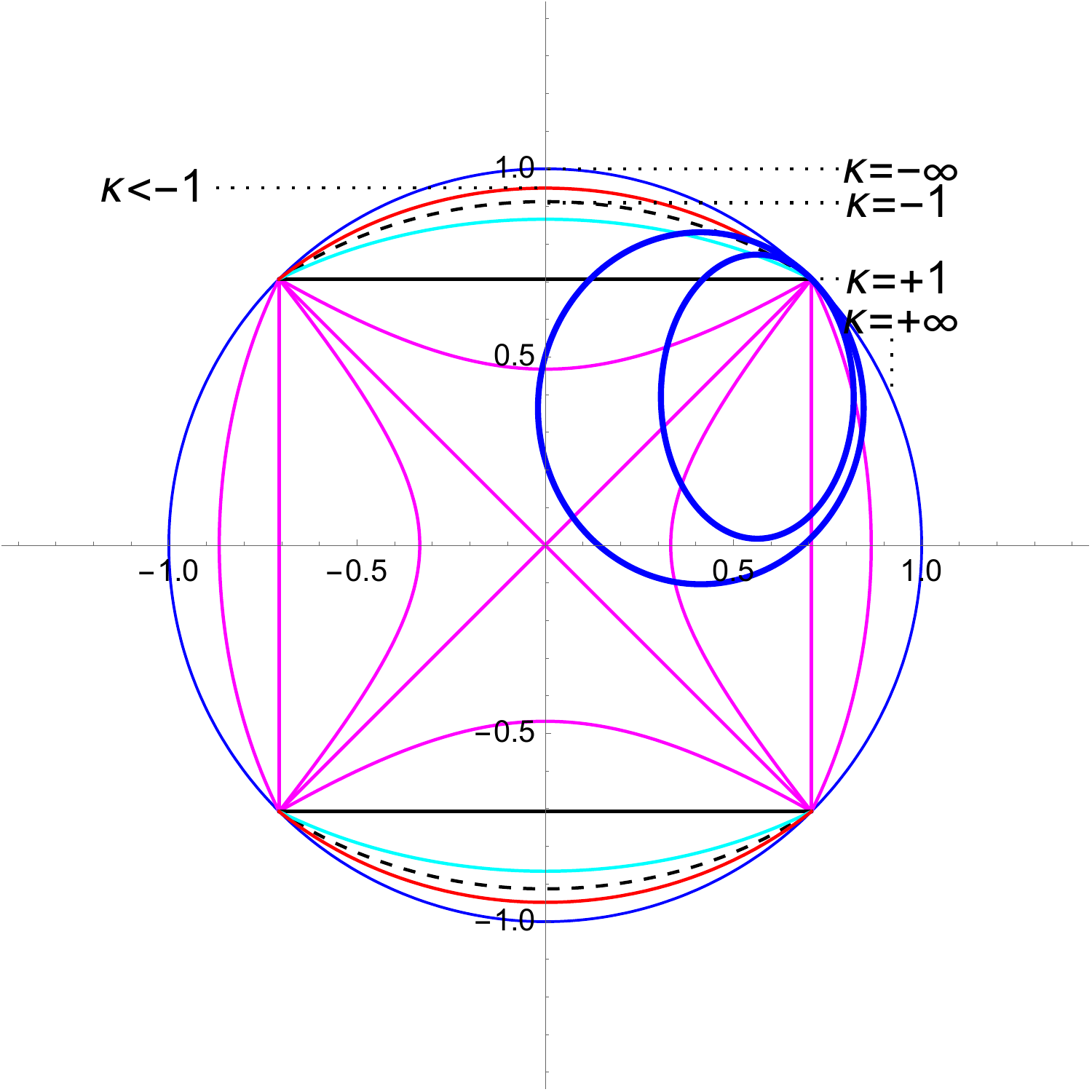}
 \includegraphics[width=4 cm]{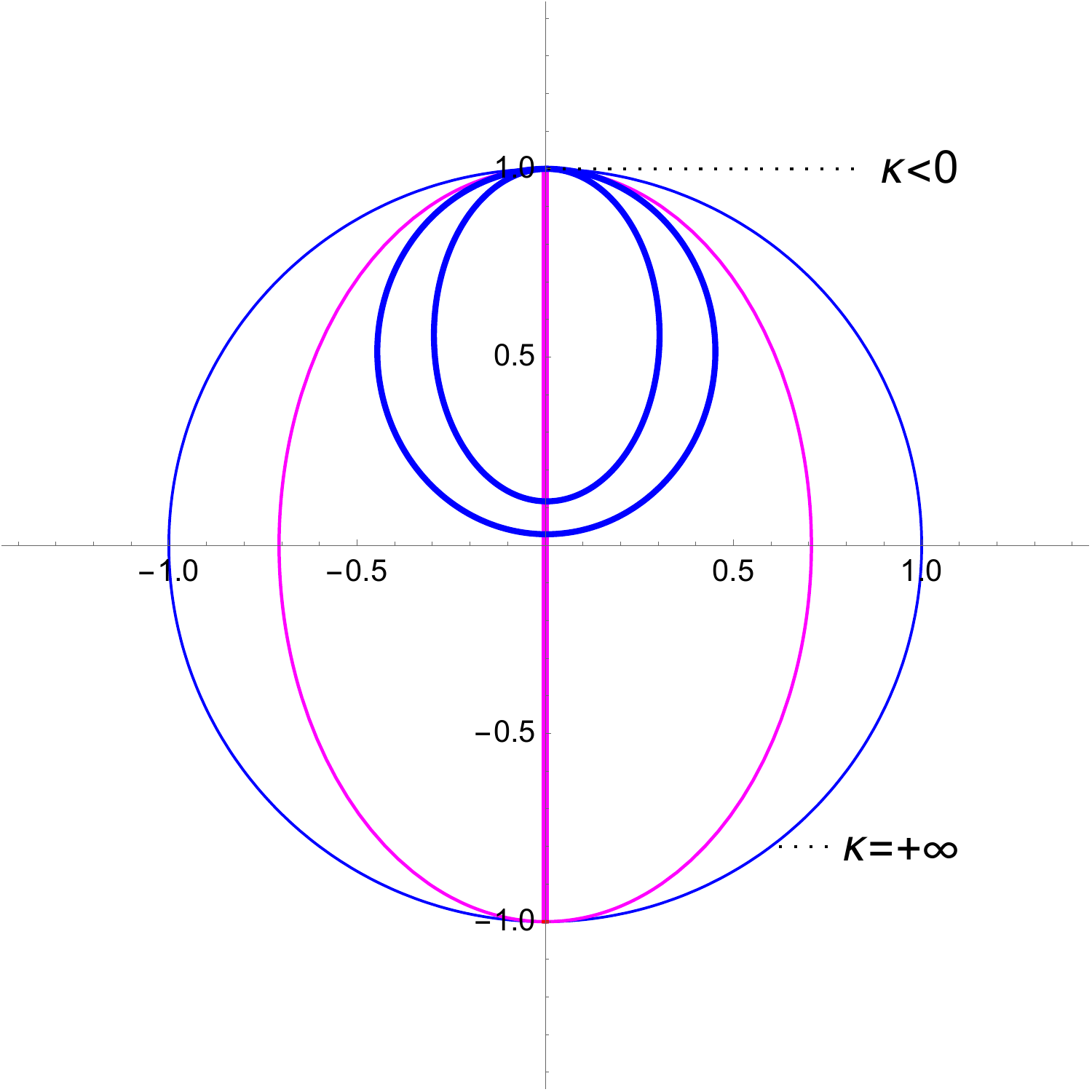}
 \end{center}
 \caption{The intersections of sections of constant $\uptau = (0,\,\pi/4,\,\pi/2)$ with various surfaces of fixed $\kappa$ (in magenta for $\kappa>1$, in black for the external horizon $\kappa=1$, in cyan for the region between the horizons: $1>\kappa>-1$, in dotted black for the internal horizon $\kappa=-1$ and in red for a typical value of $\kappa< -1$. Two intersections with surfaces of constant $v$ also are depicted (in blue, with respectively $\LR=2, \, \LL=\sqrt{2}$). They correspond to the boundary of a fundamental domain of the quotient defining the black hole space.  }\label{figsectvcst}
\end{figure}

\section{Conclusions} 
In this work, we discussed warped AdS$_3$ geometries from first principles. Starting from AdS$_3$, we considered the warped spaces that arise by deforming the space with a spacelike, timelike or lightlike (left or right) invariant  Killing vector. We illustrated their relevance as solutions of various gravity theories and also discussed generalizations to local lightlike warped geometries.
We then constructed the general expression of the metric, Eqs.\ \eqref{wmetcan}, \eqref{wmetcanPhi}, \eqref{wmetcanPhisd}, for  all quotients compatible with this warping. We discussed the causal structure of these spaces using projection diagrams, pointing out which spaces are black holes.
  We also illustrated how the comparison of invariant quantities (scalar products of specific Killing vectors) allows to relate the different choices of coordinates encountered in the literature.
Our work highlights how geometrical considerations enlighten and simplify the description of warped AdS geometries.  

 \section{Acknowledgements}
We thank Piotr Chru\'{s}ciel, St\'ephane Detournay, Mateusz Piorkowski and Friedrich Schöller for helpful and interesting discussions and Daniel  Grumiller for drawing our attention to the work \cite{Andrianopoli:2023dfm}. 
We express our gratitude to two anonymous referees whose interesting comments have improved the manuscript.
RW acknowledges support of the Fonds de la Recherche Scientifique F.R.S.-FNRS (Belgium) through 
the PDR/OL C62/5 project ``Black hole horizons: away from conformality'' (2022-2025), the
Heising-Simons Foundation under the “Observational Signatures of Quantum Gravity” collaboration grant 2021-2818 and the U.S. Department of Energy, Office of High Energy Physics,
under Award No. DE-SC0019470.
She thanks the Erwin Schrödinger Institute for hospitality, where part of this work was carried out. 
 \appendix
\section{Remarks about notations} 
\label{Notations} In this work we will encounter several metrics and the scalar products they define. We adopt the following notations to distinguish between them:
\begin{itemize}
\item When vectors (or one-forms) are considered as objects in     $M^{2, 2}$ we use a dot to denote their norms and scalar products or use the symbol $\upeta$ (Denoting, as usual, by $\eta_{AB}$  their covariant components and $\eta^{AB}$ their contravariant ones):
\begin{align}
\vec v\cdot\vec v=-(u_+\,u_-+v_+\,v_-)=\upeta(\vec v,\,\vec v)=\eta_{AB}\,v^A\,v^B
\end{align}
\item We generally use the same symbol to denotes objects in     $M^{2, 2}$ or their restriction to $\mathcal H$. When we want to insists on products of objects belonging to the tangent bundle of $\mathcal H\simeq SL(2,\,\mathbb R)$ with the bi-invariant metric (AdS$_3$) we use the notation: $
\underline{\underline g}_{(0)}(\vec v,\,\vec v)$
and when objects are expressed in components with respect to frame of right or left invariant vector or one-form fields:
\begin{align}
\vec v=v^a\,\vec r_{(a)}: \quad \vec v\cdot\vec v=\eta^{(3)}_{a\,b}\,v^a\,v^b\qquad,\qquad \oneform k=k_a\,\oneform\lambda ^a: \quad \oneform k\cdot\oneform k=\eta^{(3)}\vphantom{\eta}^{a\,b}\,k_a\,k_b\qquad,
\end{align}
\item When we consider the warped AdS$_3$ space ({\it i.e.} the right invariant but not bi-invariant metric on  $SL(2,\, \mathbb R))$ we use the notations: $
\underline{\underline g}_{(\lambda)}(\vec v,\,\vec v)
$.
\item The contraction of a vector $\vec v$   with the one-form $\oneform k$ is denoted by:
$\oneform k(\vec v)$
or in terms of components with respect to a frame and its tensorial dual:
$k_a\,v^a$
\item When we regroup vector components into matrices as in Eqs (\ref{matvec}, \ref{RLvecrep}) we have:
\begin{align}
    &\underline{\underline g}_{(0)}(\vec v(\boldsymbol{z}),\,\vec w(\boldsymbol{z}))=\ft 12\,\text{Tr}(\boldsymbol{v}\,\boldsymbol{z}^{-1}\,\boldsymbol{w}\,\boldsymbol{z}^{-1})=\frac 12\,\text{Tr}(V_R\,W_R)
\end{align}
\end{itemize}
\section{Proofs of Eqs (\ref{Clstsig}, \ref{ClstpR}, \ref{ClstpL})}\label{proofdpLR}\label{ProofsEqs}
In this Appendix, we want to give  proofs of Eqs (\ref{Clstsig}, \ref{ClstpR}, \ref{ClstpL}). Of course, they may be checked by brute force, expressing them in null-coordinates and computing derivatives. But we also could proceed as follows --- in a more elegant way.\\
\subsubsection*{Preliminaries}
The fields $\vxiR$ and $\vxiL $, expressed as linear combination of the vector fields Eqs (\ref{KVLR}), are Killing vector fields of the $M^{2,2}$ metric (\ref{M22met}):
\begin{align}
    &\mathcal{L}_{\vxiR}\upeta=0\qquad ,\qquad
    \mathcal{L}_{\vxiL}\upeta=0\qquad .\label{Liexieta}
\end{align}
Moreover the vector $\vnH $, introduced in Eq.\ (\ref{nH}) is a conformal Killing vector of the $M^{2,2}$ metric:
\begin{equation}
    \mathcal{L}_{\vnH }\upeta=-4\,\upeta\qquad \text{\it i.e.}\qquad (\mathcal{L}_{\vnH }\upeta)_{AB}=2\,\partial^2_{AB}H=-4\,\eta_{AB}\qquad .\label{Lieneta}
\end{equation}
Making use of the commutation property of the Lie derivative and the exterior differential we obtain
\begin{align}
    \vxiR(H)=0\Rightarrow \vxiR(d H)=0\qquad,\qquad \vxiL (H)=0\Rightarrow \vxiL (d H)=0\qquad .\label{xiHdH}
\end{align}
The vectors $\vxiR$  and $\vxiL $ being Killing vectors and the Lie derivative obeying the Leibniz rule with respect to contraction, we obtain (expressing Lie derivative as brackets) the commutation relations:
\begin{equation}
    [\vxiR,\,\vnH ]=[\vxiL ,\,\vnH ]=\vec 0\qquad.\label{LBxin}
\end{equation}
From  the formula (an application of the Leibniz rule), where $\vec a$, $\vec b$ and $\vec c$ are arbitrary, 
\begin{equation}
    \vec a(\vec b\cdot\vec c)=[\vec a,\,\vec b]\cdot \vec c+\mathcal{L}_{\vec a}\upeta(\vec b,\,\vec c)+\vec b\cdot [\vec a,\,\vec c]\qquad.\label{Lieprod}
\end{equation}
and the expression of $\kappa$, Eq. (\ref{defkappa}), written as $\kappa=\upeta(\vxiR,\,\vxiL )$ we obtain, using  Eqs (\ref{LBxin}, \ref{Liexieta}), that:
\begin{align}
    \vxiR(\kappa)=\vxiL (\kappa)=0\label{xikappa}
\end{align}
and as a consequence 
\begin{align}
    \vxiR(d\kappa)=\vxiL (d\kappa)=0\qquad .\label{xidkappa}
\end{align}
Notice also that from Eq. (\ref{Lieprod}) and Eqs (\ref{Lieneta}) we obtain:
\begin{align}
    \vec n_{\text{\tiny R}}(\kappa)=\vec n_{\text{\tiny R}}(\vxiR\cdot\vxiL)=-4\,\kappa \qquad,\label{nkappa}
\end{align}
while Eq. (\ref{nH}) provides:
\begin{align}
    \vec n_{\text{\tiny R}}(H)=-4\,H \qquad.\label{normnH}
\end{align}
\subsubsection*{About equation (\ref{Clstsig})}
The first two conditions expressed in Eqs (\ref{defsigma}) imply that:
\begin{align}
    \oneform{\sigma}\propto s_{\text{\tiny H}}\,dH+s_\kappa\,d\kappa\qquad .
\end{align}
To satisfy the last one  we notice that (using Eqs.\ (\ref{LBxin}))
\begin{align}
    \oneform{\sigma}(\vnH ) =& s_{\text{\tiny H}}\,\vnH (H)+s_\kappa\,\vnH (\kappa)\\
    &=s_H\,\eta^{AB}\partial_AH\,\partial_B H+s_\kappa\,\mathcal{L}_{\vnH }\upeta(\vxiR,\,\vxiL )\\
    &=-4(s_{\text{\tiny H}}\,H+s_{\kappa}\,\kappa)\qquad .
\end{align}
Thus $\oneform{\sigma}\propto H\,d\kappa-\kappa\,dH$. It remains to normalise it, by convention according to Eq.\ (\ref{normsig}):
\begin{align}
    &(H\,\partial_A\kappa-\kappa\,\partial_AH)\eta^{AB}(H\,\partial_B\kappa-\kappa\,\partial_BH) \nonumber\\
    &=H^2 \,\partial_A\kappa\, \eta^{AB}\, \partial_B\kappa-2\,H\,\kappa\,\partial_A\kappa\, \eta^{AB}\, \partial_BH-4\,\kappa^2\,\partial_AH\,\eta^{AB}\, \partial_BH
\end{align}
The last two terms are given by Eqs (\ref{nkappa}), (\ref{normnH}). 
\begin{align}
  &\partial_AH\,\eta^{AB}\, \partial_BH=\vnH \cdot\vnH =-4\,H\qquad ,\\
  &\partial_A\kappa\,\eta^{AB}\, \partial_BH=\vnH (\kappa)=-4\,\kappa\qquad .
\end{align}
To obtain the first one we notice, using the Killing equations Eqs (\ref{Liexieta}) and the absence of curvature of the flat ultra-hyperbolic metric of $M^{2,2}$, that:
\begin{align}
    &\partial_B\kappa=-2\xiR^A\partial_A\xiL_B\qquad ,\\
    &\partial_A\xiL^C\partial_B\xiL_C=\ft12\partial_{A B}^2(\xiL^C\, \xiL_C)=\ft12\,\epsL\partial_{AB}^2H\qquad ,\\
    &\partial_A\kappa\,\eta^{AB}\,\partial_B\kappa =-4\,\epsR\,\epsL\,H \qquad ,
\end{align}
which leads, after a last choice of sign, to Eq.\ (\ref{Clstsig}).

\subsubsection*{About equations (\ref{ClstpR}, \ref{ClstpL})}
Since the Lie derivatives of the metric $\upeta$ with respect to $\vxiR$ and $\vxiL $ are zero, we obtain from Eq.\ \eqref{Lieprod}:
\begin{align}
   0 &= \vxiR(\vec\sigma\cdot\vxiR)-[\vxiR,\vec\sigma]\cdot\vxiR-\vec\sigma\cdot[\vxiR,\vxiR]\nonumber\\
   &= -[\vxiR,\vec\sigma]\cdot\vxiR\qquad.
\end{align}
as $\vec\sigma\cdot \vxiR=0$ and $[\vxiR,\,\vxiR  ]=\vec 0$. Similarly, but now using $[\vxiR,\,\vxiL  ]=\vec 0$ we obtain:
\begin{equation}
0=-[\vxiL ,\vec\sigma]\cdot\vxiR\qquad.
\end{equation}
Hence on $\Delta\neq 0$, {\it i. e.} where $\vxiR$ and $\vxiL $ are linearly independent, the Lie Brackets $[\vxiR,\vec\sigma]$ and $[\vxiL ,\vec\sigma]$ are along the vectors $\vec n$ and $\vec\sigma$. \newline
From the formula:
\begin{equation}
    d\oneform{\psi}(\vec a,\,\vec b)=\vec a\left(\oneform{\psi}(\vec b)\right)-\vec b\left(\oneform{\psi}(\vec a)\right)-\oneform{\psi}([\vec a,\,\vec b])\label{dpsiab}
\end{equation}
 where $\oneform {\psi}$, a one-form, and $\vec a$, $\vec b$, vector fields, are arbitrary,  one finds thanks to  Eqs.\ \eqref{pRLxiRL}, \eqref{pRxiLsign}, \eqref{pLxiRsign}
 \begin{align}
     d\oneform{p}^R(\vxiR,\vxiL )=d\oneform{p}^R(\vxiR,\vec\sigma)=d\oneform{p}^R(\vxiL ,\vec\sigma)=0\qquad .
 \end{align}
 Using Eqs (\ref{LBxin}) in Eq. (\ref{dpsiab}) we obtain:
 \begin{align}
     d\oneform{p}^R(\vnH ,\vxiR )= d\oneform{p}^R(\vnH ,\vxiL )=0\qquad .
 \end{align}
 To compute the last component $d\oneform{p}^R(\vnH ,\vec \sigma)$ we need to know the Lie bracket of $\vnH$ and $\vec \sigma$. It results immediately from the previous results --- Eq.\ (\ref{Lieneta}) written for contravariant metric tensor, and Eqs.\ \eqref{nkappa}, \eqref{normnH} --- by noticing that the components of $\vec \sigma $ and $\oneform \sigma $ are related by:
 \begin{align}
     \sigma^A=H\,\eta^{A B}\,\sigma_B
 \end{align}
 and applying  the Leibniz rule. We obtain:
  \begin{align}
     [\vnH,\,\vec \sigma]=-4\,\vec \sigma \qquad.
 \end{align}
 Accordingly,
 \begin{align}
     d\oneform{p}^R(\vnH ,\vec \sigma)=0
 \end{align} 
which proves Eq.\ (\ref{ClstpR}). \\Equation (\ref{ClstpL}) is obtained by similar arguments.
\section{Decrypting Coordinate Systems for (warped) \texorpdfstring{AdS$_3$}{AdS3}}
\label{Hparam}
\subsection{Group structure of \texorpdfstring{$\widetilde{\text{AdS}}_3$}{AdS3t}}\label{CovSLR}
The  AdS$_3$ space naturally identifies itself with the group $SL(2,\,\mathbb R)$ whose underlying manifold is diffeomorphic to the hyperboloid Eq.\ (\ref{AdS30hyp}). For physical applications, causality imposes instead to consider its universal covering that still remains a group. We think it is interesting to provide its group structure, as it cannot be realised as a matrix group (See Refs\cite{DNF}, chapter {\bf 1}, section {\bf 3.2}, \cite{Knapp}, page 368). A standard coordinate system used on $\widetilde{\text{AdS}}_3$ consists of coordinate $\{\uptau, \,\upchi,\,\upvartheta \}$, the ones used in Eqs.\ \eqref{cylUpm}, \eqref{cylVpm},
but with the coordinate $\uptau$ running from $-\infty$ to $+\infty$ instead of on the interval $[0,\,2\,\pi]$ for the $SL(2,\, \mathbb R)$ group. In terms of these coordinates the group law on $\widetilde{SL(2,\,\mathbb R)}$ is as follows. Let us introduce a complex variable (of modulus $\vert z\vert \leq 1$): $z := \tanh(\upchi)\,e^{i\,\upvartheta }$  and label the group elements as $\{\uptau,\,z\}$.  The group composition law (denoted by $\circ$) is:
\begin{align}
    \{\uptau_2,\,z_2\}\circ \{\uptau_1,\,z_1\}:=\left\{\uptau_1+\uptau_2-\frac i2\,\ln\left(\frac{1+z_1\,\overline z_2\, e^{- i\,(\uptau_1+\uptau_2)}}{1+\overline z_1\,  z_2\, 
     e^{+ i\,(\uptau_1+\uptau_2)}}
    \right),
    \ \frac{e^{-\,i\,\uptau_2}\,z_1+e^{+i\,\uptau_1}\,z_2}{\left\vert 1+\overline z_1\, z_2\,e^{+i\,(\uptau_1+\uptau_2)}\right\vert}\right \}\qquad .
\end{align}
The identity reads $\{0,0\}$, the inverse of the element $\{\uptau,\,z\}$ is $\{-\uptau,-z\}$. The center of this group are the elements $\left\{\{ k\,\pi,\,0\}\vert k\in \mathbb Z\right\}$. \newline
The $\uptau$ and $z$ coordinates parametrise a cylinder in $\mathbb R^3$ (the three-dimensional Einstein cylinder once embedded in a four-dimensional Minkowski space). Later on, we shall use it to draw, for the black hole geometries, see Fig. \ref{figH}, several relevant surfaces such as the horizons, sections of the identification foliation and chronological singularities.
\subsection{Examples: warped \texorpdfstring{AdS$_3$}{AdS3} black holes
and NHEK geometry}\label{wAdS3bhmet}
In this Appendix 
we relate Eq.\ (\ref{wmetcanPhi}) and the commonly used expressions of (warped) Anti-de Sitter black hole metrics introduced by Ba\~nados, Teitelboim and Zanelli \cite{Banados:1992wn, Banados:1992gq, carlip19952+}, or by Anninos et {\it al.}\cite{Anninos:2008fx}.\\
BTZ black holes have been extensively analysed in \cite{Banados:1992gq}.
From now we set in this section $\epsR=\epsL=+1$.\newline
For $\lambda = 0$, by redefining 
\begin{align}
    &\tau=2\,\LR\,t\\
    &\varphi=\phi-t\\
     &\kappa=(r^2-\LR^2-\LL^2)/(2\,\LR\,\LL)
\end{align}
the metric \eqref{wmetcanPhi} reduces 
to the  BTZ black hole metric:
\begin{equation}
\label{BTZSchwarzschild}
    ds^2 = -f^2(r) dt^2 + \frac{dr^2}{f^2(r)}
    +r^2 \left( d \phi + N_\phi dt \right)^2\qquad ,
\end{equation}
where 
\begin{equation}
f^2(r) = - M +r^2 + \frac{J^2 }{4 r^2}   \,, \quad N_\phi(r) = - \frac{J}{2 r^2}\qquad .
\end{equation}
As it is well known, the parameters
\begin{align}
    M=2\,({\LR}^2+{\LL}^2)\qquad ,\label{BTZM}\\
    J=2\,(\LL^2-{\LR}^2)\qquad \label{BTZJ}
\end{align}
provide the mass  $M$ and angular momentum $J$ (see Ref. \cite{Banados:1992gq}) of the black hole geometry \eqref{BTZSchwarzschild}.
This metric constitutes a 2-parameter family of vacuum solution of the Einstein equations with cosmological constant $\Lambda = -1$. It was shown 
that \eqref{BTZSchwarzschild} describes a black hole in the parameter domain $M \geq |J|$, a condition which is always verified by the parametrisation given by Eqs.\ (\ref{BTZM}, \ref{BTZJ}). 
When  $\lambda \neq 0$, performing the coordinate transformation 
\begin{equation}
    \kappa = \frac{2 r -r_+ - r_-}{r_+ - r_-}\, , \quad 
    \tau = \frac{\nu^2+3}{4 \nu}t \, , \quad \varphi=\theta
\end{equation}
and the change of parametrisation
\begin{align}  
   &L_{\text{\tiny L}} = \frac{1}{8} \left(\nu ^2+3\right) (r_+ - r_-)\quad , \quad
   L_{\text{\tiny R}} =\frac{\left(\nu ^2+3\right) \left(\nu 
   (r_+ + r_-) -\sqrt{\left(\nu ^2+3\right) r_-
   r_+} \right )}{8 \nu }\qquad , \\
   & \lambda = \frac{3 \left(\nu ^2-1\right)}{\nu ^2+3}\quad,\quad \nu^2=3\frac{\lambda+1}{3-\lambda}
   \qquad ,
   \label{nulambda}
\end{align}
the metric Eq.\ \eqref{wmetcanPhi} reads: 
 \begin{align}
ds^2=&\,\frac  {(\nu^2+3)}4\left( dt^2+\frac{dr^2}{(\nu^2+3)(r-r_+)(r-r_-)}+ \big (2\,\nu\,r-\sqrt{r_+\,r_-\,(\nu^2+3)}\big)\,dt\,d\theta\right .\nonumber\\
&\left . +\frac r4\Big(3\,(\nu^2-1)\,r+(\nu^2+3)\,(r_++r_-)-4\,\nu\,\sqrt{r_+\,r_-\,(\nu^2+3)}\Big)\,d\theta^2\vphantom{\frac 4{(\nu^2+3)}} \right)\label{ds2APSSwithscale}\qquad 
 \end{align}
 which, up to the scale factor $\frac  {(\nu^2+3)}4=\frac 3{(3 - \lambda)}$, is the metric introduced in Ref.\cite{Anninos:2008fx}, Eq.\ (4.1).\\
This scale factor is not innocuous: it allows to bypass in Eq.\ \eqref{wmetcanPhi}, without changing the signature, the restriction imposed by the non-negative character  of $\nu^2 $ which, otherwise, restricts $\lambda$ to be less or equal to 3.

In case we put $r_+=r_-=r_\star$ in Eq. (\ref{ds2APSSwithscale}) we obtain the extremal version of the warped AdS$_3$ black hole considered in Ref. \cite{Anninos:2008fx}. This metric corresponds, before doing the identification leading to a black hole,  to a space like warping (and accordingly $\epsilon_R=+1$) with a double horizon which implies that we also have $\epsilon_L=0$. Thus, it corresponds to a configuration of type $\bs{\rmII}_a$. Indeed, performing  in Eq. (\ref{wmetcan}) the change of variables\footnote{ Here, the scale of $\kappa$ and $\beta$ are arbitrarily fixed, before the period of $\beta$ is fixed by the identification leading to the black hole.}:
\begin{align}
\kappa=(r-r_\star)\qquad,\qquad \alpha=\frac{\nu^2+3}{4\,\nu}\,\left(t+r_\star\,(\nu-\ft12 \sqrt{\nu^2+3})\,\theta\right)\qquad,\qquad \beta=\frac{\nu^2+3}{4 }\,\theta\qquad,
\end{align}
we obtain the rescaled metric (\ref{ds2APSSwithscale}) with the two horizons coinciding. The parameters $\lambda$ and $\nu$ are related as in \eqref{nulambda}.\newline
Assuming that $\theta$ is $2\,\pi$ periodic, the extremal metric is locally isometric to metric (\ref{wmetcanPhi}) with $\LL=(\nu^2+3)/4$ and 
$\LR=r_\star \,(\nu^2+3)(\nu-\ft 12\sqrt{\nu^2+3})/(4 \nu)$ and $\tau = (\nu^2 +3) t/(4 \nu)$, $\varphi = \theta$. We remind the reader that $\LL$ can be eliminated by a rescaling of $\kappa$ for the extremal black hole configuration. The extremal metric only depends on two parameters, $\lambda$ and $\LL$.

Lastly, we relate the special case $\LL = 0$ to the self-dual warped AdS$_3$ quotient which appears as the near-horizon geometry of extremal Kerr black holes.
The near horizon extremal Kerr geometry, [cf. Ref.\ \cite{Guica:2008mu}, eq. (3.7)]  is given by 
\begin{equation}
  ds^2 = 2 G J \Omega^2 (- (1+r^2) d \tau^2 + \frac{dr^2}{1+r^2} +d \theta^2 + {\tilde \Lambda}^2 (d \varphi^2 + r d \tau^2)
  ) 
\end{equation}
where $\tilde \Lambda = \frac{2 \sin\theta}{1+\cos^2\theta} \in [0, 2]$ and we set $G = J = \Omega = 1$ for simplicity. The metric induced on constant $\theta$ surfaces arises from deforming the  $\widetilde{\text{AdS}}_3$ space geometry into
\begin{equation}
   ds^2 = \frac{dr^2}{r^2+1}+2 r d\tau  d\varphi -d\tau^2+ d\varphi^2
\end{equation}
where $\varphi \in \mathbb{R}$ 
with the spacelike one-form 
\begin{equation}
\label{oneformNHEK}
    \sqrt{|1- {\tilde \Lambda}^2|} ( r  d \tau + d \varphi)\,
\end{equation}
and then identifying along the direction of $\partial_\varphi$, 
the vector dual to the one-form \eqref{oneformNHEK}.
Thus, we immediately infer that the metric induced on constant $\theta$ surfaces of the NHEK geometry corresponds to \eqref{wmetcanPhisd} with $\LL = 0$ and $\epsilon_R = 1$.

\subsection{From local to null coordinates}\label{loctonullcoord}

In this section we illustrate the various steps of the algorithm displayed in section \ref{subsec:decrypt} starting from the metric expression Eq.\ (4.1) of Ref.\cite{Anninos:2008fx}:
 \begin{align}
ds^2=& dt^2+\frac{dr^2}{(\nu^2+3)(r-r_+)(r-r_-)}+ \big (2\,\nu\,r-\sqrt{r_+\,r_-\,(\nu^2+3)}\big)\,dt\,d\theta \nonumber\\
&  +\frac r4\Big(3\,(\nu^2-1)\,r+(\nu^2+3)\,(r_++r_-)-4\,\nu\,\sqrt{r_+\,r_-\,(\nu^2+3)}\Big)\,d\theta^2\qquad. \label{ds2APSS}
 \end{align}
 \begin{enumerate}
     \item The four  Killing vector fields:\\
Two are obvious:
  \begin{align}
  &  \vec {\mathfrak k}_0= \partial_t\qquad,
  \qquad \vec {\mathfrak k_1}=\partial_\theta
  \end{align}
  and two more who require a little endeavour:
  \begin{align}
  &\vec {\mathfrak k}_2=\frac{\sqrt{r_+\,r_-\,(\nu^2+3)}(2\,r-r_+-r_-)-2\,\nu\left(r(r_++r_-)-2\,r_+\,r_-\right)}
  { \sqrt{\vert(r-r_+)(r-r_-)\vert}} \sinh\left(\ft{(\nu^2+3)(r_+-r_-)}4\,\theta\right)\,\partial_t\nonumber\\
  &\hphantom{\vec {\mathfrak k}_2=}-s_r\,(\nu^2+3)\,\sqrt{\vert(r-r_+)(r-r_-)\vert}\,(r_+-r_-)\cosh\left(\ft{(\nu^2+3)(r_+-r_-)}4\,\theta\right)\,\partial_r\nonumber\\
  &\hphantom{\vec {\mathfrak k}_2=} +\frac{2\,(2\,r-r_+-r_-)}{\sqrt{\vert(r-r_+)(r-r_-)\vert}}\sinh\left(\ft{(\nu^2+3)(r_+-r_-)}4\,\theta\right)\,\partial_\theta\\
  &\vec {\mathfrak k}_3= \frac{\sqrt{r_+\,r_-\,(\nu^2+3)}(2\,r-r_+-r_-)-2\,\nu\left(r(r_++r_-)-2\,r_+\,r_-\right)}
  { \sqrt{\vert(r-r_+)(r-r_-)\vert}} \cosh\left(\ft{(\nu^2+3)(r_+-r_-)}4\,\theta\right)\,\partial_t\nonumber\\
  &\hphantom{\vec {\mathfrak k}_2=}-s_r\,(\nu^2+3)\,\sqrt{\vert(r-r_+)(r-r_-)\vert}\,(r_+-r_-)\sinh\left(\ft{(\nu^2+3)(r_+-r_-)}4\,\theta\right)\,\partial_r\nonumber\\
  &\hphantom{\vec {\mathfrak k}_2=} +\frac{2\,(2\,r-r_+-r_-)}{\sqrt{\vert(r-r_+)(r-r_-)\vert}}\cosh\left(\ft{(\nu^2+3)(r_+-r_-)}4\,\theta\right)\,\partial_\theta
\end{align}
  with 
  \begin{align}
  s_r=\text{sign}[(r-r_+)(r-r_-)] \quad.\newline
  \end{align}
  \item  It's also obvious that $\vec {\mathfrak k}_0$ commutes with all these Killing vectors. Its musical  dual, that reads 
   \begin{align}
   \oneform \xi=dt+(\nu\,r-\ft12\,\sqrt{(\nu+3)\,r_+\,r_-})\,d\theta\label{OneFk1}\qquad ,
   \end{align}
   is proportional to the one-form   $\oneform \rho_{1}$ along which the squashing is done. 
   \item An elementary computation of the curvature shows that a (locally) AdS$_3$ geometry is obtained by subtracting from the metric, Eq. (\ref{ds2APSS}), the Kronecker product $\ft{3\,(\nu^2-1)}{4\,\nu^2}\,\oneform{\xi}\otimes \oneform \xi$. The resulting constant curvature metrics has Gauss' curvature equal to:
   \begin{equation}
       R=-\ft32\,(\nu^2+3)\qquad.
   \end{equation}
   \item Thus, to pursue we have to rescale the metric  by the factor $(\nu^2+3)/4$ to obtain a constant curvature metric whose Gauss' curvature is equal to $-6$:
    \begin{align}
& ds_{AdS_3}^2=  \frac{(\nu^2+3)^2}{16\,\nu^2}\left(d\vphantom{\frac{(\nu^2+3)^2}{16\,\nu^2}}t^2+(2\,\nu\,r-\sqrt{(\nu^2+3)\,r_+\,r_-})\,dt\,d\theta\right.\nonumber \\
& \hphantom{ds_{AdS_3}^2=} +\left.\vphantom{\frac{(\nu^2+3)^2}{16\,\nu^2}}\left(r\,\nu\,\big(\sqrt{(\nu^2+3)\,r_+\,r_-}
 -\nu(r_++r_-)\big)+\ft34\,(\nu^2-1)\,r_+\,r_-
 \right)\,d\theta^2\right)\nonumber \\
 & \hphantom{ds_{AdS_3}^2=}+\frac 1{4\,(r-r_+)(r-r_-)}dr^2
 \end{align}
\item
  This metric possesses two more Killing vectors:
  \begin{align}
      &\vec {\mathfrak k}_4=+\frac{(\sqrt{(\nu^2+3)\,r_+\,r_-}-2\,\nu\,r)}{\sqrt{\vert(r-r_+)(r-r_-)\vert}}\sinh\Big(\ft{(\nu^2+3)} {4\,\nu}\big(2\,t+{\nu(r_++r_-)-\sqrt{(\nu^2+3)\,r_+\,r_-}} \,\theta\big)\Big)\partial_t\nonumber\\
      &\hphantom{\vec {\mathfrak k}_4=}+s_r\,(\nu^2+3)\,\sqrt{\vert (r-r_+)(r-r_-)\vert}\,\cosh\Big(\ft{(\nu^2+3)} {4\,\nu}\big(2\,t+{\nu(r_++r_-)-\sqrt{(\nu^2+3)\,r_+\,r_-}} \,\theta\big)\Big)\partial_r\nonumber\\
      &\hphantom{\vec {\mathfrak k}_4=}+\frac 2{\sqrt{\vert (r-r_+)(r-r_-)\vert}}\,\sinh\Big(\ft{(\nu^2+3)} {4\,\nu}\big(2\,t+{\nu(r_++r_-)-\sqrt{(\nu^2+3)\,r_+\,r_-}} \,\theta\big)\Big)\partial_\theta\\
       &\vec {\mathfrak k}_5=-\frac{(\sqrt{(\nu^2+3)\,r_+\,r_-}-2\,\nu\,r)}{\sqrt{\vert(r-r_+)(r-r_-)\vert}}\cosh\Big(\ft{(\nu^2+3)} {4\,\nu}\big(2\,t+{\nu(r_++r_-)-\sqrt{(\nu^2+3)\,r_+\,r_-}} \,\theta\big)\Big)\partial_t\nonumber\\
      &\hphantom{\vec {\mathfrak k}_4=}-s_r\,(\nu^2+3)\,\sqrt{\vert (r-r_+)(r-r_-)\vert}\,\sinh\Big(\ft{(\nu^2+3)} {4\,\nu}\big(2\,t+{\nu(r_++r_-)-\sqrt{(\nu^2+3)\,r_+\,r_-}} \,\theta\big)\Big)\partial_r\nonumber\\
      &\hphantom{\vec {\mathfrak k}_4=}-\frac 2{\sqrt{\vert (r-r_+)(r-r_-)\vert}}\,\cosh\Big(\ft{(\nu^2+3)} {4\,\nu}\big(2\,t+{\nu(r_++r_-)-\sqrt{(\nu^2+3)\,r_+\,r_-}} \,\theta\big)\Big)\partial_\theta
  \end{align}
  \item The AdS$_3$ norms of these vectors only are locally constant (they depend on the coordinate patch). Various sets of (local) right and left invariant normed Killing vectors (in accordance with Eqs (\ref{comRL}, \ref{norm}) are given by:\\
  \begin{subequations}
  \begin{align}
  &\bullet\ \text{If } r\not\in [r_-,\,r_+]\nonumber\\
 &\vec r_1 =\frac{4\,\nu}{(\nu^2+3)}\,\vec {\mathfrak k}_0\qquad,\qquad &&\vec l_1=\frac {\epsilon_{l,1}\,}{(\nu^2+3)(r_+-r_-)}\big(8\,\vec {\mathfrak k}_1+ 4(\sqrt{(\nu^2+3)\,r_+\,r_-}-\nu\,(r_++r_-))\,\vec{\mathfrak k}_0\big)\label{r1l1rprm} \\
 &\vec r_2=\frac {2\,\epsilon_r}{(\nu^2+3)}\vec{\mathfrak k}_4\qquad,\qquad && \vec l_2= \frac {2\,{\epsilon_{l,2}}}{(r_+-r_-)(\nu^2+3)}\,\vec {\mathfrak k}_2 \\
  &\vec r_3= \frac {2\,\epsilon_r}{(\nu^2+3)}\vec{\mathfrak k}_5\qquad,\qquad && \vec l_3=\frac {2\,\epsilon_{l,1}\,\epsilon_{l,2}}{(r_+-r_-) (\nu^2+3)}\,\vec {\mathfrak k}_3\\
 &\bullet\ \text{If } r\in [r_-,\,r_+]\nonumber\\
 &\vec r_1 =\frac{4\,\nu}{(\nu^2+3)}\,\vec {\mathfrak k}_0\qquad,\qquad &&\vec l_1=\frac {\epsilon_{l,1}}{(\nu^2+3)(r_+-r_-)}\big(8\,\vec {\mathfrak k}_1+ 4(\sqrt{(\nu^2+3)\,r_+\,r_-}-\nu\,(r_++r_-))\,\vec{\mathfrak k}_0\big) \\
  &\vec r_2=\epsilon_r\,\frac 2{(\nu^2+3)}\vec{\mathfrak k}_5\qquad,\qquad && \vec l_2=\frac {2\,\epsilon_{l,2}} {(r_+-r_-) (\nu^2+3)}\,\vec {\mathfrak k}_3 \\
  &\vec r_3=\epsilon_r\,\frac 2{(\nu^2+3)}\vec{\mathfrak k}_4\qquad,\qquad && \vec l_3=\frac {2\,\epsilon_{l,1}\,\epsilon_{l,2}} {(r_+-r_-)(\nu^2+3)}\,\vec {\mathfrak k}_2\label{r3l3H}
 \end{align}
 \end{subequations}
 \begin{align}
  & \epsilon_{r}=\pm 1,\ \epsilon_{l,1}=\pm 1,\ \epsilon_{l,2}=\pm 1\label{rrel}
 \end{align}
   \newline

The null coordinate surfaces: $U_\pm=0$, $V_\pm=0$,  define the horizons.
\begin{figure}
    \centering
    \begin{subfigure}[b]{0.1\textwidth}
        \includegraphics[width=\textwidth]{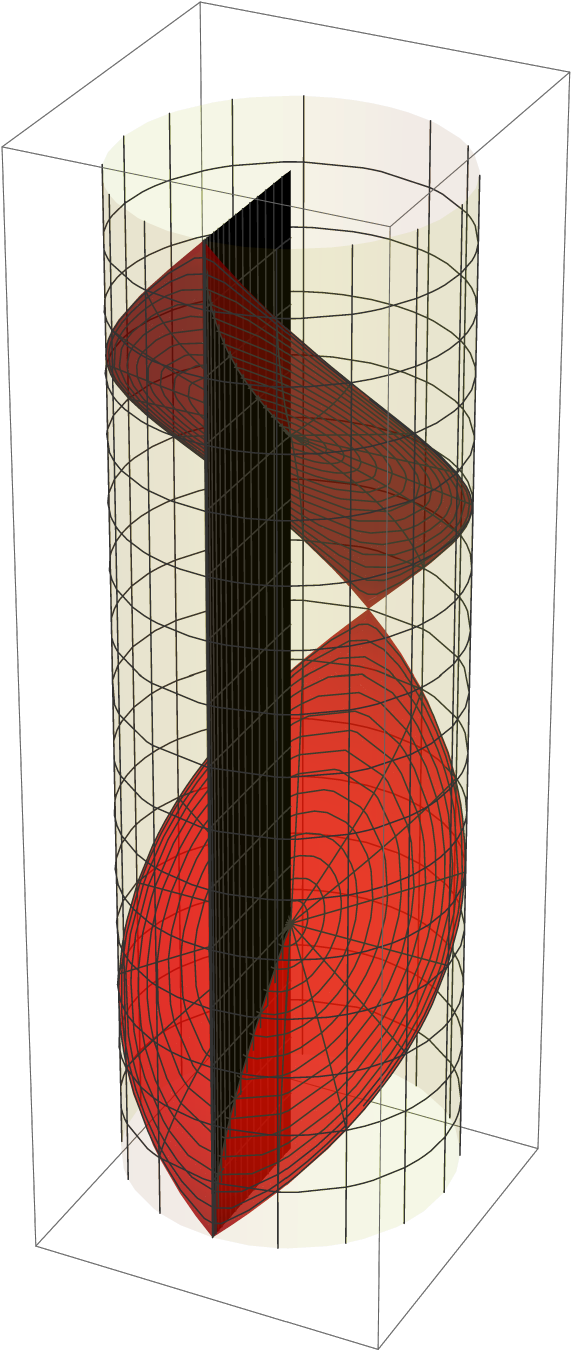}
    \end{subfigure}
    \begin{subfigure}[b]{0.1\textwidth}
        \includegraphics[width=\textwidth]{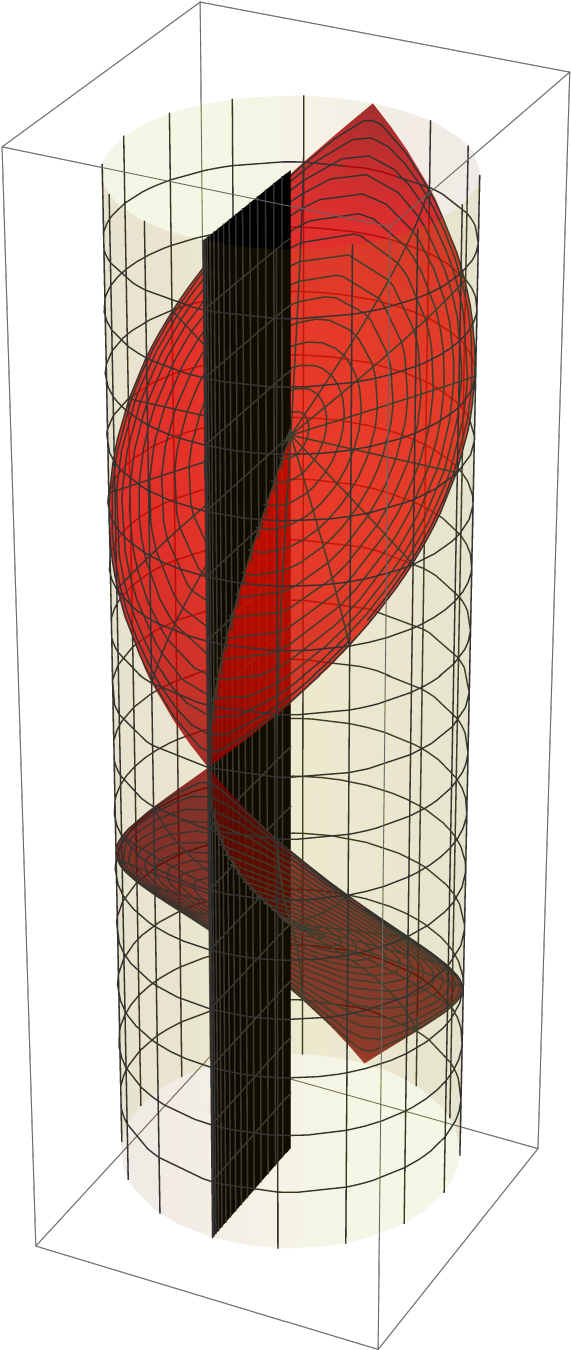}
    \end{subfigure}
    \begin{subfigure}[b]{0.1\textwidth}
        \includegraphics[width=\textwidth]{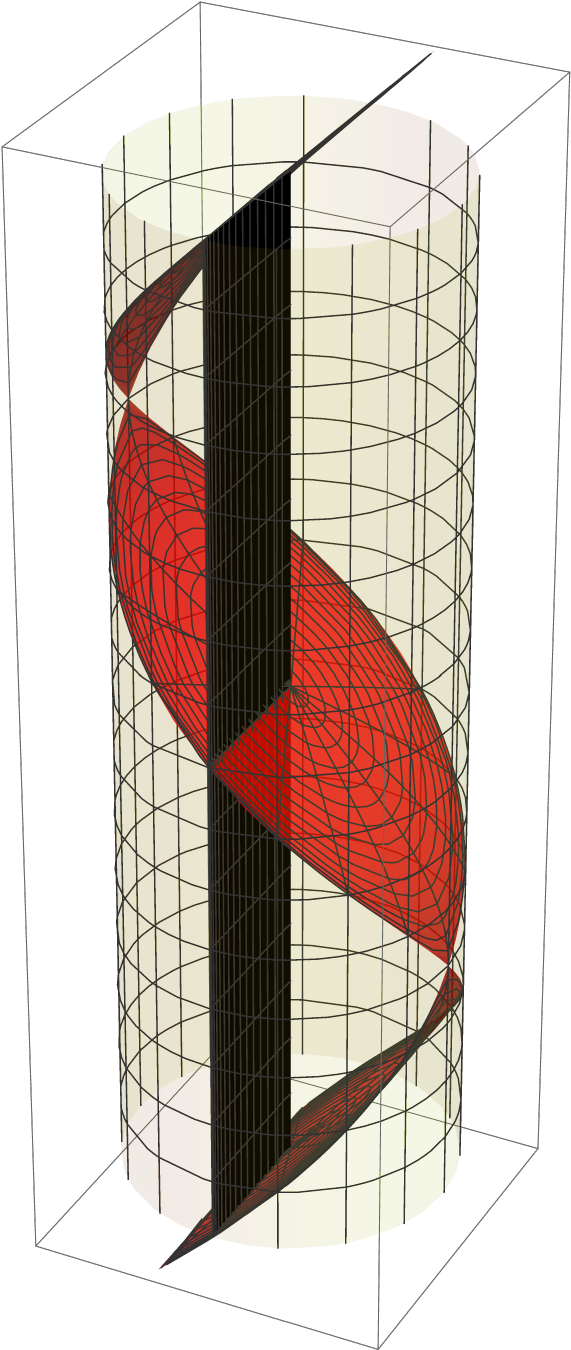}
    \end{subfigure} \begin{subfigure}[b]{0.1\textwidth}
        \includegraphics[width=\textwidth]{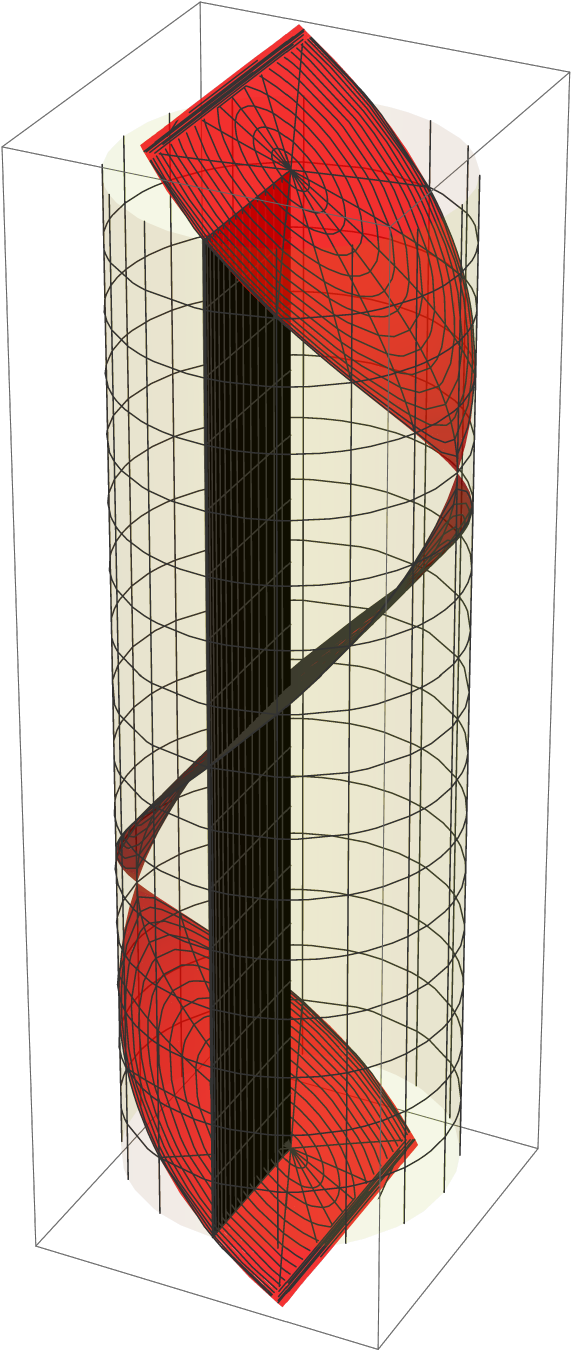}
    \end{subfigure} 
    \caption{The horizons, $U_+=0$, $U_-=0$, $V_+=0$, $V_-=0$, embedded in the  cylindrical representation of AdS$_3$; see Appendix \ref{CovSLR}. The half-planes depicted correspond to the coordinate surface $\upvartheta=0$.}\label{figH}
\end{figure}
They split the hyperboloid,  Eq.\ (\ref{AdS30hyp}), into twelve open subsets defined by the signs of $U_\pm$ and $V_\pm$. The closures of four of these domains, those where an even number of the coordinates $\{U_+,\,U_-,\,V_+,\,V_-\}$  are positive, are compact. The remaining eight\footnote{There are {\it a priori} 16 possible repartitions of signs between the four entries of the group element $\boldsymbol{z}$, Eq.\ (\ref{SL2Rmat}), but the four for which $\text{sign}[U_+\,U_-]\leq 0$ and $\text{sign}[V_+\,V_-]\leq 0$ doesn't belong to $SL(2,\mathbb R)$.} are non compact but extend to infinity.

Solving the equations, Eqs.\ (\ref{Upm2})--(\ref{UVpm}), using the previous expressions of the Killing vector fields, Eqs (\ref{r1l1rprm}--\ref{r3l3H}) with the various values of the coefficients $\epsilon_r$, $\epsilon_{l,1}$ and $\epsilon_{l,2}$ we obtain with:
\begin{subequations}
\begin{align}
& \Phi_-:=\ft{(\nu^2+3)}{4\,\nu}(t+\theta(\nu\,r_- -\ft{1}2\sqrt{(\nu^3+3)\,r_+\,r_-}))\label{Phip}\\
& \Phi_+:=\ft{(\nu^2+3)}{4\,\nu}(t+\theta(\nu\,r_+-\ft{1}2\sqrt{(\nu^3+3)\,r_+\,r_-}))\label{Phim} \\
    &U_\pm=\frac{\mathcal U_\pm}{\sqrt{r_+-r_-}} \qquad,\qquad 
    V_\pm=\frac{\mathcal V_\pm}{\sqrt{r_+-r_-}}\nonumber
\end{align}
\end{subequations}
the following solutions of Eqs (\ref{Upm2}--\ref{UVpm}):\newline
For $r$ not between the horizons:
\begin{table}[h]
    \centering
    \begin{tabular}{|ccc|c|cccc|}
    \hline
        $\epsilon_r$&$\epsilon_{l,1}$&$\epsilon_{l,2}$& $r\not\in[r_-,\,r_+]$&$\mathcal{U_+}$&$\mathcal{U_-}$&$\mathcal{V_+}$&$\mathcal{V_-}$ \\
           \hline
         1&1&1 & $r_->r$&$\pm\sqrt{r_- -r}\,e^{\Phi_+}$&$\mp\sqrt{r_- -r}\,e^{-\Phi_+}$&$\pm\sqrt{r_+ -r}\,e^{\Phi_-}$&$\pm\sqrt{r_+ -r}\,e^{-\Phi_-}$\\
         -1&-1&-1 & $r>r_+$&$\pm\sqrt{r-r_+ }\,e^{\Phi_-}$&$\mp\sqrt{r-r_+ }\,e^{-\Phi_-}$&$\pm\sqrt{r-r_-}\,e^{\Phi_+}$&$\pm\sqrt{r-r_- }\,e^{-\Phi_+}$\\
             1&-1&-1 & $r_->r$&$\pm\sqrt{r_+ -r}\,e^{\Phi_-}$&$\pm\sqrt{r_+ -r}\,e^{-\Phi_-}$&$\mp\sqrt{r_- -r}\,e^{\Phi_+}$&$\pm\sqrt{r_- -r}\,e^{-\Phi_+}$\\
         -1&1&1& $r>r_+$&$\pm\sqrt{r-r_- }\,e^{\Phi_+}$&$\pm\sqrt{r-r_- }\,e^{-\Phi_+}$&$\mp\sqrt{r-r_+}\,e^{\Phi_-}$&$\pm\sqrt{r-r_+ }\,e^{-\Phi_-}$\\
             -1&1&-1 & $r_->r$&$\pm\sqrt{r_- -r}\,e^{\Phi_+}$&$\mp\sqrt{r_- -r }\,e^{-\Phi_+}$&$\mp\sqrt{r_+-r}\,e^{\Phi_-}$&$\mp\sqrt{r_+-r }\,e^{-\Phi_-}$\\
          1&-1&1 & $r>r_+$&$\pm\sqrt{r_+ -r}\,e^{\Phi_-}$&$\pm\sqrt{r_+ -r}\,e^{-\Phi_-}$&$\mp\sqrt{r_- -r}\,e^{\Phi_+}$&$\pm\sqrt{r_- -r}\,e^{-\Phi_+}$\\
            -1&-1&1 & $r_->r$&$\pm\sqrt{r_+ -r}\,e^{\Phi_-}$&$\pm\sqrt{r_+ -r}\,e^{-\Phi_-}$&$\pm\sqrt{r_- -r}\,e^{\Phi_+}$&$\mp\sqrt{r_- -r}\,e^{-\Phi_+}$\\ 
           1&1&-1 & $r>r_+$&$\pm\sqrt{r-r_- }\,e^{\Phi_+}$&$\pm\sqrt{r-r_- }\,e^{-\Phi_+}$&$\pm\sqrt{r-r_+}\,e^{\Phi_-}$&$\mp\sqrt{r-r_+ }\,e^{-\Phi_-}$\\
          \hline
    \end{tabular}
    \caption{Local parametrisations of the non-compact regions of the (warped) AdS$_3$ space(s), those extending from horizon surfaces up to infinity.}
    \label{tab:paramNonComp}
\end{table}
 and for $r$ between the horizons:
\begin{table}[h]
    \centering
    \begin{tabular}{|ccc|cccc|}
    \hline
        $\epsilon_r$&$\epsilon_{l,1}$&$\epsilon_{l,2}$& $\mathcal{U_+}$&$\mathcal{U_-}$&$\mathcal{V_+}$&$\mathcal{V_-}$ \\
        \hline
          -1&1&1 & $\pm\sqrt{r -r_-}\,e^{\Phi_+}$&$\pm\sqrt{r-r_-}\,e^{-\Phi_+}$&$\mp\sqrt{r_+ -r}\,e^{\Phi_-}$&$\mp\sqrt{r_+ -r}\,e^{-\Phi_-}$\\
            1&-1&1 & $\pm\sqrt{r_+ -r}\,e^{\Phi_-}$&$\pm\sqrt{r_+ -r}\,e^{-\Phi_-}$&$\mp\sqrt{r-r_- }\,e^{\Phi_+}$&$\mp\sqrt{r-r_- }\,e^{-\Phi_+}$\\
          1&1&-1 &  $\pm\sqrt{r-r_- }\,e^{\Phi_+}$&$\pm\sqrt{r-r_- }\,e^{-\Phi_+}$&$\pm\sqrt{r_+-r}\,e^{\Phi_-}$&$\pm\sqrt{r_+-r }\,e^{-\Phi_-}$\\
          -1&-1&-1 & $\pm\sqrt{ r_+ - r}\,e^{\Phi_-}$&$\pm\sqrt{r_+ - r }\,e^{-\Phi_-}$&$\pm\sqrt{r-r_-}\,e^{\Phi_+}$&$\pm\sqrt{r-r_- }\,e^{-\Phi_+}$\\
          \hline
    \end{tabular}
    \caption{Local parametrisations of the compact regions of the (warped) AdS$_3$ space(s), those bounded by horizon surfaces.}
    \label{tab:paramComp}
\end{table}
As we see the various ranges of positive or negative values of $U_\pm$ and $V_\pm$ are all parametrised in two distinct ways. This reflects the isometry between a domain corresponding to a range of values of $t,\,r,\,\theta$ and the one where $r$ is replaced by $r'=r_++r_--r$ (the correspondence between the values of $t$ and $\theta$ being given by Eqs.\ (\ref{Phip}), (\ref{Phim}), so that $\Phi_\pm(t,\,\theta)=\Phi_\mp(t',\,\theta')$. It also allows to vary the $r$ coordinate continuously across all the space when crossing horizons.
\end{enumerate}

\begin{figure}[h]
    \centering
    \begin{subfigure}[b]{0.2\textwidth}
        \includegraphics[width=\textwidth]{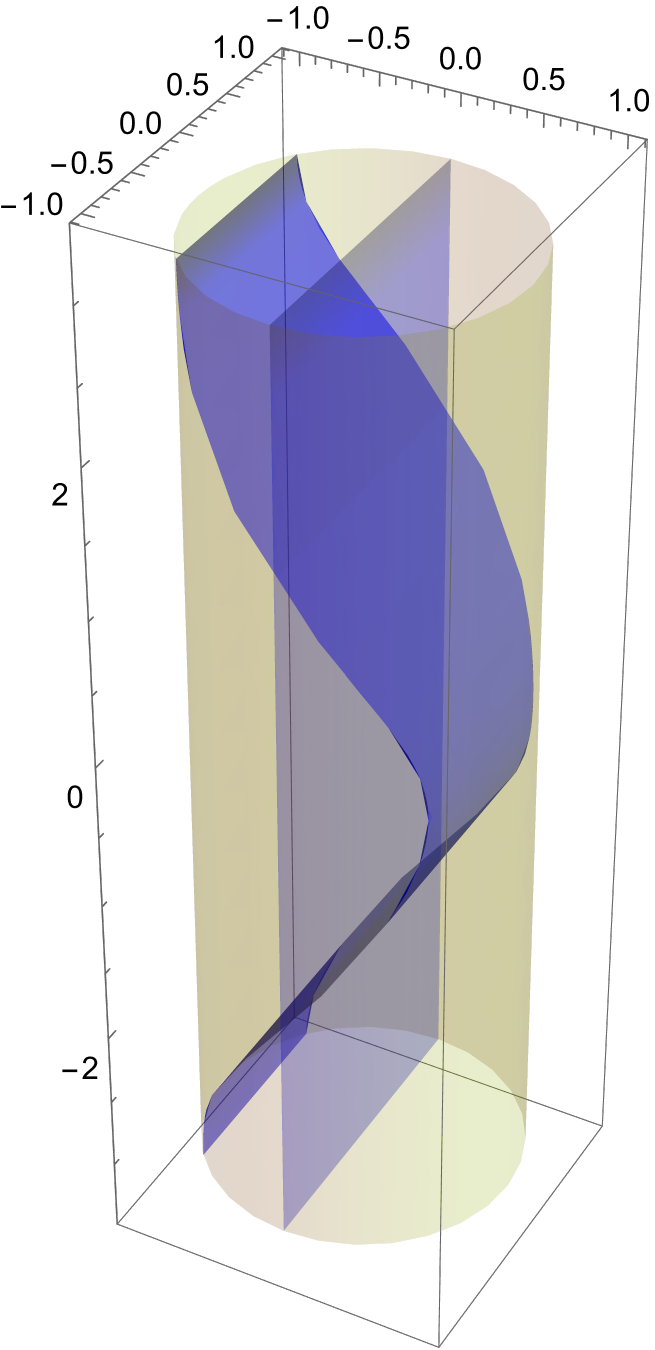}
    \end{subfigure}
    \begin{subfigure}[b]{0.2\textwidth}
        \includegraphics[width=\textwidth]{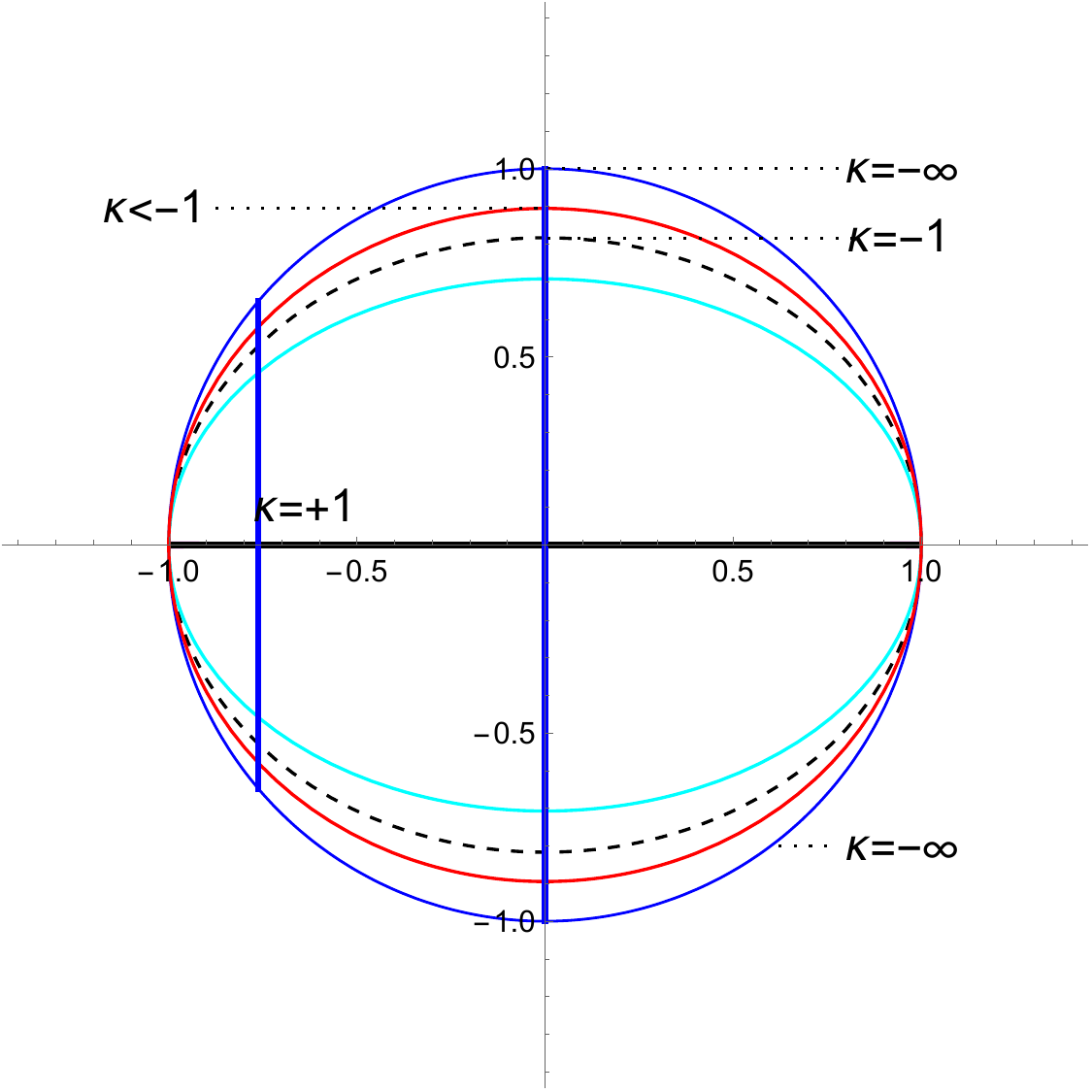}
    \end{subfigure}
    \begin{subfigure}[b]{0.2\textwidth}
        \includegraphics[width=\textwidth]{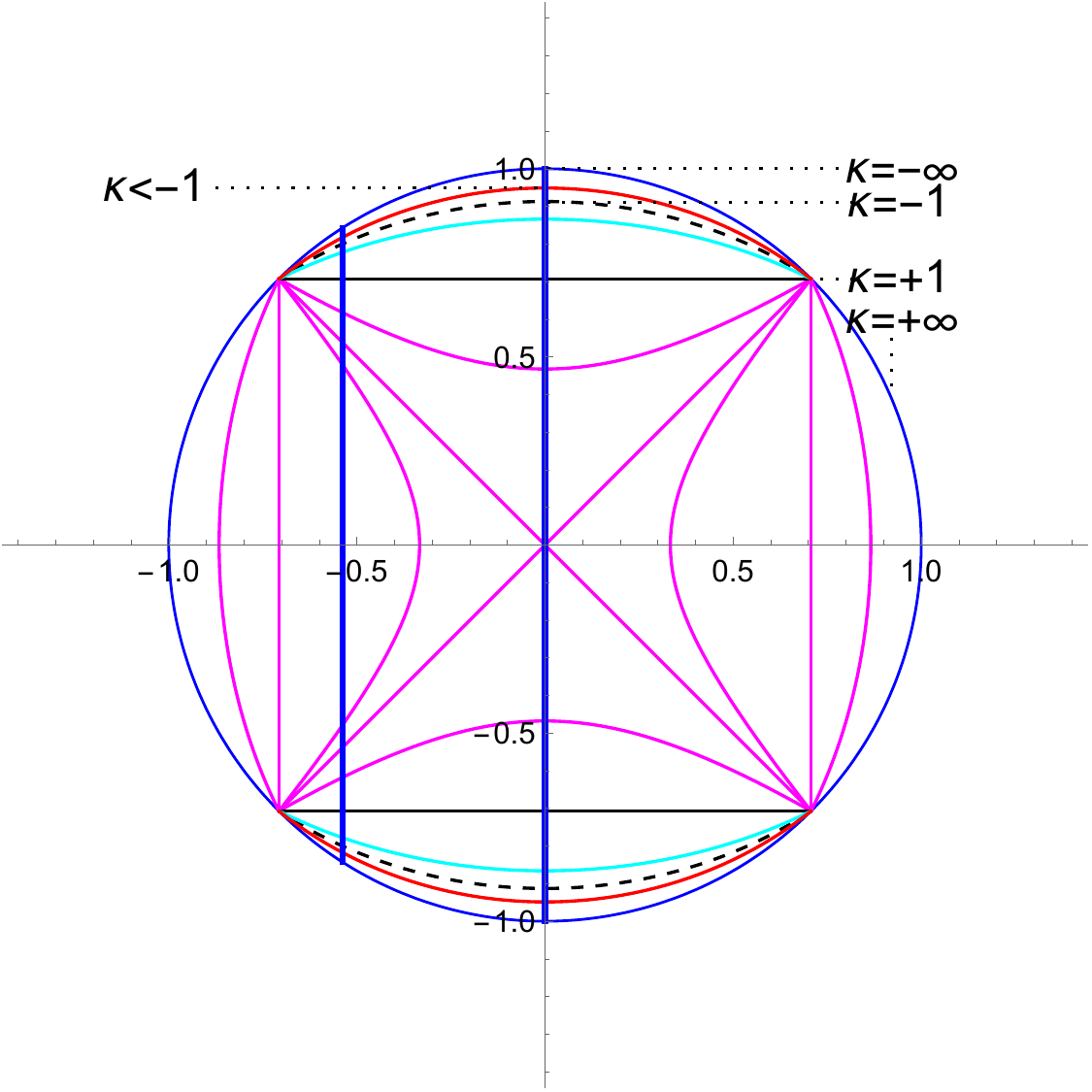}
    \end{subfigure} \begin{subfigure}[b]{0.2\textwidth}
        \includegraphics[width=\textwidth]{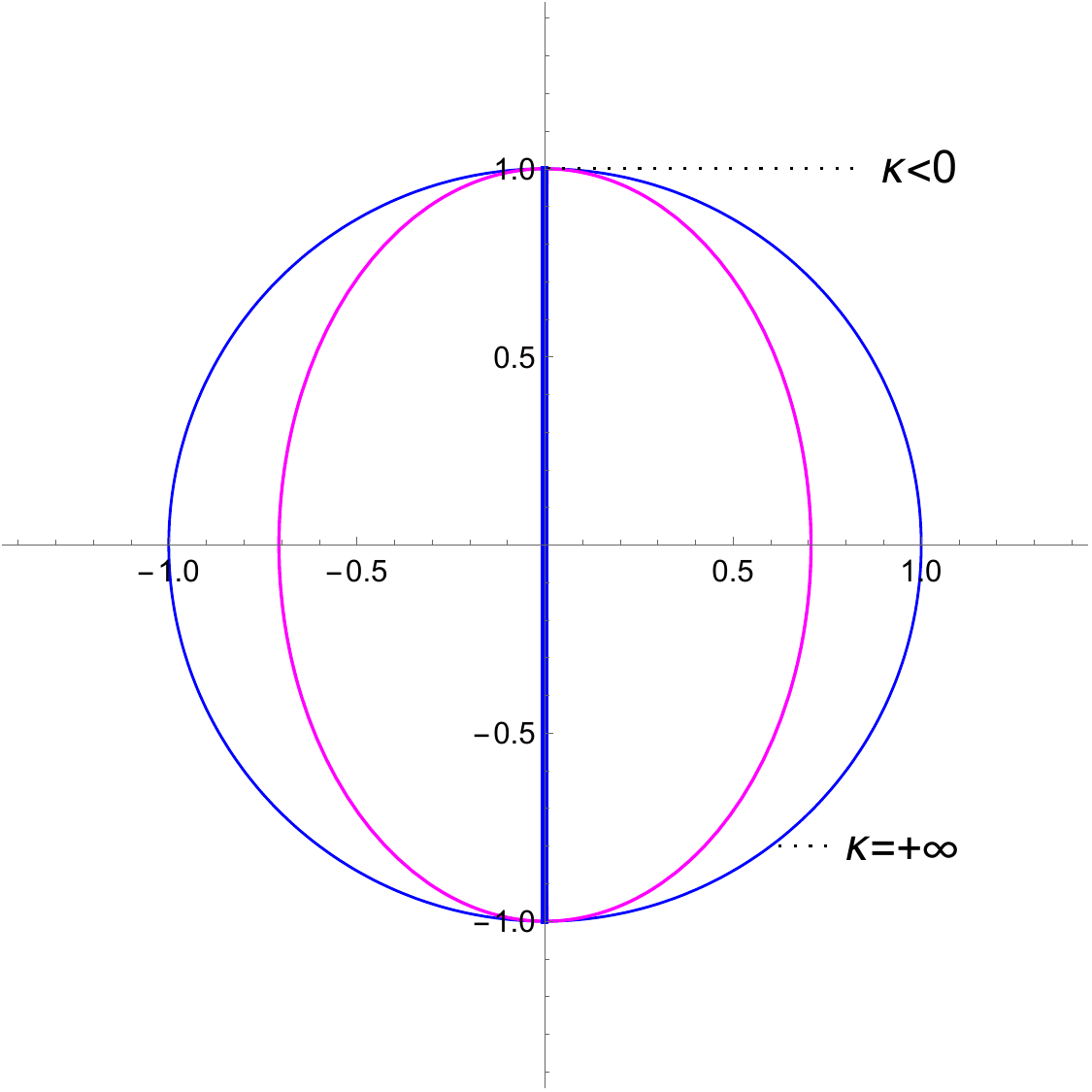}
    \end{subfigure} 
    \caption{The first picture displays the median plane ($\upvartheta=0,\, \upvartheta=\pi$) of $\widetilde{SL(2,\,\mathbb R)}$ and its image under $Exp[\vec\partial]$. The next three pictures represent $\uptau=0,\,\pi/4,\,\pi/2$ sections of the first ones, on which the constant $\kappa$ sections considered in Figs. \ref{figsectvcst} are represented.  The evolution with respect to $\uptau$ of a fundamental domain also is drawn. As shown by Eq. \eqref{sectau2} this fundamental domain can be described as the region between a diameter and a parallel chord of the circular section of the cylinder representation of AdS$_3$. The collapse of this domain,  depicted for $\uptau=\pi/2$, reflects the occurrence of a topological singularity. These singularities are also seen on the first left figure where at $\uptau=\pm\pi/2$ the median plane coincides with its image.}\label{Jnul}
    \label{figlast}
\end{figure}

To make an end we display hereafter, see Fig. \ref{figlast}, the image of the median plane of $\widetilde{SL(2,\,\mathbb R)}$ 
and one of its images under the identification $\boldsymbol{z}\mapsto Exp[\vec\partial]\,\boldsymbol{z}$ when $\LR=\LL$: $U_\pm\mapsto e^{\pm\LR}\,U_\pm,\, V_\pm\mapsto V_\pm$, defining a fundamental domain corresponding to non-rotating warped AdS$_3$ black hole structure. We also provide some $\uptau$-constant sections of  them, analogs to those considered in Fig.\ \ref{figsectvcst}. Their equations, once more, is particularly simple:
\begin{equation}
x=\frac{\cos[\uptau]\,\sinh[\LR]}{\sqrt{\cos^2[\uptau]\,\sinh^2[\LR]+\sin^2[\uptau]}}\label{sectau2}\qquad.
\end{equation}
%\begin{mcbibliography}
\bibliographystyle{utphys}
\bibliography{MyBiblio.bib}

\providecommand{\href}[2]{#2}\begingroup\raggedright\begin{thebibliography}{10}

\bibitem{Guica:2008mu}
M.~Guica, T.~Hartman, W.~Song, and A.~Strominger, ``{The Kerr/CFT Correspondence},'' \href{http://dx.doi.org/10.1103/PhysRevD.80.124008}{{\em Phys.Rev.} {\bfseries D80} (2009) 124008},
\href{http://arxiv.org/abs/0809.4266}{{\ttfamily arXiv:0809.4266 [hep-th]}}.
%%CITATION = ARXIV:0809.4266;%%.

\bibitem{Anninos:2008fx}
D.~Anninos, W.~Li, M.~Padi, W.~Song, and A.~Strominger, ``{Warped AdS(3) Black Holes},'' \href{http://dx.doi.org/10.1088/1126-6708/2009/03/130}{{\em JHEP} {\bfseries 0903} (2009) 130},
\href{http://arxiv.org/abs/0807.3040}{{\ttfamily arXiv:0807.3040 [hep-th]}}.
%%CITATION = ARXIV:0807.3040;%%.

\bibitem{horowitz1995new}
G.~T. Horowitz and A.~A. Tseytlin, ``{A New class of exact solutions in string theory},'' \href{http://dx.doi.org/10.1103/PhysRevD.51.2896}{{\em Phys. Rev. D} {\bfseries 51} (1995) 2896--2917}, \href{http://arxiv.org/abs/hep-th/9409021}{{\ttfamily arXiv:hep-th/9409021}}.

\bibitem{kiritsis1995infrared}
E.~Kiritsis and C.~Kounnas, ``{Infrared regularization of superstring theory and the one loop calculation of coupling constants},'' \href{http://dx.doi.org/10.1016/0550-3213(95)00156-M}{{\em Nucl. Phys. B} {\bfseries 442} (1995) 472--493}, \href{http://arxiv.org/abs/hep-th/9501020}{{\ttfamily arXiv:hep-th/9501020}}.

\bibitem{Rooman:1998xf}
M.~Rooman and P.~Spindel, ``{G\"odel metric as a squashed anti-de Sitter geometry},'' \href{http://dx.doi.org/10.1088/0264-9381/15/10/024}{{\em Class.Quant.Grav.} {\bfseries 15} (1998) 3241--3249},
\href{http://arxiv.org/abs/gr-qc/9804027}{{\ttfamily arXiv:gr-qc/9804027 [gr-qc]}}.
%%CITATION = GR-QC/9804027;%%.

\bibitem{Israel:2003cx}
D.~Israel, ``{Quantization of heterotic strings in a G\"odel / anti-de Sitter space-time and chronology protection},'' \href{http://dx.doi.org/10.1088/1126-6708/2004/01/042}{{\em JHEP} {\bfseries 0401} (2004) 042},
\href{http://arxiv.org/abs/hep-th/0310158}{{\ttfamily arXiv:hep-th/0310158 [hep-th]}}.
%%CITATION = HEP-TH/0310158;%%.

\bibitem{Israel:2004vv}
D.~Israel, C.~Kounnas, D.~Orlando, and P.~M. Petropoulos, ``{Electric/magnetic deformations of S**3 and AdS(3), and geometric cosets},'' \href{http://dx.doi.org/10.1002/prop.200410190}{{\em Fortsch.Phys.} {\bfseries 53} (2005) 73--104},
\href{http://arxiv.org/abs/hep-th/0405213}{{\ttfamily arXiv:hep-th/0405213 [hep-th]}}.
%%CITATION = HEP-TH/0405213;%%.

\bibitem{Detournay:2015ysa}
S.~Detournay and C.~Zwikel, ``{Phase transitions in warped AdS$_{3}$ gravity},'' \href{http://dx.doi.org/10.1007/JHEP05(2015)074}{{\em JHEP} {\bfseries 05} (2015) 074}, \href{http://arxiv.org/abs/1504.00827}{{\ttfamily arXiv:1504.00827 [hep-th]}}.

\bibitem{Bengtsson:2005zj}
I.~Bengtsson and P.~Sandin, ``{Anti de Sitter space, squashed and stretched},'' \href{http://dx.doi.org/10.1088/0264-9381/23/3/022}{{\em Class. Quant. Grav.} {\bfseries 23} (2006) 971--986}, \href{http://arxiv.org/abs/gr-qc/0509076}{{\ttfamily arXiv:gr-qc/0509076}}.

\bibitem{Banados:1992gq}
M.~Banados, M.~Henneaux, C.~Teitelboim, and J.~Zanelli, ``{Geometry of the (2+1) black hole},'' \href{http://dx.doi.org/10.1103/PhysRevD.48.1506}{{\em Phys.Rev.} {\bfseries D48} (1993) 1506--1525},
\href{http://arxiv.org/abs/gr-qc/9302012}{{\ttfamily arXiv:gr-qc/9302012 [gr-qc]}}.
%%CITATION = GR-QC/9302012;%%.

\bibitem{Nutku:1993eb}
Y.~Nutku, ``{Exact solutions of topologically massive gravity with a cosmological constant},'' \href{http://dx.doi.org/10.1088/0264-9381/10/12/022}{{\em Class. Quant. Grav.} {\bfseries 10} (1993) 2657--2661}.

\bibitem{Gurses:1994bjn}
M.~G\"urses, ``{Perfect Fluid Sources in 2+1 Dimensions},'' \href{http://dx.doi.org/10.1088/0264-9381/11/10/017}{{\em Class. Quant. Grav.} {\bfseries 11} no.~10, (1994) 2585}.

\bibitem{Bouchareb:2007yx}
A.~Bouchareb and G.~Clement, ``{Black hole mass and angular momentum in topologically massive gravity},'' \href{http://dx.doi.org/10.1088/0264-9381/24/22/018}{{\em Class.Quant.Grav.} {\bfseries 24} (2007) 5581--5594},
\href{http://arxiv.org/abs/0706.0263}{{\ttfamily arXiv:0706.0263 [gr-qc]}}.
%%CITATION = ARXIV:0706.0263;%%.

\bibitem{Anninos:2010pm}
D.~Anninos, G.~Compere, S.~de~Buyl, S.~Detournay, and M.~Guica, ``{The Curious Case of Null Warped Space},'' \href{http://dx.doi.org/10.1007/JHEP11(2010)119}{{\em JHEP} {\bfseries 1011} (2010) 119},
\href{http://arxiv.org/abs/1005.4072}{{\ttfamily arXiv:1005.4072 [hep-th]}}.
%%CITATION = ARXIV:1005.4072;%%.

\bibitem{Chrusciel:2012gz}
P.~T. Chrusciel, C.~R. Olz, and S.~J. Szybka, ``{Space-time diagrammatics},'' \href{http://dx.doi.org/10.1103/PhysRevD.86.124041}{{\em Phys. Rev. D} {\bfseries 86} (2012) 124041}, \href{http://arxiv.org/abs/1211.1718}{{\ttfamily arXiv:1211.1718 [gr-qc]}}.

\bibitem{Jugeau:2010nq}
F.~Jugeau, G.~Moutsopoulos, and P.~Ritter, ``{From accelerating and Poincare coordinates to black holes in spacelike warped AdS$_3$, and back},'' \href{http://dx.doi.org/10.1088/0264-9381/28/3/035001}{{\em Class. Quant. Grav.} {\bfseries 28} (2011) 035001}, \href{http://arxiv.org/abs/1007.1961}{{\ttfamily arXiv:1007.1961 [hep-th]}}.

\bibitem{DKLSW2024}
S.~Detournay, S.~Kanuri, A.~Lupsasca, P.~Spindel, Q.~Vandermiers, and R.~Wutte, ``{Photon Rings and Quasi-Normal Modes for generic Warped AdS$_3$ black holes ({\it Work in progress})},''.

\bibitem{gilmore2008lie}
R.~Gilmore, {\em Lie groups, physics, and geometry: an introduction for physicists, engineers and chemists}.
\newblock Cambridge University Press, 2008.

\bibitem{gilmore1974lie}
R.~Gilmore, {\em Lie groups, Lie algebras, and some of their applications}.
\newblock John Wiley and Sons, Inc., New York, 1974.

\bibitem{rawnsley2012universal}
J.~Rawnsley, ``On the universal covering group of the real symplectic group,'' {\em Journal of Geometry and Physics} {\bfseries 62} no.~10, (2012) 2044--2058.

\bibitem{garrett}
P.~B. Garrett, ``Sporadic isogenies to orthogonal groups,''
\newblock 2013.
\newblock \url{https://api.semanticscholar.org/CorpusID:14897263}.

\bibitem{Bieliavsky:2002ki}
P.~Bieliavsky, M.~Rooman, and P.~Spindel, ``{Regular Poisson structures on massive nonrotating BTZ black holes},'' \href{http://dx.doi.org/10.1016/S0550-3213(02)00867-2}{{\em Nucl. Phys.} {\bfseries B645} (2002) 349--364},
\href{http://arxiv.org/abs/hep-th/0206189}{{\ttfamily arXiv:hep-th/0206189 [hep-th]}}.
%%CITATION = HEP-TH/0206189;%%.

\bibitem{Spindel:1986ic}
P.~Spindel, ``{Gravity before supergravity},'' in {\em {1984 NATO ASI on Supersymmetry}}, pp.~455--533.
\newblock 9, 1986.

\bibitem{Helgason}
S.~Helgason, {\em Differential Geometry, Lie Groups, and Symmetric Spaces}, vol.~34 of {\em Graduate Studies in Mathematics}.
\newblock American Mathematical Society, Providence, U.S.A., 2001.

\bibitem{krasnov2000holography}
K.~Krasnov, ``{Holography and Riemann surfaces},'' \href{http://dx.doi.org/10.4310/ATMP.2000.v4.n4.a5}{{\em Adv. Theor. Math. Phys.} {\bfseries 4} (2000) 929--979}, \href{http://arxiv.org/abs/hep-th/0005106}{{\ttfamily arXiv:hep-th/0005106}}.

\bibitem{krasnov2002analytic}
K.~Krasnov, ``{Analytic continuation for asymptotically AdS 3-D gravity},'' \href{http://dx.doi.org/10.1088/0264-9381/19/9/306}{{\em Class. Quant. Grav.} {\bfseries 19} (2002) 2399--2424}, \href{http://arxiv.org/abs/gr-qc/0111049}{{\ttfamily arXiv:gr-qc/0111049}}.

\bibitem{krasnov2003holomorphic}
K.~Krasnov, ``On holomorphic factorization in asymptotically ads 3d gravity,'' {\em Classical and Quantum Gravity} {\bfseries 20} no.~18, (2003) 4015.

\bibitem{HawkingEllis}
S.~Hawking and G.~Ellis, {\em The Large Scale Structure of Space-Time}.
\newblock Cambridge Monographs on Mathematical Physics. Cambridge University Press, Cambridge, U.K., 1973.

\bibitem{duff1999ads3}
M.~J. Duff, H.~Lu, and C.~N. Pope, ``{AdS(3) x S**3 (un)twisted and squashed, and an O(2,2,Z) multiplet of dyonic strings},'' \href{http://dx.doi.org/10.1016/S0550-3213(98)00810-4}{{\em Nucl. Phys. B} {\bfseries 544} (1999) 145--180}, \href{http://arxiv.org/abs/hep-th/9807173}{{\ttfamily arXiv:hep-th/9807173}}.

\bibitem{Knapp}
A.~Knapp, {\em Lie Groups Beyond an Introduction}, vol.~140 of {\em Progress in Mathematics}.
\newblock Birkh\"auser, Boston, U.S.A., 2nd~ed., 2002.

\bibitem{Detournay:2005fz}
S.~Detournay, D.~Orlando, P.~M. Petropoulos, and P.~Spindel, ``{Three-dimensional black holes from deformed anti-de Sitter},'' \href{http://dx.doi.org/10.1088/1126-6708/2005/07/072}{{\em JHEP} {\bfseries 07} (2005) 072},
\href{http://arxiv.org/abs/hep-th/0504231}{{\ttfamily arXiv:hep-th/0504231 [hep-th]}}.
%%CITATION = HEP-TH/0504231;%%.

\bibitem{banados2006three}
M.~Banados, G.~Barnich, G.~Compere, and A.~Gomberoff, ``{Three dimensional origin of Godel spacetimes and black holes},'' \href{http://dx.doi.org/10.1103/PhysRevD.73.044006}{{\em Phys. Rev. D} {\bfseries 73} (2006) 044006}, \href{http://arxiv.org/abs/hep-th/0512105}{{\ttfamily arXiv:hep-th/0512105}}.

\bibitem{Deser:1981wh}
S.~Deser, R.~Jackiw, and S.~Templeton, ``{Topologically Massive Gauge Theories},''
\href{http://dx.doi.org/10.1016/0003-4916(82)90164-6}{{\em Annals Phys.} {\bfseries 140} (1982) 372--411}.
%%CITATION = APNYA,140,372;%%.

\bibitem{chow2010classification}
D.~D. Chow, C.~Pope, and E.~Sezgin, ``Classification of solutions in topologically massive gravity,'' {\em Classical and Quantum Gravity} {\bfseries 27} no.~10, (2010) 105001.

\bibitem{Bergshoeff_2009a}
E.~A. Bergshoeff, O.~Hohm, and P.~K. Townsend, ``Massive gravity in three dimensions,'' \href{http://dx.doi.org/10.1103/physrevlett.102.201301}{{\em Physical Review Letters} {\bfseries 102} no.~20, (May, 2009) }. \url{http://dx.doi.org/10.1103/PhysRevLett.102.201301}.

\bibitem{Bergshoeff_2009b}
E.~A. Bergshoeff, O.~Hohm, and P.~K. Townsend, ``More on massive 3d gravity,'' \href{http://dx.doi.org/10.1103/physrevd.79.124042}{{\em Physical Review D} {\bfseries 79} no.~12, (June, 2009) }. \url{http://dx.doi.org/10.1103/PhysRevD.79.124042}.

\bibitem{Andrianopoli:2023dfm}
L.~Andrianopoli, B.~L. Cerchiai, R.~Noris, L.~Ravera, M.~Trigiante, and J.~Zanelli, ``{New torsional deformations of locally AdS3 space},'' \href{http://dx.doi.org/10.1103/PhysRevD.108.044011}{{\em Phys. Rev. D} {\bfseries 108} no.~4, (2023) 044011}, \href{http://arxiv.org/abs/2305.17168}{{\ttfamily arXiv:2305.17168 [hep-th]}}.

\bibitem{PhysRevLett.133.031602}
L.~Andrianopoli, R.~Noris, M.~Trigiante, and J.~Zanelli, ``Supersymmetric states in anti--de sitter $\mathit{D}=3$ supergravity with chiral torsion,'' \href{http://dx.doi.org/10.1103/PhysRevLett.133.031602}{{\em Phys. Rev. Lett.} {\bfseries 133} (Jul, 2024) 031602}. \url{https://link.aps.org/doi/10.1103/PhysRevLett.133.031602}.

\bibitem{1965cngg.conf..222K}
R.~P. {Kerr} and A.~{Schild}, ``{A new class of vacuum solutions of the Einstein field equations},'' in {\em IV Centenario Della Nascita di Galileo Galilei, 1564-1964}, p.~222.
\newblock Jan., 1965.

\bibitem{Kerr:1965wfc}
R.~P. Kerr and A.~Schild, ``{Some algebraically degenerate solutions of Einstein\textquoteright{}s gravitational field equations},'' {\em Proc. Symp. Appl. Math.} {\bfseries 17} (1965) 199.

\bibitem{geiges2004contactgeometry}
H.~Geiges, ``Contact geometry.'' \url{https://arxiv.org/abs/math/0307242}, 2004.

\bibitem{Clement:1994sb}
G.~Clement, ``{Particle - like solutions to topologically massive gravity},'' \href{http://dx.doi.org/10.1088/0264-9381/11/9/001}{{\em Class. Quant. Grav.} {\bfseries 11} (1994) L115--L120}, \href{http://arxiv.org/abs/gr-qc/9404004}{{\ttfamily arXiv:gr-qc/9404004}}.

\bibitem{minguzzi2008causal}
E.~Minguzzi and M.~S{\'a}nchez, ``{The causal hierarchy of spacetimes},'' {\em {Recent developments in pseudo-Riemannian geometry}} {\bfseries 4} (2008) 299–358, \href{http://arxiv.org/abs/gr-qc/0609119}{{\ttfamily arXiv:gr-qc/0609119}}.

\bibitem{PhysRevD.75.125015}
M.~Ba\~nados, A.~T. Faraggi, and S.~Theisen, ``$\mathcal{N}=2$ supergravity in three dimensions and its g\"odel supersymmetric background,'' \href{http://dx.doi.org/10.1103/PhysRevD.75.125015}{{\em Phys. Rev. D} {\bfseries 75} (Jun, 2007) 125015}. \url{https://link.aps.org/doi/10.1103/PhysRevD.75.125015}.

\bibitem{Deger_2014}
N.~Deger, A.~Kaya, H.~Samtleben, and E.~Sezgin, ``Supersymmetric warped ads in extended topologically massive supergravity,'' \href{http://dx.doi.org/10.1016/j.nuclphysb.2014.04.011}{{\em Nuclear Physics B} {\bfseries 884} (July, 2014) 106–124}. \url{http://dx.doi.org/10.1016/j.nuclphysb.2014.04.011}.

\bibitem{OColgain:2015jlg}
E.~\'O~Colg\'ain, ``{Warped AdS$_3$, dS$_3$ and flows from $\mathcal{N} = (0,2)$ SCFTs},'' \href{http://dx.doi.org/10.1103/PhysRevD.91.105029}{{\em Phys. Rev. D} {\bfseries 91} no.~10, (2015) 105029}, \href{http://arxiv.org/abs/1501.04355}{{\ttfamily arXiv:1501.04355 [hep-th]}}.

\bibitem{ADMW}
A.~Aggarwal, S.~Detournay, A.~Somerhausen, and R.~Wutte, ``{Supersymmetry for Warped AdS$_3$ black holes ({\it Work in progress})},''.

\bibitem{Detournay:2019xgl}
S.~Detournay, W.~Merbis, G.~S. Ng, and R.~Wutte, ``{Warped Flatland},'' \href{http://dx.doi.org/10.1007/JHEP11(2020)061}{{\em JHEP} {\bfseries 11} (2020) 061}, \href{http://arxiv.org/abs/2001.00020}{{\ttfamily arXiv:2001.00020 [hep-th]}}.

\bibitem{Walker:1970}
M.~Walker, ``Block diagrams and the extension of timelike two-surfaces,'' \href{http://dx.doi.org/10.1063/1.1665393}{{\em J. Mathematical Phys.} {\bfseries 11} (1970) 2280--2286}. \url{https://doi.org/10.1063/1.1665393}.

\bibitem{chrusciel2020geometry}
P.~Chrusciel, {\em {Geometry of Black Holes}}.
\newblock International Series of Monographs on Physics. Oxford University Press, 4, 2023.

\bibitem{Bieliavsky:2003de}
P.~Bieliavsky, S.~Detournay, M.~Herquet, M.~Rooman, and P.~Spindel, ``{Global geometry of the 2+1 rotating black hole},'' \href{http://dx.doi.org/10.1016/j.physletb.2003.07.055}{{\em Phys. Lett.} {\bfseries B570} (2003) 231--236},
\href{http://arxiv.org/abs/hep-th/0306293}{{\ttfamily arXiv:hep-th/0306293 [hep-th]}}.
%%CITATION = HEP-TH/0306293;%%.

\bibitem{DNF}
S.~Doubrovine~B., Novikov and A.~Fomenko, {\em Géométrie contemporaine: Méthodes et applications, vol. 2}.
\newblock MIR, Moscow, 1979.

\bibitem{Banados:1992wn}
M.~Banados, C.~Teitelboim, and J.~Zanelli, ``{The black hole in three-dimensional space-time},'' \href{http://dx.doi.org/10.1103/PhysRevLett.69.1849}{{\em Phys.Rev.Lett.} {\bfseries 69} (1992) 1849--1851},
\href{http://arxiv.org/abs/hep-th/9204099}{{\ttfamily arXiv:hep-th/9204099 [hep-th]}}.
%%CITATION = HEP-TH/9204099;%%.

\bibitem{carlip19952+}
S.~Carlip, ``{The (2+1)-Dimensional black hole},'' \href{http://dx.doi.org/10.1088/0264-9381/12/12/005}{{\em Class. Quant. Grav.} {\bfseries 12} (1995) 2853--2880}, \href{http://arxiv.org/abs/gr-qc/9506079}{{\ttfamily arXiv:gr-qc/9506079}}.

\end{thebibliography}\endgroup
%\end{mcbibliography}
\end{document}